%% file: MAIN.tex
\def\huawei{0}

\if\huawei1
\documentclass[acmsmall,nonacm, review=false]{acmart}
\else
\documentclass[acmsmall, review=false]{acmart}
\fi

\usepackage{amsfonts}
\usepackage{array}
\usepackage{lipsum}
\usepackage{subcaption}
\usepackage{longtable}
\usepackage{xspace}
\usepackage{algorithm}
\usepackage[noend]{algpseudocode}
\usepackage{amsmath}

\usepackage{subcaption}
\usepackage{relsize}
\usepackage{xcolor}
\usepackage{mathtools}
\usepackage{multirow}

\newcolumntype{L}[1]{>{\raggedright\let\newline\\\arraybackslash\hspace{0pt}}m{#1}}
\newcolumntype{C}[1]{>{\centering\let\newline\\\arraybackslash\hspace{0pt}}m{#1}}
\newcolumntype{R}[1]{>{\raggedleft\let\newline\\\arraybackslash\hspace{0pt}}m{#1}}
\usepackage{multirow}

\newcommand{\APPR}{SAFE\xspace} %

\definecolor{mygreen}{rgb}{0.0, 0.2, 0.13}

\usepackage{marginnote}
\usepackage[backgroundcolor=white,bordercolor=blue,linecolor=blue]{todonotes}

\newcommand{\MINOR}[2]{#2}

\newcommand{\MAJORBEGIN}{\color{black}}
\newcommand{\MAJOREND}{\color{black}}

\newcommand{\MAJOR}[2]{#2}

\newcommand{\IEE}{IEE\xspace}

\newcommand{\CloseDNN}{OC\xspace}

\newcommand{\OC}{OC\xspace}

\newcommand{\EquationsSize}{\small}

\newcommand{\GD}{GD\xspace}
\newcommand{\HPD}{HPD\xspace}

\usepackage[pagewise]{lineno}

\usepackage{pifont}%

\begin{document}

\title{Supporting Safety Analysis of Image-processing DNNs through Clustering-based Approaches}

\author{Mohammed~Oualid~Attaoui}
\affiliation{%
  \institution{SnT Centre, University of Luxembourg}
  \streetaddress{JFK 29}
  \city{Luxembourg}
  \country{Luxembourg}}
\email{mohammed.attaoui@uni.lu}

\author{Hazem Fahmy}
\affiliation{%
  \institution{SnT Centre, University of Luxembourg}
  \streetaddress{JFK 29}
  \city{Luxembourg}
  \country{Luxembourg}}
\email{hazem.fahmy@uni.lu}

\author{Fabrizio Pastore}
\affiliation{%
  \institution{SnT Centre, University of Luxembourg}
  \streetaddress{JFK 29}
  \city{Luxembourg}
  \country{Luxembourg}}
\email{fabrizio.pastore@uni.lu}

\author{Lionel Briand}
\authornote{This work was done while affiliated with University of Luxembourg and University of Ottawa}
\affiliation{%
  \institution{Lero Centre, University of Limerick}
  \streetaddress{Tierney building}
  \city{Limerick}
  \country{Ireland}}
  \affiliation{%
  \institution{School of EECS, University of Ottawa}
  \city{Ottawa}
  \country{Canada}}
\email{lionel.briand@uni.lu}

\begin{abstract}

The adoption of deep neural networks (DNNs) in safety-critical contexts is often prevented by the lack of effective means to explain their results, especially when they are erroneous.
In our previous work, we proposed a white-box approach (HUDD) and a black-box approach (SAFE) to automatically characterize DNN failures. They both identify clusters of similar images from a potentially large set of images leading to DNN failures. However, the analysis pipelines for HUDD and SAFE were instantiated in specific ways according to common practices, deferring the analysis of other pipelines to future work.

In this paper, we report on an empirical evaluation of 99 different pipelines for root cause analysis of DNN failures. They combine transfer learning, autoencoders, heatmaps of neuron relevance, dimensionality reduction techniques, and different clustering algorithms. 
Our results show that the best pipeline combines transfer learning, DBSCAN, and UMAP.
It leads to clusters almost exclusively capturing images of the same failure scenario, thus facilitating root cause analysis. Further, it generates distinct clusters for each root cause of failure, thus enabling engineers to detect all the unsafe scenarios. Interestingly, these results hold even for failure scenarios that are only observed in a small percentage of the failing images. 

\end{abstract}

\ccsdesc[500]{Software and its engineering~Software defect analysis}
\ccsdesc[500]{Computing methodologies~Machine learning}

\keywords{DNN Explanation, DNN Functional Safety Analysis, DNN Debugging, Clustering, Transfer Learning}

\maketitle

\input{introduction.tex}

\input{background.tex}

\input{approach.tex}

\input{evaluation.tex}

\input{threats}

\input{related.tex}

\input{conclusion.tex}

\begin{acks}
This project has received funding from IEE Luxembourg, Luxembourg’s National Research Fund (FNR) under grant BRIDGES2020/IS/14711346/FUNTASY, and NSERC of Canada under the Discovery and CRC programs. Authors would like to thank Thomas Stifter from IEE for his valuable support. The experiments presented in this paper were carried out using the HPC facilities of the University of Luxembourg (see \url{http://hpc.uni.lu}).
\end{acks}

\bibliographystyle{ACM-Reference-Format}
\bibliography{BlackBoxExplanation}

\input{appendix}

\end{document}

%% file: introduction.tex
\section{Introduction}
Deep neural networks (DNNs) have achieved extremely high predictive accuracy in various domains, such as computer vision~\cite{attaoui2022regions, reddy2022deep}, autonomous driving~\cite{tian2018deeptest, li2021testing}, and natural language processing~\cite{nikhath2022intelligent, ekinci18poet}. Despite their superior performance, the lack of explainability of DNN models remains an issue in many contexts. While they can approximate complex and arbitrary functions, studying their structure often provides little or no insight into the underlying prediction mechanisms. There seems to be an intrinsic tension between Machine Learning (ML) performance and explainability. Often the highest-performing methods (for example, Deep Learning) are the least explainable, and the most explainable (for example, decision trees) are the least accurate~\cite{gunning2019darpa}. 

For DNNs to be trustworthy, in many critical contexts where they are used, we must understand why they behave the way they do~\cite{carvalho2019machine}. Explanation methods aim at making neural network decisions trustworthy~\cite{gilpin2018explaining}. Several explanation methods are proposed in the literature (see Section~\ref{sec:related}). In our work, because of our focus on safety analysis, we focus on explanation methods for root cause analysis, that is identifying the underlying reason of a DNN failure (root cause) which is, in our context, an incorrect DNN prediction. 
\MAJOR{R3.2}{More precisely, we aim to identify root causes in terms of characteristics of the input images leading to failures; in other words, we are interested in identifying the different scenarios in which the DNN may fail. Such characterization is the first step towards retraining the DNN. 
} 

Root cause analysis techniques based on unsupervised learning have proven their effectiveness~\cite{ter2022comparing, weitz2018towards}. These methods group failure samples (e.g., data collected during hardware testing) 
without requiring diagnostic labels, such that the samples in each cluster share similar root causes.

Our previous work is the first application of unsupervised learning to perform root cause analysis targeting DNN failures. Precisely, we proposed two DNN explanation methods: 
\emph{SAFE} (Safety Analysis based on Feature Extraction)~\cite{attaoui2022black} and \emph{HUDD} (Heatmap-based Unsupervised Debugging of DNNs)~\cite{fahmysupporting}. They both process a set of failure-inducing images and generate clusters of similar images. Commonalities across images in each cluster provide information about the root cause of the failure. 
\MAJOR{R3.2}{Further, the identified root causes support safety analysis because they help identify possible countermeasures to address the problem.}
For example, 
applying our approaches to failure-inducing images for a DNN that classifies car seat occupancy may include
a cluster of images with child seats containing a bag; such cluster may help engineers determine that bags inside child seats are likely to be misclassified.
\MAJOR{R3.2}{Possible countermeasures could be to retrain the DNN using more child seats with objects or, if it does not work, integrating additional components that make the approach safer (e.g., radar technology~\cite{Vitasense}).}
Both SAFE and HUDD also support the identification of additional images to be used to retrain the DNN.

HUDD and SAFE differ with respect to the kind of data used to perform clustering and the pipeline of steps they rely on.
HUDD applies clustering based on internal DNN information; precisely, for all failure-inducing images, it generates heatmaps capturing the relevance of DNN neurons on the DNN output. Finally, it applies a hierarchical clustering algorithm relying on a distance metric based on the generated heatmaps.
SAFE is black-box as it does not rely on internal DNN information. It generates clusters based on the visual similarity across failure inducing images. To this end it relies on feature extraction based on transfer learning, dimensionality reduction, and the DBSCAN clustering algorithm.

\emph{SAFE} and \emph{HUDD} rely on a pipeline that has been configured in specific ways according to best practices. However, several variants exist for each component of both approaches (e.g., different transfer learning models, different clustering algorithms).

In this paper, we aim to evaluate these pipeline variants for both SAFE and HUDD. Therefore, we propose an empirical evaluation of 99 alternative configurations for SAFE and HUDD (pipelines). These pipelines were obtained using different combinations of feature extraction methods, clustering algorithms, and dimensionality reduction techniques; in addition, 
we assessed the effect of fine tuning the transfer learning models used by feature extraction methods.
\MAJOR{R1.3}{Consistent with HUDD and SAFE, our pipelines support the characterization of DNNs tested at the level of models, not systems. Model-level testing, also called offline testing~\cite{Haq:EMSE:2021}, concerns testing DNN models in isolation, whereas system-level testing, also called online testing~\cite{Haq:EMSE:2021}, targets the system integrating the DNN (e.g., an autonomous driving system tested within a simulator~\cite{Haq:EMSE:2021}). Supporting system-level testing is part of future work.}

For our empirical evaluation we considered six case study subjects, two of which were provided by our industry partner in the automotive domain, IEE Sensing~\cite{IEE}. Our subjects' applications include head pose classification, eye gaze detection, drowsiness detection, steering angle prediction, unattended child detection, and car position detection.

We present a systematic and extensive evaluation scheme for these pipelines, which entails generating failure causes that resemble realistic scenarios (e.g., poor lighting conditions or camera misconfiguration). \MAJOR{R1.1}{Since in these scenarios the causes of failures are known a priori, such an evaluation scheme enables us to objectively analyze and evaluate the performance of pipelines while controlling the frequency of such failure scenarios.}

Our empirical results suggest that the best pipelines support and facilitate the process of functional safety analysis such that they 
1) can generate root-cause clusters (RCCs) that group together a very high proportion of images capturing a same root cause ($94.3\%$, on average), 2) can capture most of the root causes of failures for all case study subjects ($96.7\%$, on average),
and 3) 
\MINOR{R2.1}{ can perform well (i.e., are reliable) in the presence of rare failure instances in a dataset} 
(i.e., when some causes of failures affect less than 10\% of the failure-inducing images). 
\MAJOR{2.1a}{In our approach, the root causes of failures are determined by engineers after inspecting the identified clusters. Although such a solution still requires human involvement, it simplifies an engineer's task\footnote{In our previous work~\cite{attaoui2022black}, we conducted a user study demonstrating that the inspection of five random images per cluster is sufficient for an analyst to correctly identify the root causes of failures.}; indeed, it is unlikely that a human can manually identify similarities across a large set of images leading to DNN failures. Further, though our previous work (i.e., SEDE \cite{fahmy2023simulator}) aims to improve the degree of automation by automatically deriving expressions capturing commonalities in failure-inducing images, in this paper, we tackle an orthogonal problem:  assessing which pipelines lead to clusters with better purity and coverage. One possible future work is the integration of the best analysis pipeline with SEDE.} %

The remainder of this paper is organized as follows. In Section~\ref{sec:background}, we briefly present the main features and limitations of SAFE and HUDD, along with other feature extraction models (Autoencoders and Backpropagation-based Heatmaps). In Section~\ref{sec:approach}, we describe the different models and algorithms we use in our evaluated pipelines. In Section~\ref{sec:empirical}, we present the research questions, the experiment design and results, including a comparison between 99 pipelines. In Section~\ref{sec:related}, we discuss and compare related work. Finally, we conclude this paper in Section~\ref{sec:conclusion}.

%% file: background.tex
\section{Background}
\label{sec:background}
This section provides an overview of our previous work that inspired this research. We focus on clustering methods, heatmap-based DNN Explanations, the HUDD and SAFE DNN explanation methods, and Autoencoders.

\subsection{Clustering}
\label{sec:background:clustering}
Clustering is a data analysis method that mines essential information from a dataset by grouping data into several groups called \emph{clusters}. In clustering, similar data points are grouped into the same cluster, while non-similar data points are put into different clusters. There are two main objectives in data clustering; the first objective is to minimize the dissimilarity within the cluster, and the second objective is to maximize the inter-cluster dissimilarity. HUDD and SAFE rely on hierarchical agglomerative clustering (HAC~\cite{rao2021hierarchical}) and density-based clustering (DBSCAN~\cite{ester1996density}), respectively. 
In HAC, each observation starts in its own cluster and pairs of clusters are iteratively merged to minimize an objective function (e.g., error sum of squares~\cite{Ward}). DBSCAN works by considering dense regions as clusters; it is detailed in Section~\ref{sec:approach}.

\subsection{Heatmap-based DNN Explanations}

Approaches that aim to explain DNN results have been developed in recent years~\cite{GARCIA2018}. 
Most of these concern the generation of heatmaps that capture the importance of pixels in image predictions. They include black-box~\cite{Petsiuk2018rise,Dabkowski17} and white-box
approaches~\cite{Montavon2019,Selvaraju17,Zeiler14,DB15a,Zhou16}. 
Black-box approaches generate heatmaps for the input layer and do not provide insights regarding internal DNN layers.
White-box approaches rely on the backpropagation of the relevance score computed by the DNN~\cite{Montavon2019,Selvaraju17,Zeiler14,DB15a,Zhou16}.

In this Section, we focus on a white-box technique called  Layer-Wise Relevance Propagation (LRP)~\cite{Montavon2019} because it has been integrated into HUDD.
LRP was selected because it does not present the shortcomings of other heatmap generation approaches~\cite{fahmysupporting}.

\begin{figure}[tb]
\includegraphics[width=0.8\textwidth]{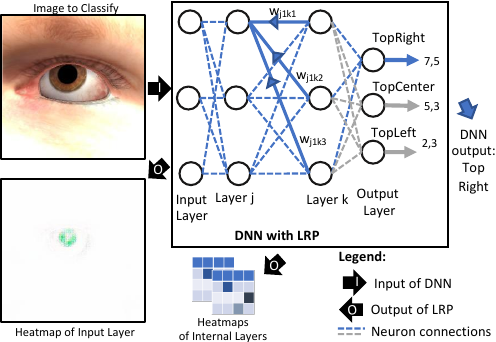}
\caption{Layer-Wise Relevance Propagation.}
\label{fig:LRP}
\end{figure}

LRP redistributes the relevance scores of neurons in a higher layer to those of the lower layer. 
Figure~\ref{fig:LRP} illustrates how LRP operates on a fully connected network used to classify inputs. 
In the forward pass, the DNN receives an input and generates an output (e.g., classifies the gaze direction as TopLeft) while recording  the activations of each neuron.
In the backward pass, LRP generates \emph{internal heatmaps} for a DNN layer $k$, which consists of a matrix with the relevance scores computed for all the neurons of layer $k$.

\begin{figure}[tb]
    \includegraphics[width=0.7\textwidth]{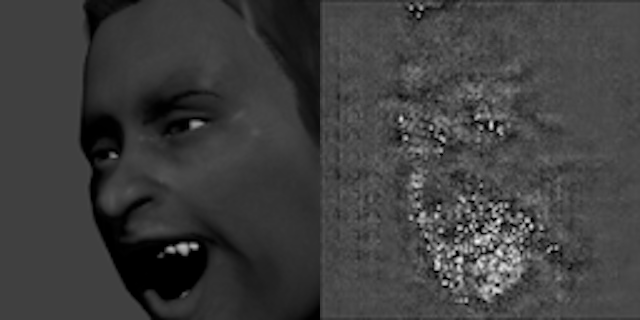}
    \caption{An example image of HPD subject (on the left) and applied LRP (on the right) showing that the mouth had a large influence on the DNN behavior.  }
    \label{fig:hpd_lrp}
\end{figure}

The heatmap in Figure~\ref{fig:LRP} shows that the pupil and part of the eyelid, which are the non-white parts in the heatmap, had a significant effect on the DNN output. Furthermore, the heatmap in Figure~\ref{fig:hpd_lrp} shows that the mouth and part of the nose are the input pixels that mostly impacted on the DNN output.

A heatmap is a matrix with entries in $\mathbb{R}$, i.e., it is a triple $(N,M,f)$ where $N,M \in \mathbb{N}$ and $f$ is a map $[N] \times [M] \rightarrow \mathbb{R}$. We use the syntax $H[i,j]_x^L$ to refer to an entry in row $i$ (i.e., $i < N$) and column j (i.e., $j < M$) of a heatmap $H$ computed on layer $L$ from an image $x$. The size of the heatmap matrix (i.e., the number of entries) is $N \cdot M$, with $N$ and $M$ are determined by the dimensions of the DNN layer $L$. For convolution layers, $N$ represents the number of neurons in the feature map, whereas $M$ represents the number of feature maps. For example, the heatmap for the eighth layer of AlexNet has size $169 \times 256$ (convolution layer), while the the heatmap for the tenth layer has size $4096 \times 1$ (linear layer).

\subsection{Heatmap-based Unsupervised Debugging of DNNs (HUDD)}
 \label{sec:background:hudd}

\begin{figure}[H]
\includegraphics[width=0.8\textwidth]{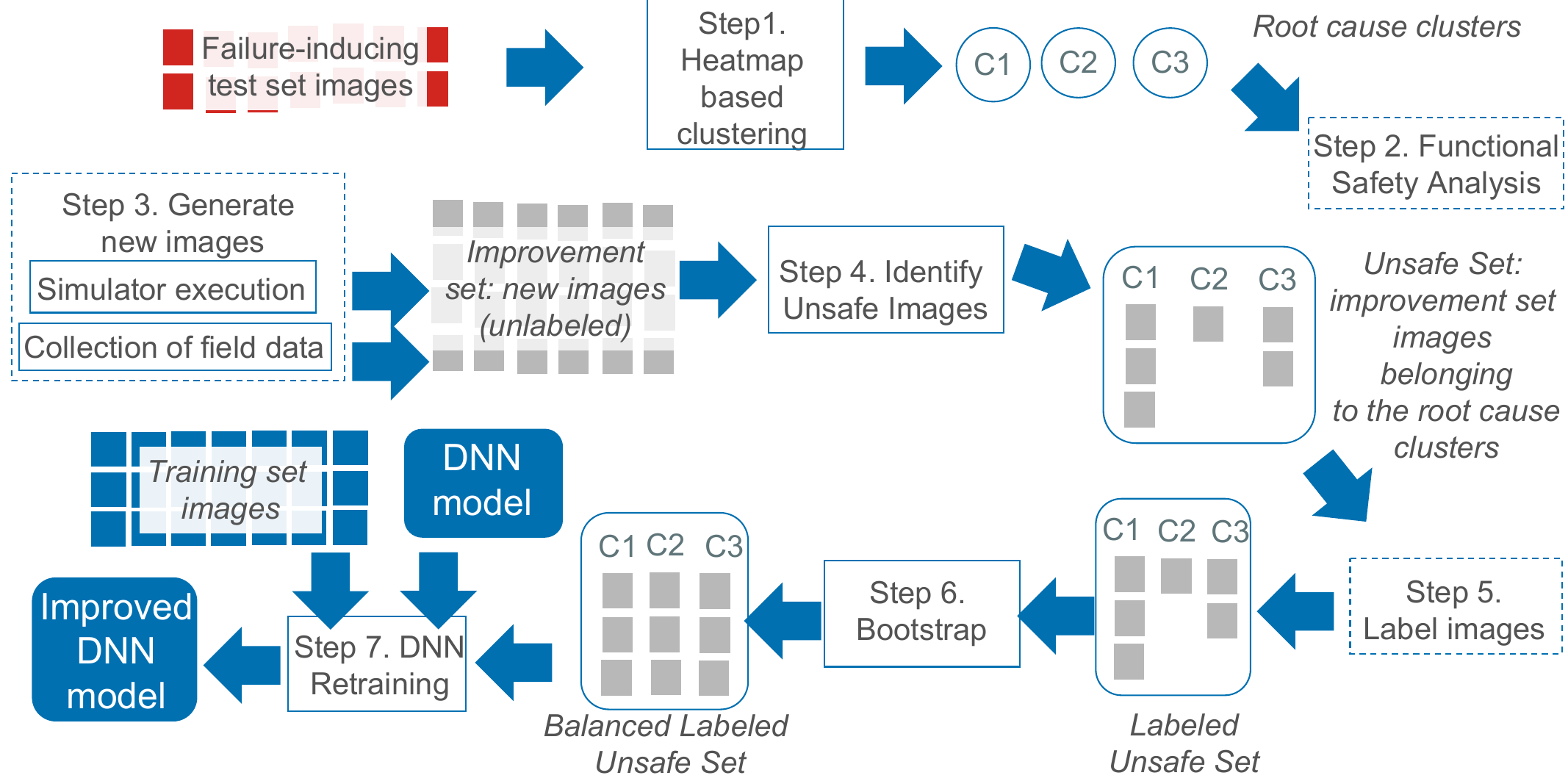}
\caption{Overview of HUDD.}
\label{fig:HUDD}
\end{figure}

Although heatmaps may provide useful information to determine the characteristics of an image that led to an erroneous result from the DNN, they are of limited applicability because, to determine the cause of all DNN errors observed in the test set, engineers may need to visually inspect all the error-inducing images, which is practically infeasible. To overcome such limitations, we recently developed HUDD~\cite{fahmysupporting}, a technique that facilitates the explanation and removal of the DNN errors observed in a test set. HUDD generates clusters of images that lead to a DNN error because of the same root cause.
The root cause is determined by the engineer who visualizes a subset of the images belonging to each cluster and identifies the commonality across each image (e.g., for a Gaze detection DNN, all the images present a closed eye).
To further support DNN debugging, HUDD automatically retrains the DNN by selecting a subset from a pool of unlabeled images that will likely lead to DNN errors because of the same root causes observed in the test set.

Figure~\ref{fig:HUDD} provides an overview of HUDD, which consists of six steps. In Step 1, root cause clusters are identified by relying on a hierarchical clustering algorithm applied to heatmaps generated for each failure inducing image.  Step 2 involves a visual inspection of clustered images. 
In this step, engineers visualize a few representative images for each RCC; the inspection enables the engineers to determine which are the commonalities across the images in each cluster and, therefore, determine the failure root cause. Example root causes include the presence of an object inside a child seat (as reported in the Introduction) or a face turned left thus making an eye not visible and causing misclassification in a gaze detection system. HUDD's Step 2 supports functional safety analysis because each failure root cause represents a usage scenario in which the DNN is likely to fail, and, based on domain knowledge, engineers can determine the likelihood of each failure scenario, its safety impact, and possible countermeasures, as required by functional safety analysis standards \cite{ISO24765, ISO26262}. For example, objects inside child seats might be very common but they lead to false alarms not hazards; misclassified gaze may instead prevent the system from determining that the driver is not pay attention to the road. Countermeasures include the retraining of the DNN, which is supported by HUDD's Step 3.
In Step 3, a new set of images, referred to as the \textit{improvement set}, is provided by the engineers to retrain the model. In Step 4, HUDD automatically selects a subset of images from the improvement set called the \textit{unsafe set}. The engineers label the images in the unsafe set in Step 5. Finally, in Step 6, HUDD automatically retrains the model to enhance its prediction accuracy.

\paragraph{\textbf{Heatmap-based Clustering in HUDD}} Clustering based on heatmaps is a key component of HUDD, an its functioning is useful to understand some of the pipelines considered in this paper. HUDD relies on LRP to generate an heatmap for every internal layer of the DNN, for each failure-inducing image. However, since distinct DNN layers lead to entries defined on different value ranges~\cite{MONTAVON2017DTD}, to enable the comparison of clustering results across different layers, we generate normalized heatmaps by relying on min-max normalization~\cite{DataMiningBook}.

For each DNN layer $L$, a distance matrix is constructed using the generated heatmaps; it captures the distance between every pair of failure-inducing image in the test set. 
The distance between a pair of images $\langle a,b \rangle$, at layer $L$, is computed as follows:
\begin{equation}
\EquationsSize
\mathit{heatmapDistance}_L(a,b)=\mathit{EuclideanDistance}(\tilde{H}^L_a,\tilde{H}^L_b)
\end{equation}
where $\tilde{H}^L_x$ is the heatmap computed for image $x$ at layer $L$. 
$\mathit{EuclideanDistance}$ is a function that computes the euclidean distance between two $N \times M$ matrices according to the formula 
\begin{equation}
\EquationsSize
\mathit{EuclideanDistance}(A,B)=\sqrt{ \sum_{i=1}^{N} \sum_{j=1}^{M}  (A_{i,j} - B_{i,j})^{2} }
\end{equation}
where $A_{i,j}$ and $B_{i,j}$ are the values in the cell at row $i$ and column $j$ of the matrix.

HUDD applies the HAC clustering algorithm multiple times, once for every DNN layer. 
For each DNN layer, HUDD selects the optimal number of clusters using the knee-point method applied to the weighted average intra-cluster distance ($\mathit{WICD}$).
$\mathit{WICD}$ is defined according to the following formula:
\begin{equation}
\EquationsSize
\label{eq:WICD}
\mathit{WICD}(L_l)=\frac{\sum^{|L_l|}_{j=1}\bigg( ICD(L_l,C_j)*\frac{|C_j|}{|C|} \bigg) }{|L_l|} 
\end{equation}
where $L_l$ is a specific layer of the DNN, $|L_l|$ is the number of clusters in the layer $L_l$, $ICD$ is the intra-cluster distance for cluster $C_i$ belonging to layer $L_l$, 
$|C_j|$ represents the number of elements in cluster $C_j$, whereas $|C|$ represents the number of images in all the clusters.

In Formula~\ref{eq:WICD}, $\mathit{ICD}(L_l,C_j)$ is computed as follows:
\begin{equation}
\EquationsSize
\label{eq:ICD}
\mathit{ICD}(L_l,C_j)=\frac{\sum^{N_j}_{i=0}\mathit{heatmapDistance}_{L_{l}}(p^a_i,p^b_i)}{N_j}
\end{equation}
where $p_i$ is a unique pair of images in cluster $C_j$, and $N_j$ is the total number of pairs it contains. The superscripts $a$ and $b$ refer to the two images of the pair to which the distance formula is applied.

HUDD then select the layer $L_m$ with the minimal $\mathit{WICD}$. 
By definition, the clusters generated for layer $L_m$ are the ones that maximize cohesion and we therefore anticipate that they will group together images that exhibit similar characteristics.

In our study, we rely on HUDD as a feature extraction method; precisely, we use the heatmaps generated by the layer selected by HUDD as features.

\subsection{Safety Analysis based on Feature Extraction (SAFE)}

\begin{figure}[b]
\includegraphics[width=0.8\textwidth]{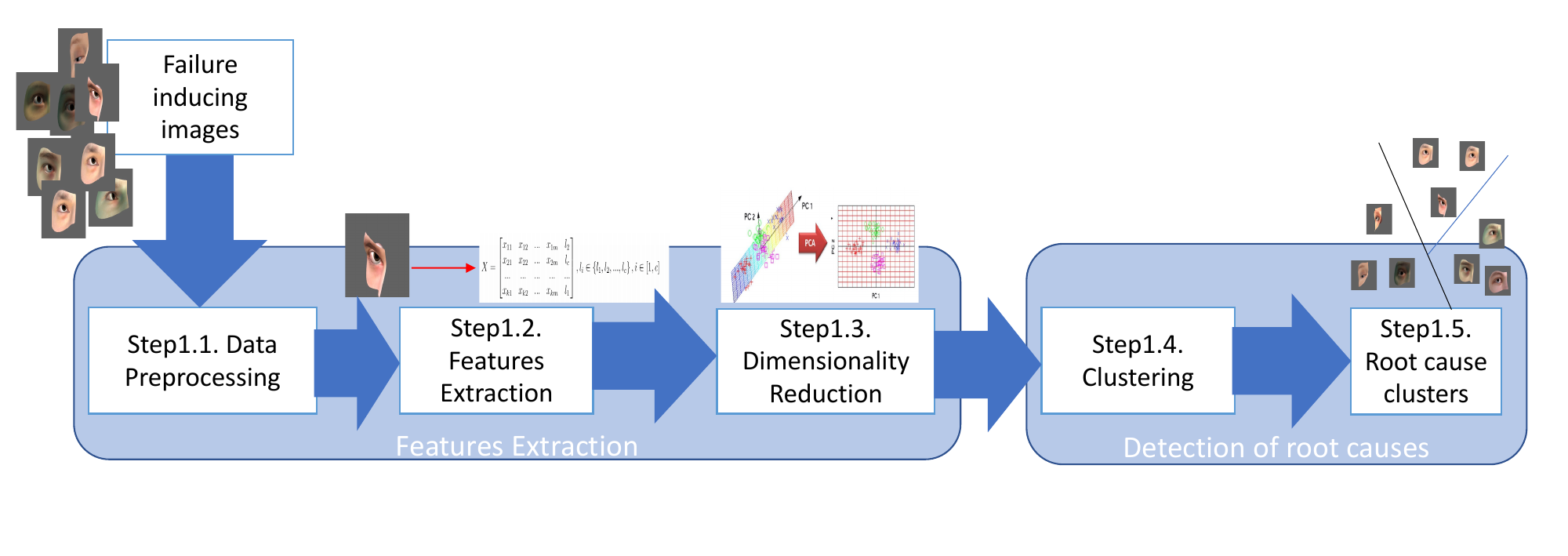}
\caption{Generation of root cause clusters with \APPR}
\label{fig:step1}
\end{figure}

SAFE is based on a combination of a transfer learning-based feature extraction method, a clustering algorithm, and a dimensionality reduction technique.
The workflow of SAFE matches HUDD's, except for Step 1 and Step 4. In SAFE's Step 1 RCCs are identified by relying on non-convex clustering (DBSCAN) applied to features extracted from failure-inducing images; HUDD, instead, applies hierarchical clustering to heatmaps. In Step 4, SAFE selects the improvement step using a procedure that relies on DBSCAN's outputs.

The pipelines evaluated in this paper had been inspired by the pipeline implemented in SAFE's Step 1, which consists of three stages (see Figure \ref{fig:step1}): \emph{Feature Extraction}, \emph{Dimensionality Reduction}, and \emph{Clustering}. In this paper we investigate variants of the SAFE pipeline using different combinations of these components. Additionally, we introduce a fine-tuning stage where we fine-tune the pre-trained transfer learning models to generate more domain-specific models. Excluding clustering, which was introduced in Section~\ref{sec:background:clustering}, the components of SAFE's pipeline are briefly described below.

\subsubsection{Transfer Learning and Feature Extraction}
\label{sec:back:FeatureExtraction}
To maximize the accuracy of image-processing DNNs in a cost-effective way, engineers often rely on the transfer learning approach, which consists of transferring knowledge from a generic domain, usually ImageNet~\cite{IMAGENET}, to another specific domain, (e.g., safety analysis, in our case). In other terms, we try to exploit what has been learned in one task and  generalize it to another task. Researchers have demonstrated the efficiency of transfer learning from ImageNet to other domains \cite{talo2019automated}. %

Transfer learning-based \emph{Feature Extraction} is an efficient method to transform unstructured data into structured raw data to be exploited by any machine learning algorithm. In this method, the features are extracted from images using a pre-trained CNN model~\cite{dif2021transfer}. 

The standard  CNN architecture comprises three types of layers: convolutional layers, pooling layers, and fully connected layers.
The convolutional layer is considered the primary building block of a CNN. This layer extracts relevant features from input images during training. Convolutional and pooling layers are stacked to form a hierarchical feature extraction module. The CNN model receives an input image of size $(224,224,3)$. This image is then passed through the network's layers to generate a vector of features. The feature extraction process, for each image, generates raw data represented by a $2D$ matrix (denoted as $X$) formalized below:  
	{
		\begin{equation}
			X = \begin{bmatrix}
				x_{11} & x_{12}  & ...  &   x_{1m} & l_1\\ 
				x_{21} &  x_{22}  & ...  & x_{2m} & l_2 \\ 
				... & ...   & ...   & ...  & ... \\ 
				x_{k1} & x_{k2}  & ...  & x_{km} & l_k \\
			\end{bmatrix}, l_i \in \left \{ C_1,C_2,...,C_{c}  \right \} 
		\end{equation}
	}
	where $C_i$ represent the class categories, $c$ is the number of categories, $m = N \times N $ is the number of features, and $k$ is the size of the dataset. 

SAFE uses the VGG16 model pre-trained on the ImageNet dataset as a feature extraction method.

\subsubsection{Dimensionality Reduction}
Dimensionality reduction aims at approximating data in high-dimensional vector spaces \cite{gorban2010principal}. It is important in our context since we extract a high number of features from the images (512 to 2048). In SAFE, we used the Principal Component Analysis (PCA) dimensionality reduction method to reduce the number of features from 2048 to 100.

\subsection{Autoencoders}
\label{bg:ae}
Autoencoders (AE) are unsupervised artificial neural networks that learn how to compress and encode the data before reconstructing it from the compressed encoded version to a representation that resembles the original input as much as possible. AEs can extract features that can be used to improve downstream tasks, such as clustering or supervised learning, that benefit from dimensionality reduction and higher-level features. In other words, AEs try to learn an approximation to the identity function and, by placing various constraints on the network’s architecture and activation functions, they extract useful representations~\cite{forest2021deep}. 

\begin{figure}[H]
 \centering
 \includegraphics[width=.8\textwidth]{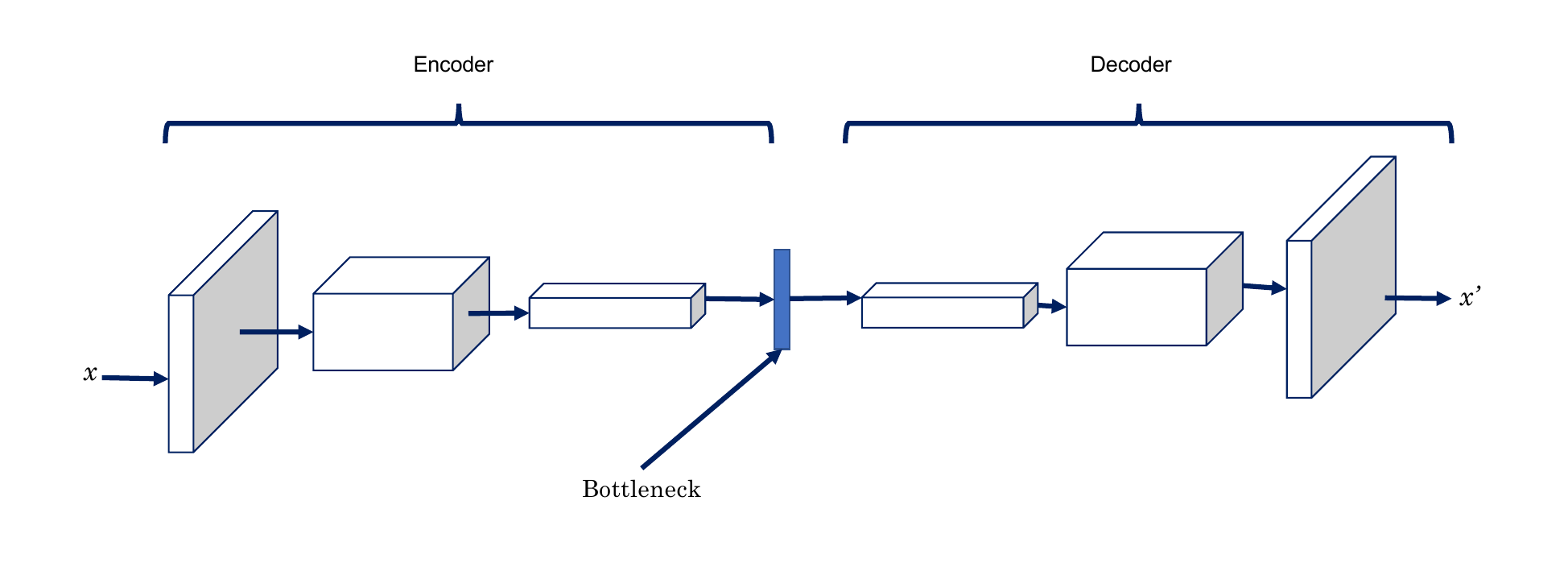}
 \caption{Autoencoder Architecture}\label{fig:autoencoder}
 \end{figure}

Figure~\ref{fig:autoencoder} illustrates the neural network architecture of a simple AE. It consist of four main components:
\begin{itemize}
    \item \textbf{Encoder:} learns how to compress the input data and reduce its dimensions into an encoded representation.
    \item \textbf{Bottleneck:} contains the encoded representation of the input data (i.e., the extracted features vector).
     \item \textbf{Decoder:} reconstructs the input data from the encoded version (retrieved from the \emph{Bottleneck}) such that it resembles the original input data as much as possible.
     \item \textbf{Reconstruction Loss:} the difference between the \emph{Encoder}'s input and the reconstructed version (the \emph{Decoder}'s output). The objective is to minimize such loss during training.
\end{itemize}
The objective of an AE's training process is to minimize its reconstruction loss, measured as either the mean-squared error or the cross-entropy loss between original inputs and its constructed inputs.

%% file: approach.tex
\section{The proposed pipelines}
\label{sec:approach}
\begin{figure}[h!]
\includegraphics[width=\textwidth]{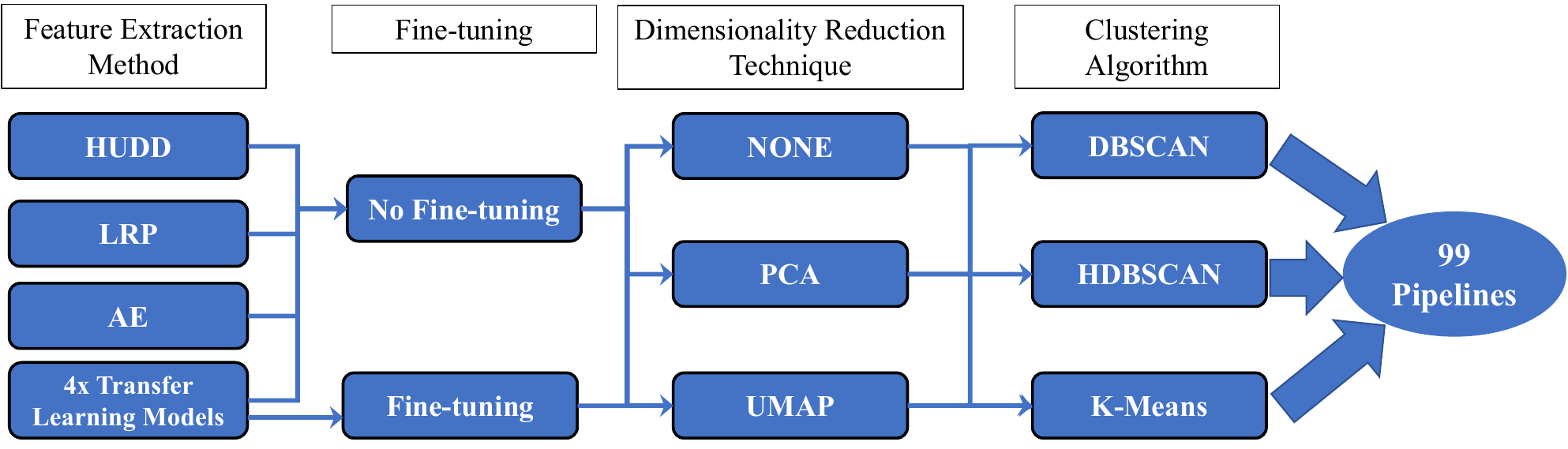}
\caption{Pipelines evaluated in our experiments.}
\label{fig:pipelines}
\end{figure}

This section presents the different pipelines that can be used to implement variants of SAFE and HUDD. The evaluated pipelines differ from the original SAFE and HUDD variants with respect to four components: Feature Extraction, Dimensionality Reduction, Clustering, and Fine-Tuning.
Each pipeline is a combination of a feature extraction method (FE), a dimensionality reduction technique (DR), and a clustering algorithm (CA). When feature extraction is based on transfer learning, we distinguish between models that are fine-tuned and not fine-tuned  (FT/NoFT); feature extraction approaches not based on transfer learning cannot be fine-tuned.
We refer to each pipeline with the pattern \emph{FE/\{FT,NoFT\}/DR/CA}, with each keyword being replaced with the name of the selected method.
We depict in 
Figure~\ref{fig:pipelines}
all the pipelines evaluated in our study; the different components are described in the following sections. 

\subsection{Feature Extraction}
\subsubsection{Feature Extraction based on Transfer Learning} 
Several DNN architectures to extract features based on transfer learning have been proposed:
Inception-V3~\cite{szegedy2015going}, VGGNet~\cite{simonyan2014very}, ResNet-50~\cite{ he2016deep}, and Xception~\cite{chollet2017xception}. These DNNs were trained on ImageNet~\cite{deng2009imagenet}, which is a dataset with more than 14 millions annotated images.
The number of extracted features depends on the selected DNN architecture; Inception-V3, VGGNet-16, ResNet-50, and Xception generate $2048$, $512$, $2048$, and $2048$ features, respectively. 
They are described in the following paragraphs.

\begin{itemize}

\item  \textbf{VGG-16:}
VGG-16 is a Convolution Neural Network (CNN) architecture and the winner of the ILSVR (Imagenet) competition in 2014. VGG-16 focuses on convolution layers of $3 \times 3$ filters with a stride of $1$ and always uses the same padding and maxpooling layer of $2 \times 2$ with a stride of $2$ instead of having a large number of hyper-parameters. It follows this arrangement of convolution and max pool layers consistently throughout the whole architecture. VGG-16 has two fully connected layers followed by a softmax layer as an output. The network has an image input size of $224 \times 224$.

\item \textbf{ResNet-50:} ResNet \cite{ he2016deep} is a CNN based on residual blocks. This architecture aims to solve vanishing gradient problems in deep neural networks. During the backpropagation process, the gradient diminishes dramatically in deep networks. Small values of gradients prevent the weights from changing their values, which slows the training process. To solve this issue, ResNet introduces residual blocks. These building blocks present skip connections between the previous convolutional layer's input and the current convolutional layer's output. Similar to VGG-16, the network has an image input size of $224 \times 224$. 

\item \textbf{Inception-V3:} Inception-V3 is a refined version of Inception \cite{szegedy2016rethinking}. This network proposes additional variants of Inception blocks to reduce the number of multiplications in the convolution and minimize computational complexity. These variants are based on two factorizations: factorization into smaller convolutions and factorization into asymmetric convolutions. The network has an image input size of $224 \times 224$.  

\item \textbf{Xception:} Xception is a pre-trained CNN that is $71$ layers deep and can classify images into $1,000$ different classes such as animals, objects, and humans. This allowed the model to learn various feature representations for a wide range of images.
The Xception's input size is $299 \times 299$.
\end{itemize}

\subsubsection{Fine-tuning}
   \MAJOR{R1.4}{Fine-tuning is a strategy to improve the performance of pre-trained models~\cite{fine-tuning-ref1}. It aims to benefit from the knowledge gained from a source task and generalize it to a target task. 
    Fine-tuning requires a pre-trained model to be prepared for reuse (i.e., the final layers may be removed and replaced with more appropriate ones), and configured to enable knowledge transfer. Knowledge transfer is achieved by freezing} the shallow layers (close to the input), which learn more generic features (edges, shapes, and textures), and retraining the deeper layers (i.e., we let the DNN algorithm update the weights of the layers close to the output), which learn more specific features from the input data~\cite{fine-tuning-ref1}. 
   
   To fine-tune a pretrained DNN model,
   we follow four steps:

\begin{enumerate}
    \item  Create a new model whose layers (along with their weights) are cloned from the pre-trained model, except for the output layer.
    \item Add a new fully connected output layer with a number of outputs equal to the number of classes in the target dataset, and initialize its weights with random values.
    \item Freeze shallow layers in the network, which are responsible for the feature extraction process (to guarantee that all the important features, previously learned by the pre-trained model, are not eliminated).
    \item Start training the new model on the target dataset, where the weights of all the non-frozen layers will keep updating using the backpropagation process. For the termination criterion, we use 100 epochs or until the loss stops improving (whichever criterion is met first).
\end{enumerate}

\subsubsection{Feature Extraction based on Autoencoders}

As explained in Section \ref{bg:ae}, 
an \emph{Encoder} plus a \emph{Decoder} make up an autoencoder (AE).
The input is compressed by the \emph{Encoder}, and the \emph{Decoder} reconstructs the input using the \emph{Encoder}'s compressed version (at the \emph{Bottleneck}).

Since AEs extract only the few input features necessary to aid the reconstruction of the output, the encoder might ignore other features which are not prioritized. For example in case of face images, the AE can discard the color of the skin because it is a non-prioritized feature to the AE.

However, the encoder often learns useful properties of the data~\cite{Goodfellow-et-al-2016}.
The model can then receive input data from any domain, and a fixed-length feature vector obtained at the \emph{Bottleneck} can be used for clustering. Such a vector offers a compressed version of the input data representation containing sufficient information about this data.

\subsubsection{Feature Extraction based on Heatmaps}

In our work, we rely on heatmaps as an additional method for feature extraction. Since heatmaps represent the  relevance of each neuron on DNN outputs, failure-inducing inputs sharing 
the same underlying cause should show similar heatmaps.

Precisely, we rely on two different methodologies for extracting features using heatmaps, we refer to them as \emph{LRP} and \emph{HUDD}, according to the name of the technique driving feature extraction.
LRP and HUDD have been introduced in Section~\ref{sec:background}. Feature extraction based on LRP, which generates heatmaps for internal layers but does not integrate a mechanism to select the most informative layer,  considers the heatmap computed by the LRP technique for the input layer. Feature extraction based on HUDD, instead, considers the heatmap generated for the DNN internal layer selected by HUDD as the best for clustering.

\subsection{Dimensionality Reduction}
\label{sec:dim_red}
Several dimensionality reduction techniques exist in the literature. In this paper, we rely on two state-of-the-art techniques: Principal Component Analysis (PCA)~\cite{pearson1901liii} and Uniform Manifold Approximation and Projection (UMAP)~\cite{mcinnes2018umap-software}. PCA is used for its simplicity of implementation and because it does not require much time and memory resources. UMAP is used for its effectiveness when applied before clustering. UMAP groups data points based on relative proximity, which optimizes the clustering results. PCA and UMAP are described below.

\subsubsection{Principal Component Analysis (PCA)}

To reduce dimensionality, PCA creates a 2-dimensional matrix of variables and observations. Then, for this matrix, it constructs a variable space with a dimension corresponding to the number of variables available. Finally, it projects each data point onto the first few maximum variance directions in the variable space. This procedure allows PCA to obtain a lower-dimensional data representation while maximizing data variation. 
The first principal component can equivalently be defined as the direction that maximizes the variance of the projected data~\cite{shlens2014tutorial}. 
In our context, we reduce the features for all our evaluated pipelines to $10$ components. We empirically obtained this number in a preliminary investigation conducted with one of our case study subjects (i.e., HPD). Precisely, we executed a clustering algorithm (K-means) multiple times; each execution was performed with a set of features obtained by applying PCA with a different number of components. We evaluated all the clustering solutions using the Silhouette Index~\cite{rousseeuw1987silhouettes} (see Section~\ref{sec:clusteringAlgo}) and chose the number of components yielding the highest index value.

\subsubsection{Uniform Manifold Approximation and Projection (UMAP)} 

Uniform Manifold Approximation and Projection (UMAP) is a dimension reduction technique that can be used for visualization but also for general non-linear dimension reduction. UMAP is fast, and scaling well in terms of both dataset size and dimensionality. The main limitation of UMAP is that it doesn't preserve the density of the data, which is, instead, better preserved by PCA.

First, UMAP forms a weighted graph representation between each pair of data points, where the edge weights are the probability of two data points being connected to each other. This graph is obtained by extending a radius outward each data point such that two data points are connected if their radii overlap. 
However, since an underestimation of such a radius can lead to the generation of small, isolated clusters, and its overestimation can lead to connecting all data points together, UMAP selects such a radius locally. The radius selection is performed based on the distance from each data point to its '$n-th$' neighbor. 
Finally, UMAP decreases the likelihood of two data points getting connected past the first neighbor (as the radius grows larger), which preserves the balance between the high-dimensional and low-dimensional representations. Once the high-dimensional graph is constructed, UMAP optimizes the layout of a low-dimensional representation to be as similar as possible. The general idea is to initialize the low-dimensional data points and then move them around until they form clusters that have the same structure as the high-dimensional data, preserving the connectedness of the data points. UMAP calculates Similarity Scores (distances) in the high dimensional graph to help identify the clustered points and tries to preserve that clustering in the low dimensional graph.

\MAJOR{R1.5}{Since UMAP can keep the structure of the data, even in a 2-dimensional (2D) space, we reduce the number of features to two. 
We ran an ablation study where we evaluated clustering results obtained for one of our case study subjects (HPD presented in Section~\ref{sec:subj}), considering both the original high-dimensional feature space and its reduced 2D counterpart. The average silhouette index, which we employ to measure the quality of the clusters, was similar in both cases. Such consistency indicates that the 2D representation sufficiently captures the relevant structure of the data, yielding cluster qualities comparable to those achieved in higher dimensions.
}

\subsection{Clustering algorithms}
\label{sec:clusteringAlgo}

In this study, we rely on three well-known clustering algorithms, K-means~\cite{Macqueen67}, DBSCAN~\cite{ester1996density}, and HDBSCAN~\cite{mcinnes2017hdbscan} described below. These three clustering algorithms were chosen after preliminary experiments including also the Hierarchical Agglomerative Clustering (HAC) \cite{Murtagh2014} and the Mean Shift algorithm \cite{fukunaga1975estimation}. \MAJOR{R1.8}{For our study, we selected algorithms that were already integrated in HUDD and SAFE (i.e., HAC and DBSCAN) but also natural extensions of SAFE (i.e., relying on HDBSCAN) and mean-based algorithms (i.e., K-means and Mean Shift).}
When generating clusters for one of our subjects (i.e., HPD, see Section~\ref{sec:subj}), HAC and Mean Shift yielded much lower values of the Silhouette Index than the DBSCAN, HDBSCAN, and K-means algorithms; therefore, we discarded HAC and Mean Shift from our selection.
Since a clustering algorithm may require the manual selection of  parameters' values, such as the number of clusters (K-means) or the minimum distance between data points (DBSCAN), we rely on an internal evaluation metric (the Silhouette Index~\cite{rousseeuw1987silhouettes}) and the \emph{knee-point method}~\cite{SatopaaKNEE11} to automate the selection of such values.

The \emph{Silhouette Index} is a standard practice in cluster analysis that maximizes cohesion (i.e., how closely related objects are in a cluster) and separation (i.e., how well-separated a cluster is from other clusters). 

The \emph{knee-point method} automates the \emph{elbow method} heuristics~\cite{Thorndike1953} by fitting a spline to the raw data using univariate interpolation, normalizing min/max values of the fitted data, and selecting the knee-points at which the curve most significantly deviates from the straight line segment that connects the first and last data point. We rely on the \emph{knee-point method} to automatically select the optimal number of clusters for the K-means algorithm.

\subsubsection{K-means:}
\label{sec:clust:kmeans}
 K-means is a well-known clustering algorithm. It takes a number $K$ as input and divides the data into $K$ clusters based on the distance calculated from the data points to the center of the cluster. This algorithm's main function is to minimize the distance between the data points and their cluster center as much as possible. In the original K-means algorithm, the number of clusters ($K$) is set manually, which can affect the quality of the clusters since we don't have any prior knowledge of the data (i.e., in our context, engineers cannot know in advance how many root causes of failures should be identified). 
 
\begin{figure}[t]
    \includegraphics[width=0.7\textwidth]{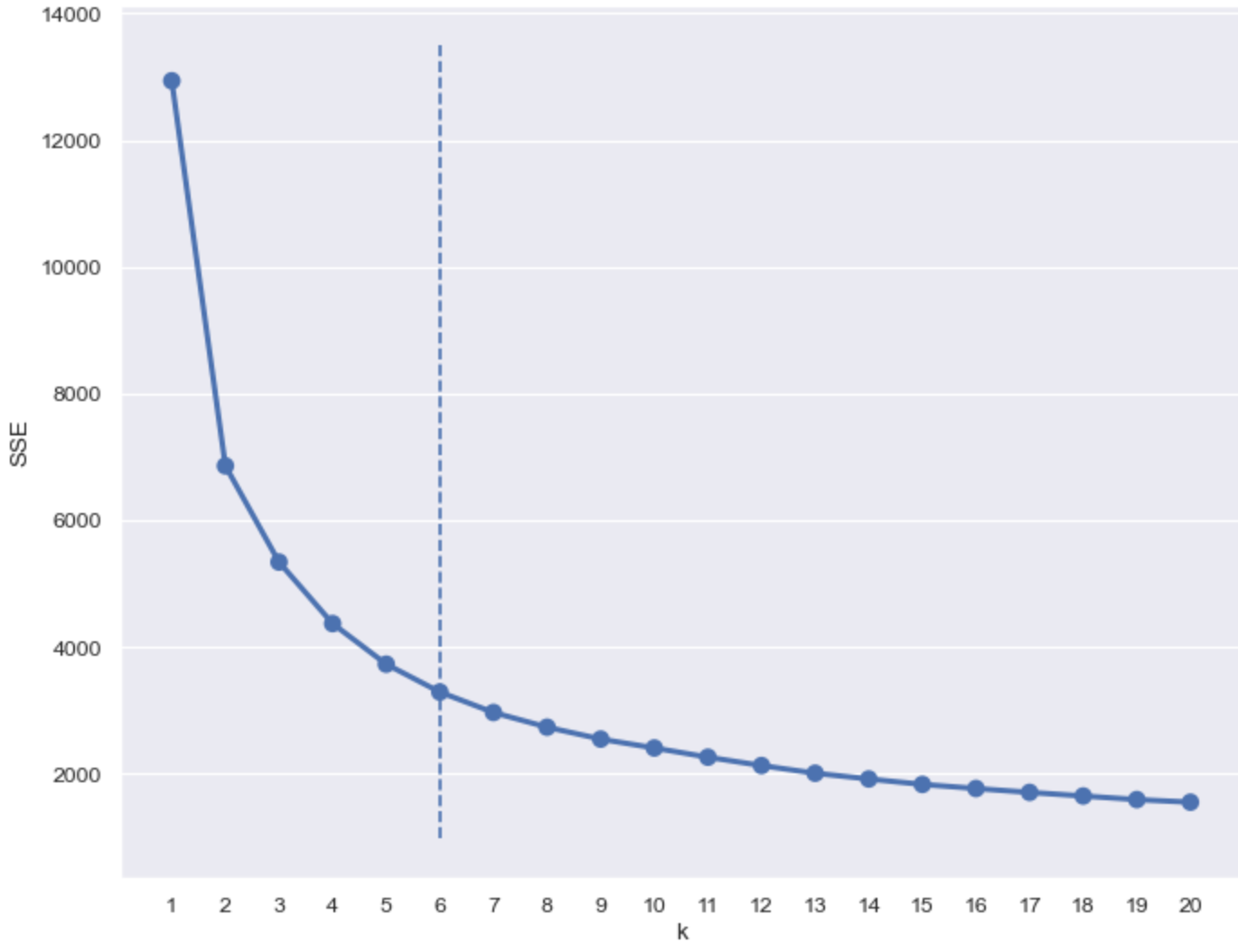}
    \caption{Approximating the optimal number of clusters $K$ using the Knee-point method. In this case the optimal $K$ is equal to 6. }
    \label{approx_k}
\end{figure}

To select an optimal value of $K$, we rely on the knee-point method. Precisely,
we cluster the data with different values of $K$ (in our evaluation, we consider the range $[5-35]$). For each clustering result, we compute the within-cluster sum of squared errors (SSD), which is the sum of the distances of each point to its cluster center. We then apply the knee-point approach to these SSDs and their respective $K$. 
 Figure \ref{approx_k} shows an example of $K$-approximation using the knee-point method.

\subsubsection{DBSCAN:} 
DBSCAN (Density-Based Spatial Clustering of Applications with Noise)~\cite{ester1996density}, is an algorithm that defines clusters using local density estimation. It can be divided into four steps: 
\begin{enumerate}

\item The $\epsilon$-neighborhood of a data point is determined as the set of data points that are at most $\epsilon$ distant from it. 
\item If a data point has a number of neighbors, above a configurable threshold (called \emph{MinPts}), it is then considered a core point, and a high-density area has been detected. 
\item 
Since core points can be in each other's neighborhoods, a cluster consists of the set of core points that can be reached through their $\epsilon$-neighborhoods and all the data points in these $\epsilon$-neighborhoods.
\item Any data point that is not a core point and does not have a core point in its neighborhood is considered noise. 
\end{enumerate}

To obtain clusters using DBSCAN, we need to select two configuration parameters:
(1) the distance threshold, $\epsilon$, to determine the $\epsilon$-neighborhood of each data point, and (2) the minimum number of neighbors, \emph{MinPts}, needed for a data point to be considered a core point. For the identification of the values for $\epsilon$ and \emph{MinPts}, we rely on the same strategy integrated in SAFE, described below.

We determine the optimal value for $\epsilon$ by first computing the Euclidean distance from each data point to its closest neighbor. Then, we identify the optimal $\epsilon$ value as the knee-point of the curve obtained by considering those distances in ascending order.

To select an optimal \emph{MinPts} value, we execute DBSCAN multiple times with varying $MinPts$ values and with an $\epsilon$ equal to the optimal value determined above. We then select the clustering configuration that corresponds to the highest Silhouette Index value. %

\subsubsection{HDBSCAN:}
\label{sec:clust:hdbscan}
HDBSCAN (Hierarchical Density-Based Spatial Clustering of Applications with Noise) is an extension of DBSCAN to solve its main limitation: selecting a global $\epsilon$. DBSCAN uses a single global $\epsilon$ value  to determine the clusters. When the clusters have varying densities, using one global value can lead to a suboptimal partitioning of the data. 
Instead, HDBSCAN overcomes such a limitation by relying on different $\epsilon$ values for each cluster, thus finding clusters of varying densities.

HDBSCAN first builds a hierarchy using varying $\epsilon$  to figure out which clusters end up merging together and in which order. Based on the hierarchy of the clusters, HDBSCAN selects the most persisting clusters as final clusters. Cluster persistence represents how long a cluster stays without splitting when decreasing the value of $\epsilon$. After selecting a cluster, all its descendants are ignored.

\begin{figure}[H]
    \includegraphics[width=\textwidth]{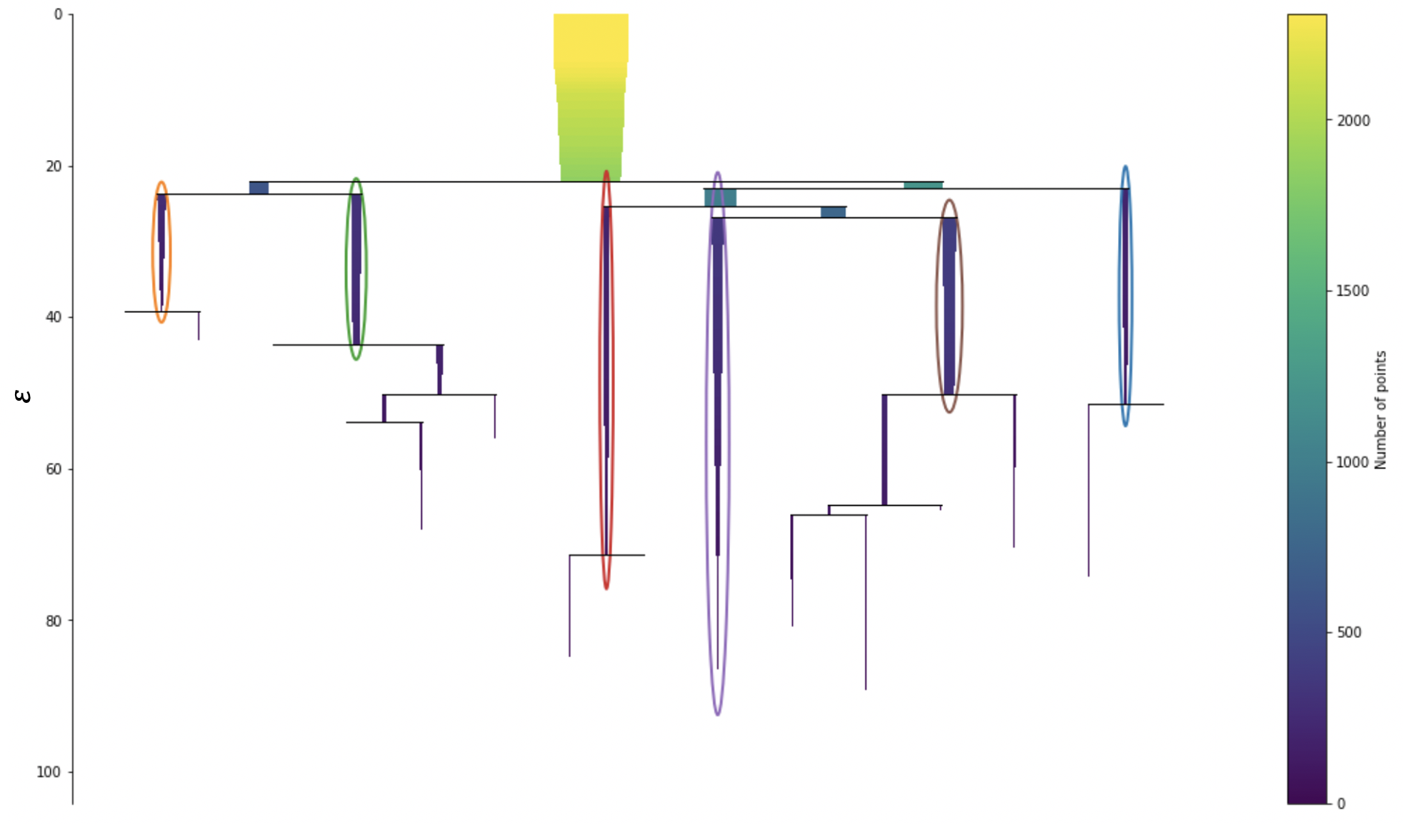}
    \caption{Example clusters selected by HDBSCAN.  }
    \label{hdb_example}
\end{figure}

Figure \ref{hdb_example} shows an example of the clusters' hierarchy found by HDBSCAN. The $y$-axis represents the values of $\epsilon$. Vertical bars represent clusters; the color and width of each vertical bar depict the size of the cluster. We can notice that certain clusters split after the value of $\epsilon$ is increased, while others persist. HDBSCAN decides which subclusters to select based on their persistence. The persistence of a subcluster is captured by the length of the colored vertical bars in the plot. 
HDBSCAN selects the clusters having the highest persistence. The unselected data points are considered noise. 
In our example, only 6 clusters are selected (circled bars); they are the longest vertical bars in the hierarchy.

%% file: evaluation.tex
\section{Empirical Evaluation}
\label{sec:empirical} 

In this section, we aim to evaluate the pipelines presented in Section~\ref{sec:approach}. 
A pipeline leads to the generation of clusters of images that are visually inspected by safety engineers to determine the root cause captured by each. We assume that a root cause can be described in terms of the commonalities across the images in a cluster; each root cause is thus a distinct scenario in which the DNN may fail (hereafter, \emph{failure scenario}).
The pipeline that best support such process should be the one requiring minimal effort towards accurate identification of root causes. 
Therefore, the best pipeline is the one that generates clusters
having a high proportion of similar images (to facilitate the identification of the root cause, based on analyzing similarities across images in a cluster),  enable the detection of all the root causes of failures, and is \MINOR{R2.1}{is reliable in the presence of rare root causes} of a particular root cause (to avoid ignoring infrequent but unsafe failure scenarios).
Based on the above, we defined three research questions to drive our empirical evaluation:

\textbf{RQ1} \emph{Which pipeline generates root cause clusters with the highest purity?}
We define a \emph{pure} cluster as one that contains only images representing the same failure scenario. Such clusters are expected to be easier to interpret; indeed, the engineer should more easily determine the root cause of failures if all the images share the same characteristics. Therefore, the best pipeline is the one that leads to clusters with the highest degree of purity. The purity of a cluster is computed as the maximum proportion of images belonging to a same failure scenario in this cluster. %

\textbf{RQ2} \emph{Which pipelines generate root cause clusters covering more failure scenarios?}
This research question investigates to which extent the different pipelines miss failure scenarios. Ideally, all failure scenarios should be captured by one or more clusters.
We say that a failure scenario is covered by a cluster if a majority of the images in the cluster belong to the scenario; indeed, commonalities shared by most of the images in a cluster should be noticed by engineers during visual inspection. We aim to determine which pipeline maximizes such coverage.

\textbf{RQ3} \emph{How is the quality of the generated root cause clusters affected by infrequent failure scenarios?}
Some failure scenarios may be infrequent but are nevertheless important to identify \MAJOR{R1.15}{as they may lead to severe hazards once the DNN is deployed in the field.} Ideally, a pipeline should be able to produce high-quality clusters even when a small number of images belong to one or more failure scenarios. In this research question, 
we vary the number of images belonging to failure scenarios and study how the effectiveness of pipelines---purity and coverage of the generated clusters---is affected.

\MAJOR{R3.7}{\textbf{RQ4} \emph{How do pipelines perform with failure scenarios that are not synthetically injected?} The only way to know what are the failures scenarios affecting our subject DNNs, for RQ1 to RQ3, is to rely on test set images presenting alterations (e.g., blurriness) that DNN cannot process (e.g., because it was not trained on such images). However, the results observed with injected failure scenarios may not generalize to pre-existing failure scenarios (i.e., scenarios that the original DNN cannot properly handle despite being trained for them). This research question assesses if the pipelines that perform best with injected failure scenarios also perform best with pre-existing failure scenarios and vice-versa.}

\vspace{1em}
To perform our empirical evaluation, \MAJOR{R1.6}{we implemented our pipelines' components using different libraries. 
Feature extraction based on LRP was implemented with PyTorch~\cite{PyTorch}, Tensorflow~\cite{tensorflow2015}, and Keras~\cite{chollet2015keras}  as an extension of the DNNs under test whereas transfer learning models were implemented using Tensorflow and Keras. The clustering algorithms and the dimensionality reduction methods rely on the Scikit-Learn library~\cite{scikit-learn}.} 
All the experiments were carried out on an Intel Core i9 processor running macOS with 32GB RAM.
Additionally, in our experiments, we relied on the LRP implementation provided by LRP authors~\cite{WIFS2017} for well-known types of layers (i.e., MaxPooling, AvgPooling, Linear, and Convolutional layers).

\subsection{Subjects of the study}
\label{sec:subj}

To evaluate our pipelines, we consider four different DNNs that process synthetic images in the automotive domain. These DNNs support gaze detection, drowsiness detection, headpose detection, and
unattended child detection, which are subjects of ongoing innovation projects at \IEE Sensing, our industry partner developing sensing components for automotive. Additionally, we consider two DNNs that process real-world images to support autonomous driving: steering angle prediction and car position detection.

The gaze detection DNN (\GD) performs gaze tracking; it can be used to determine a driver's focus and attention. It divides gaze directions into eight categories: TopLeft, TopCenter, TopRight, MiddleLeft, MiddleRight, BottomLeft, BottomCenter, and BottomRight. The drowsiness detection DNN (\CloseDNN{}) has the same architecture as the gaze detection DNN and relies on the same dataset, except that it predicts whether the driver's eyes are open or closed.

The head-pose detection DNN (\HPD) is an important cue for scene interpretation and computer remote control, such as in driver assistance systems. It determines the pose of a driver's head in an image based on nine categories: straight, rotated left, rotated left, rotated top left, rotated bottom right, rotated right, rotated top right, tilted, and headed up.

The unattended child detection DNN is trained with the  \emph{Synthetic dataset for Vehicle Interior Rear seat Occupancy} detection  (SVIRO) \cite{cruz2020sviro}. SVIRO is a dataset generated by IEE Sensing that represents scenes in the passenger compartment of ten different vehicles. The dataset has been used to train DNNs performing rear seat occupancy detection using a camera system. 
\MAJOR{R1.7}{The original IEE's DNN classifies SVIRO images into seven classes: adult, child, infant, child seat (empty or occupied), and infant seat (empty or occupied). However, the trained IEE DNN cannot be made publicly available for replication studies; therefore, in our study, we use SVIRO to retrain IEE's DNN from scratch with only three output classes (i.e., \emph{empty seats}, \emph{children/infants not accompanied by adults}, and \emph{the presence of an adult}). For our classification task, we relabeled the SVIRO dataset as follows. A seat is empty when there is an object, an empty child/infant seat, or nothing. We consider the presence of a child/infant and the presence of an adult as distinct classes. IEE's DNN architecture is opensource~\cite{DiasDaCruz2022Uncertainty}, it follows a VGG-19 architecture and we retrained it for $2,000$ epochs, with a batch size of 64.}

\MAJOR{R3.6}{Steering Angle Prediction (SAP) datasets are commonly used in autonomous driving or vehicle control systems \cite{du2019self}. These datasets are designed to train machine learning models to predict the appropriate steering angle for a given input image. The steering angle is a crucial parameter that determines the direction in which a vehicle should turn. The images can represent different perspectives of the road ahead, including images from a front-facing camera, multiple camera angles, or even side or rear cameras. For example, an image in the dataset could show the view of the road ahead from the driver's perspective.}

For Steering Angle Prediction, we rely on the pre-trained Autumn DNN model~\cite{autumn}, which follows the DAVE-2 architecture~\cite{bojarski2016end} provided by NVIDIA. It is a DNN to automate steering commands of self-driving vehicles~\cite{udacity}; it predicts the angle at which the wheel should be turned. It has been trained on a dataset of road images captured by a dashboard camera mounted in the car.

\MAJOR{R3.6}{Car Position Detection (CPD) DNNs are used by most Advanced-Driver Assistance Systems (ADAS) to predict the positions of the cars in the scene \cite{wang20213d}. For example, a dataset could include images captured from different angles or heights, representing various driving scenarios like urban environments, highways, or parking lots. The goal is to predict the position of each car on the scene.} We rely on the CenterNet DNN~\cite{duan2019centernet}, which is an accurate DNN used by most competition-winning approaches for object detection~\cite{pku}. It has been trained on images from the ApolloScape dataset~\cite{apollo} collected using a dashboard camera to estimate the absolute position of vehicles with respect to the ego-vehicle.

For each subject DNN, we apply our pipelines to a set of failure-inducing images. 
Such sets consist of (1) images belonging to a provided test set and leading to a DNN failure and (2) test set images that were not leading to a DNN failure but had been modified to cause a DNN failure; the latter are images with injected root causes of failures and are described in Section~\ref{sec:injected}. 
In classifier DNNs (i.e., OC, GD, HPD, and SVIRO) a failure occurs in the presence of an image being incorrectly classified.
For SAP and CPD, 
which are regression DNNs, we set a threshold to determine DNN failures. 
For SAP, we observe a DNN failure when the squared error between the predicted and the true angle is above $0.18$ radian ($10.3^{\circ}$), which is deemed to be an acceptable error 
in related work~\cite{tian2018deeptest}. 
For CPD, since it tackles a multi-object detection problem, we report a DNN failure when the result contains at least one \emph{false positive} (i.e., the distance between the predicted position of the car and the ground truth is above $10$ pixels~\cite{song2019apollocar3d}).

In Table~\ref{tab:dnns}, we provide details about the case study subjects used to evaluate our pipelines. For each subject, we report the source of the dataset (e.g., the simulator used to generate the data), the training and test set sizes, the accuracy of the DNN on the original test set, the number of failure-inducing images and the number of images for each injected root cause (they are detailed in Section~\ref{sec:injected}). 
\input{tables/tableDNNs}

We fine-tune the pipelines relying on transfer learning using the test sets of the respective case studies. We use the resulting fine-tuned model to extract the features from the failure-inducing sets. We train on the test sets because the number of images in each set is sufficient for the model to learn the features. 
Also, we train the autoencoders on the training set, and use the test set of the respective case study to validate the results. The termination criterion is $50$ epochs unless we reach an early stopping point (the model stops improving). After training, we use only the encoder part to extract the features from the images in the failure inducing set.

\subsection{Injected Failure Scenarios}
\label{sec:injected}

To assess the ability of different pipelines to generate clusters that are pure and cover all the root causes of failures, we need to know the root causes of failures in the test set. Since such root causes may vary (e.g., lack of sufficient illumination, presence of a shadow in a specific part of the image) and it is not possible to objectively demonstrate that a failure cause has been correctly captured by a cluster (e.g., some readers may not agree that certain images show lack of sufficient illumination), to avoid introducing bias and subjectivity in our results, we modify a subset of the provided test set images so that they will fail because of known root causes of failures. In total, we considered nine different root causes to be injected in our test set images and refer to them as \emph{injected failure scenarios} (i.e., failure scenarios with injected root causes).

We derive an image belonging to an injected failure scenario by modifying a test set image according to the specific root cause we aim to inject; for example, by covering the mouth of a person with a mask. To ensure that a modified image leads to a DNN failure because of the injected root cause, we modify only test set images that, before modification, lead to a correct DNN output.

\begin{figure}[H]
 \centering
 \includegraphics[width=\textwidth]{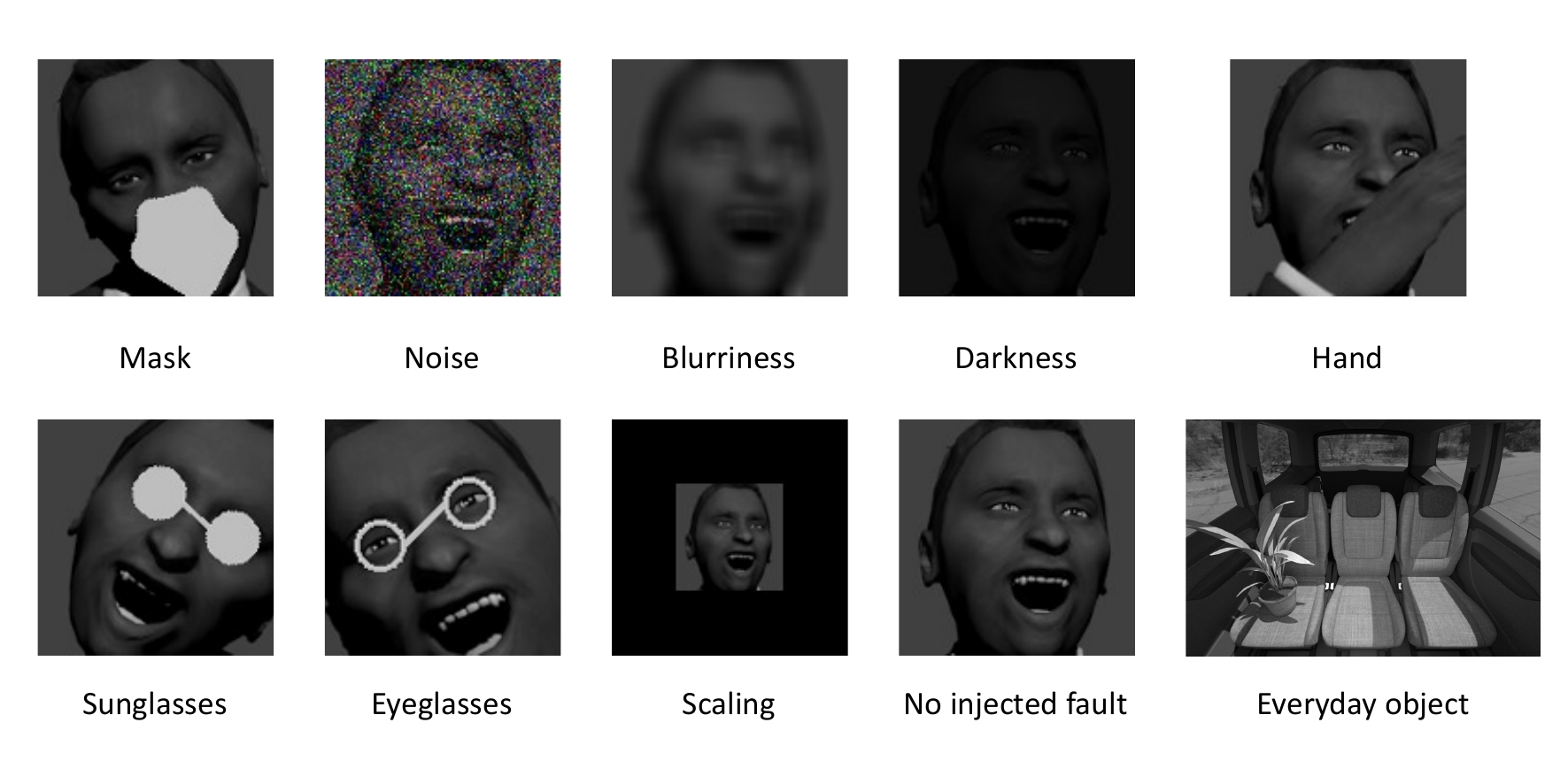}
 \caption{Injected failure scenarios in our study.}\label{fig:inj_faults}
 \end{figure}

Figure \ref{fig:inj_faults} illustrates the different injected failure scenarios. Below, we describe the nine root causes considered in our study:
\begin{itemize}
    \item \textbf{Hand:} The presence of a hand blocking the full view of the driver's face could affect the DNN result, leading it to mispredict the driver's head direction. We simulate a hand that is partially covering the face appearing in the image. 
    
    \item \textbf{Mask:} Similar to \emph{Hand}, the presence of a mask covering the nose or the mouth may affect a DNN that recognizes the driver's head pose. Using image key points, we drew the shape of a white mask to simulate a mask covering the nose and the mouth.
    
    \item \textbf{Sunglasses:} As for the \emph{Mask}, we use the eyes' key points to draw sunglasses covering the driver's eyes.
    
    \item \textbf{Eyeglasses:} Different from the \emph{Sunglasses}, we draw glasses with the eyes  being still visible.
    
    \item \textbf{Noise:} A noisy image is one that contains random perturbations of colors or brightness. \MAJOR{R1.2}{In real-world automotive systems, such a failure scenario occurs due to a defective camera or a high signal-to-noise ratio (SNR) in the communication channel between different electronic control units (ECUs), resulting in a noisy input. Also, some image compression algorithms, particularly those used in certain file formats like JPEG, can introduce artifacts and noise into the image during the compression process \cite{liu2020, hasinoff2016burst}. Related work has considered this failure scenario to assess the  fault tolerance of DNNs \cite{zheng2021mindfi}. }
    We use the Scikit-Image library~\cite{scikit-image} to add Gaussian Noise, a statistical noise with a probability density function equal to a normal distribution, also known as Gaussian Distribution.
    
    \item \textbf{Blurriness:} \MAJOR{R1.2}{This scenario can occur because of camera shake, especially when the camera is integrated into the car. Motion blur can also happen when capturing moving objects such as cars and pedestrians. This failure scenario was used to evaluate DNN robustness \cite{tian2018deeptest}.} 
    We use the Pillow library~\cite{clark2015pillow} to add blurriness to images using a radius of 30 pixels. 
    
    \item \textbf{Darkness:} In practice, poor lighting conditions could make the DNN fail because it cannot clearly recognize what is depicted in a relatively dark image. \MAJOR{R1.2}{This failure scenario was used in a related work to evaluate DNN robustness \cite{pei2017deepxplore}.} We use the Pillow library~\cite{clark2015pillow} to decrease the brightness of images by a factor of $0.3$; we selected $0.3$ because it is the lowest value introducing failures in our subject results. 
    
    \item \textbf{Scaling:} Such a failure scenario mimics the situation where a camera is misconfigured, leading to rescaled images being fed to the DNN. We reduce the size of an image by a value based on the image size (i.e., large 1200px $\times$ 1200px images are scaled by 400px, small 320px $\times$ 320px images by 70px).
    and insert a black background using the Pillow library~\cite{clark2015pillow}. 
    \MAJOR{R1.2}{Camera malfunctions or technical issues with the zooming mechanism can result in a scaling failure. Scaling was also used in the literature for data augmentation  \cite{chen2021scale}.} 
    \item \textbf{Everyday Object:} For the SVIRO dataset, we introduce, in the car's rear seat, an everyday object (e.g., a washing machine or a handbag) never observed in the training set, thus simulating the effect of an incomplete training dataset. \MAJOR{R1.2}{Such objects capture the case of an unseen label during the training, which is a commonly used faulty scenario \cite{humbatova2021deepcrime}.} 

\end{itemize}

For regression DNNs (SAP and CPD), we randomly selected $90$ images to be copied and modified for each failure scenario.
For classifier DNNs, for each failure scenario, we randomly selected $10$ images for each class label.
\subsection{\MAJOR{R3.2}{}Pre-existing Failure Scenarios}
\label{sec:pre-existing}
\MAJORBEGIN
Since it is usually not possible to achieve perfect accuracy through training, our DNNs, like any machine learning model, are affected by failure scenarios whose effects are visible in the original test set (e.g., borderline images that are misclassified because they are very similar to the ones belonging to another class).
In other words, some failure scenarios could already be identified in the original test set and we refer to such scenarios as  \emph{pre-existing failure scenarios}.

Unfortunately, it is not possible to identify pre-existing failure scenarios in a test set because commonalities across failure-inducing images might be partially perceptible (e.g., shadows on faces) and, consequently, it might be difficult to precisely determine the causes for such failures. Therefore, we cannot perform an accurate assessment of our pipelines on pre-existing failure scenarios. However, for some of the subject DNNs classifying simulator images, it is possible to make assumptions on some of the possible causes of DNN errors; such causes can be expressed in terms of simulator parameters leading to borderline cases that are likely hard to classify by a DNN. 
We refer to such parameters as \emph{failure-inducing parameters}. For each failure-inducing parameter, it is possible to identify one or more \emph{unsafe values}. We then generate images that are likely to cause a DNN failure by configuring the simulator with a value for a failure-inducing parameter close to an unsafe value. 

\input{tables/tableBoundaries}

In our previous work~\cite{attaoui2022black}, we have identified a set of failure-inducing parameters affecting the HPD, OC, and GD DNNs; they are listed in Table~\ref{tab:boundary2}. 
For GD, we identified unsafe values related to the angle of the eye gaze (8 values) and the openness of the eye (1 value) because they all may affect gaze detection results. For OC, we consider the openness of the eye (1 unsafe value), which directly affects classification, and values characterizing an unrealistic image, with a pupil below the eyelid  (i.e., a distance between the pupil and the bottom eyelid below -16 pixels) or above the eyelid (i.e., a distance between the top eyelid and the pupil below -16 pixels).
For HPD, we consider the Horizontal and Vertical Headpose parameters which represent the classification classes of the DNN (8 unsafe values).
For Gaze Angle,
Openness, Headpose-X, and Headpose-Y, the value of a failure-inducing parameter is considered close to an unsafe value if
the difference between them is below 25\% of the length of the subrange including the average value.
For PupilToBottom, and TopToPupil, the value of a failure-inducing parameter is considered close to an unsafe value if it is below or equal to it. 
Table~\ref{tab:pre-existing-distribution} provides the list of failure inducing scenarios for each subject DNN; basically, we have one failure scenario for each unsafe value except for the unsafe values of PupilToBottom and TopToPupil, which capture the same unsafe scenario (i.e., unrealistic image). Table~\ref{tab:pre-existing-distribution} also reports the number of failure-inducing test set images belonging to each pre-existing failure scenario; note that an image can belong to one or more pre-existing failure scenarios and it was not possible to associate every image to a failure scenario. For example, this may happen because the DNN failure is due to the rendering of the image (e.g., a shadow may affect how the shape of the nose is perceived by the DNN), which is not controllable through simulator parameters but is the result of complex interactions among them (e.g., illumination direction, head orientation, light intensity).

\input{tables/pre-existing-distribution}

For the HPD, OC, and GD DNNs, we could determine unsafe values for each failure-inducing parameter, because we know what are the simulator parameter values used to generate each image.
For the SVIRO case study, we could not identify failure-inducing parameters because we only have access to the dataset, not the parameters associated with each image.
Therefore, the possible reasons for misclassification (e.g., object size) cannot be directly mapped to the information provided to us, which is coarse grained (e.g., presence of an object on the seat).

Since we cannot know what are all the failure scenarios in our case study subjects,
we do not compare our pipelines based on pre-existing failure scenarios. However, for completeness, in Section~\ref{sec:pre-existing-results}, we report on the performance of our pipelines  with such failure scenarios affecting the OC, GD, and HPD DNNs.

For the experiments with injected failure scenarios (i.e., experiments assessing RQ1 to RQ3), we still include  images belonging to pre-existing failure scenarios into the dataset since they are usually observed for any DNN and, therefore, should be considered when generating RCCs.  However, clusters that do not include any image belonging to an injected failure scenario are assumed to capture root causes related to pre-existing failure scenarios and, therefore, are ignored for computing purity and coverage (details are provided in the next Sections). 

For RQ1-3, since we cannot make assumptions about the distribution of pre-existing and other failure scenarios, we include the same number of images for pre-existing failure scenarios and injected failure scenarios (see Table \ref{tab:dnns}). For the experiments assessing pipelines with pre-existing failure scenarios (i.e., RQ4), instead, to be realistic, we consider the whole set of failure-inducing test images belonging to a pre-existing failure scenario.

\MAJOREND
\subsection{RQ1: Which pipeline generates root cause clusters with the highest purity?}
\label{sec:evaluation:RQ1}

\subsubsection{Design and measurements.} 
\label{sec:evaluation:RQ1:design}
A pure cluster includes only images presenting the same root cause (i.e., common cause leading to a DNN failure); for example, a hand covering a person's mouth. Pure clusters simplify root cause analysis because they should make it easier for an engineer to determine the commonality across images and therefore the cause of failures.

Since the likely root cause of the failure in our \emph{injected failure scenarios} is known, we focus on these scenarios to respond to RQ1.
For each RCC, we compute the proportion of images belonging to each injected failure scenario. Therefore, we measure the purity $P$ of a cluster $C$ (hereafter, $P_C$) as the highest proportion of images belonging to one injected failure scenario $f \in F$ assigned to cluster $C$, where $F$ is the set of all failure scenarios. $P_C$ is computed as follows:

\begin{equation}
 \mathit{P}_C = \max_{f \in F}\left(\frac{C_f}{|C|}\right)
\end{equation}

The proportion of a failure scenario $f$ in a cluster $C$ is computed as the number of images belonging to $f$ assigned to cluster $C$ ($C_f$), divided by the size of  cluster $C$.

Clusters that do not include any image belonging to an injected failure scenario are assumed to capture root causes due to pre-existing failure scenarios and, consequently, are excluded from our analysis.

We study the purity distribution across RCCs generated for the different case study subjects. Since, ideally, we would like to obtain pure clusters, the best pipeline is the one that maximizes the average purity across the generated RCCs.

\subsubsection{\MAJOR{R1.18}{Methodology}}
\label{rq1_methodology}
We use the Conditional Inference Tree (CTree) algorithm~\cite{hothorn2015ctree} to generate a decision tree with a maximum depth set to $4$ (we have four components in a pipeline) and a minimum split set to $10$ (i.e., the weight of a node to be considered for splitting).
The dataset used to build the tree consists of the components of each pipeline as attributes, and the purity of the generated clusters as the predicted outcome. The dataset size is equal to 99, the number of pipelines. \MAJOR{R1.9}{We rely on decision trees because they enable us to determine how the different pipeline components contribute to the results (i.e., precision); the manual inspection of the configurations leading to the highest precision would not have enabled us to determine which components contribute most to precision.}

Each node of the tree represents a feature of the pipeline. Leaves (terminal nodes) depict box plots representing distributions of the average purity across RCCs generated by the pipelines belonging to each leaf. Each point in the box plot is the average purity of one pipeline (i.e., the average of the purity of all the RCCs generated across all case study subjects). 
To split a node, the CTree algorithm first identifies the feature with the highest association (covariance) with the response variable (purity, in our case). Precisely, it relies on a permutation test of independence (null hypothesis) between any of the features and the response~\cite{PermutationTests}; the feature with the lowest significant p-value is then selected ($\mathit{alpha} = 0.05$, in our case). Once a feature has been selected, a binary split is then performed by identifying the value that maximizes the test statistics across the two subsets.
Since we are in the presence of multiple hypotheses (assume $m$, for each node), to prevent a Type I error, for each feature $j$, CTree computes its Bonferroni-adjusted~\cite{wright1992adjusted} $p\text{-value}_j$ as 

$$\text{adjusted}\ p\text{-value}_j = 1 - ( 1 - p\text{-value}_j )^m $$

\subsubsection{Results} 
\label{sec:res:RQ1}
\begin{figure}[H]
 \centering
 \includegraphics[width=\textwidth]{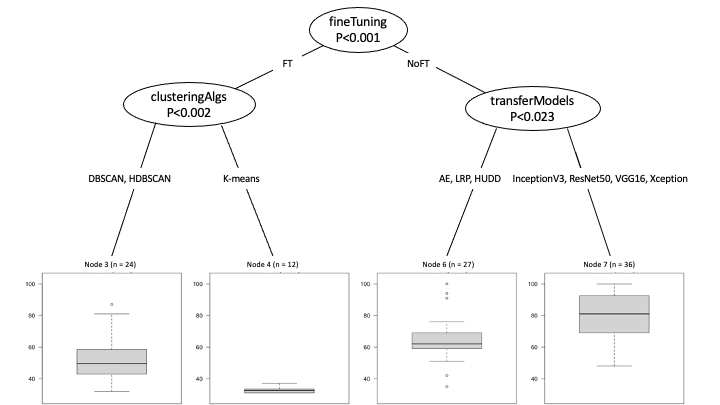}
 \caption{Decision Tree illustrating how the different features of a pipeline determine the average purity of root-cause clusters.}\label{fig:rq1_average_tree}
 \end{figure}

Figure~\ref{fig:rq1_average_tree} depicts a regression tree illustrating how the different components of a pipeline (feature extraction methods, fine-tuning, dimensionality reduction techniques and clustering algorithms) determine the purity of the clusters generated by a pipeline. We notice that the pipelines with fine-tuned models (Node $3$ and $4$) generate lower-purity clusters than those without any fine-tuning (Node $6$ and $7$)\MAJOR{R1.10}{, which can be explained by the fine-tuning dataset not including the injected failure scenarios. For our approach, the objective of fine-tuning is to learn features that are specific for the context of use; please recall that our transfer learning models are based on ImageNet and we rely on them for feature extraction. However, we perform fine-tuning using the test set, which is smaller than the training set and thus leading to a quicker process. Further, to simulate a realistic usage, we did not include the injected failures into the dataset used for fine-tuning; indeed, since our injected root causes aim to capture scenarios not foreseen at training time, it would be unrealistic to consider such scenarios for fine-tuning.
Finally, fine-tuning with images including injected failures (e.g., noise) may affect the quality of fine-tuning.
Because of the choices above, fine-tuning leads to features that do not capture the injected faults but the characteristics of images without faults. As a result, in our experiments, images are clustered based only on their pre-existing fault (e.g., borderline class) instead of the injected faults. ImageNet models, instead, may capture features that are useful to cluster injected faults  (e.g., the presence of everyday objects in SVIRO), but such features are then forgotten as an effect of catastrophic forgetting during fine-tuning~\cite{chen2019catastrophic}, thus leading to clustering results that are worse for fine-tuned transfer-learning models.}

The pipelines using non-fine-tuned transfer learning models as a feature extraction method (Node $7$) generate purer clusters (min = $50\%$, median = $80\%$, max = $96\%$) than the pipelines using an autoencoder model, HUDD, or LRP (Node $6$) (min = $50\%$, median = $65\%$, max = $70\%$). The purpose of the Autoencoder model is to provide a condensed representation of the image to be used for reconstruction. This is done by ignoring the features that the model considers insignificant and only keeping the features that help the encoder reconstruct the image accurately. Therefore, \MAJOR{R1.11}{a possible explanation for our result is that} since the autoencoder is trained on the training set, the injected faults are ignored. Given that clustering is based on the condensed representation, the generated clusters are less pure than the ones generated by the pipelines with transfer learning models. \MAJOR{R1.11}{Note that without empirical assessment, it is not possible to know in advance how autoencoders support clustering; indeed, injected faults may mask certain autoencoder features (e.g, presence of non-black pixels around the borders for scaled images) that turn out to be useful for clustering.}

As for HUDD and LRP, it seems that their main limitation is that heatmaps cannot capture the presence of root causes affecting all the pixels in an image (i.e., the result of noise, blurriness, darkness, scaling). Heatmaps mainly capture which pixels of the image drive the DNN output, thus leading clustering to group images where the same pixels affected the output.
For instance, the DNN's response to a blurred image with a shadow on the mouth could be different from that of another blurred image with a shadow on the eyes, thus leading to different clusters for these images although they represent the same injected failure scenario (blurriness). 

Finally, we notice that the pipelines using HDBSCAN and DBSCAN (Node $3$) as a clustering algorithm yield purer clusters (min = $25\%$, median = $40\%$, max = $80\%$) than those using K-means (Node $4$, min = $22\%$, median = $27\%$, max = $29\%$). This is because K-means faces difficulty dealing with non-convex clusters. A cluster is convex if, for every pair of points belonging to it, it also includes every point on the straight line segment between them
~\cite{kriegel2011density}, which gives the cluster a \MAJOR{R1.12}{hyperspherical} form. Nevertheless, in many practical cases, the data leads to clusters with arbitrary, non-convex shapes. Such clusters, however, cannot be appropriately detected by a centroid-based algorithm (e.g., K-means), as they are not designed for arbitrary-shaped clusters.

DBSCAN and HDBSCAN are density-based clustering algorithms. They consider high-density regions as clusters (see Section~\ref{sec:background}). The root cause clusters generated by DBSCAN and HDBSCAN are arbitrary-shaped and more homogeneous (i.e., clusters with higher within-cluster similarity) with very similar images. In contrast, a convex cluster generated by K-means tends to be less dense and can group rather dissimilar images. As a result, a convex cluster is less pure than a non-convex one.

We report the significance of these results in Table \ref{tab:rq1_stat}, including the values of the Vargha and Delaney's $\hat{A}_{12}$ effect size and the $p$-values resulting from performing a Mann-Whitney U-test to compare the average purity of the pipelines using transfer learning models (Node $7$ in the decision tree) and the pipelines represented by the other nodes. 
Typically, an $\hat{A}_{12}$ effect size above $0.56$ is considered practically significant with higher thresholds for medium ($0.64$) and large ($0.71$) effects~\cite{Kitchenham2017}, thus suggesting the effect sizes between the pipelines using transfer learning models and other pipelines are large. Further, $p$-values suggest these differences are statistically significant. 

 \input{tables/VDA_results}

Finally, in Table~\ref{tab:rq1_top}, we report the pipelines that generated clusters with an average purity above $90\%$ across all case study subjects, along with the purity obtained for each subject; the complete results obtained for all pipelines appear in Appendix~\ref{appendix:RQ1}, Table~\ref{tab:rq1_appendix}.
An average purity of $100\%$ means that all the clusters generated by the pipeline are pure. 
Interestingly, all the pipelines in Table~\ref{tab:rq1_top} belong to Node 7 in Figure~\ref{fig:rq1_average_tree}, thus confirming our main finding. Five of these seven best pipelines, rely on UMAP, without fine-tuning but with a transfer learning model, which is therefore our suggestion to perform root cause analysis.
The best result is obtained with ResNet-50 combined with UMAP and DBSCAN. 

\input{tables/RQ1_Purities_top}

 \subsection{RQ2: Which pipelines generate root cause clusters covering more failure scenarios?}
\label{sec:evaluation:RQ2}

\subsubsection{Design and measurements.} This research question investigates the extent to which our pipelines identify all failure scenarios. We compare pipelines in terms of the percentage of injected failure scenarios being covered by at least one RCC. A failure scenario is covered by an RCC if it enables the engineer to determine the root cause of the failure.
Precisely, when images belonging to a failure scenario $f$ represent a sufficiently large share of images in a cluster $C$, it is easier for an engineer to notice that $f$ is a likely root cause. %
Therefore, we assume that an injected failure scenario $f$ is covered by a cluster $C$ if the cluster $C$ contains at least $90\%$ of images with $f$. Since this threshold is relatively high, our results can be considered conservative. 

Given that our injected failure scenarios are represented by the same number of images in the failure-inducing test set, every failure scenario has the same likelihood of being observed. Therefore, we expect to obtain RCCs corresponding to each failure scenario.
\subsubsection{\MAJOR{R1.18}{Methodology}}
We follow the same methodology as for RQ1 (see Section \ref{rq1_methodology}) but we construct a decision tree considering, for each pipeline, the average coverage achieved across case study subjects instead of the average purity.

\subsubsection{Results.} 
\begin{figure}[H]
 \centering
 \includegraphics[width=\textwidth]{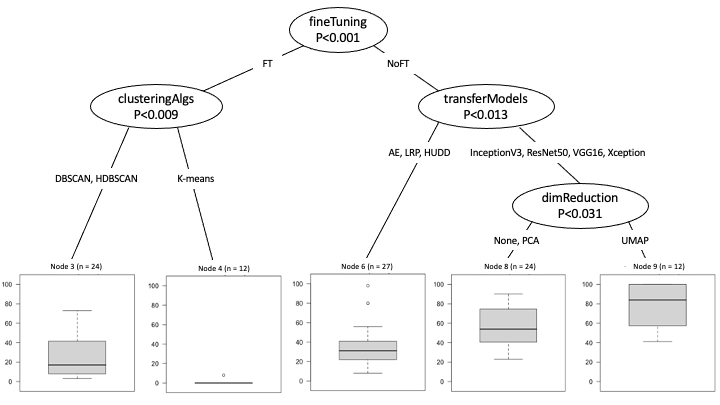}
 \caption{Decision Tree illustrating how the different features of a pipeline determine the coverage of the failure scenarios.)}\label{fig:rq2_average_tree}
 \end{figure}

Figure~\ref{fig:rq2_average_tree} shows a decision tree illustrating how the different components of a pipeline determine the coverage of failure scenarios.  

Each leaf node depicts a box plot with the distribution of the percentages of  failure scenarios covered by the set of pipelines that include the components listed in the decision nodes.

For instance, Node $9$ provides the distribution of the percentage of failure scenarios covered by the RCCs generated by pipelines using UMAP as a dimensionality reduction technique and non-fine-tuned transfer learning models as feature extraction methods ($12$ pipelines). Ideally, the root-cause clusters generated by a pipeline should cover $100\%$ of the failure scenarios. 

The decision tree in Figure \ref{fig:rq2_average_tree} confirms RQ1 results. The pipelines without fine-tuning (Nodes $6$, $8$ and $9$) outperform the pipelines with fine-tuning (Nodes $3$ and $4$). The pipelines with transfer learning models (Nodes $8$ and $9$) generate clusters that cover more failure scenarios than those generated by the pipelines using HUDD, LRP, and AE (Node $6$). Also, the pipelines using the DBSCAN and HDBSCAN clustering algorithms (Node $3$) yield better results than the ones using K-means (Node $4$).

Further, the decision tree in Figure \ref{fig:rq2_average_tree} gives us more insights into which dimensionality reduction method is more effective. We notice that the root-cause clusters generated by the pipelines using UMAP (Node $9$) lead to a better distribution (min = $45\%$, median = $85\%$, max = $100\%$) than the pipelines using PCA or not using any dimensionality reduction (Node $8$, min = $25\%$, median = $55\%$, max = $90\%$). This is because UMAP yields a better separation of the clusters (i.e., less clusters overlap) compared to PCA. When using UMAP, all the data points converge towards their closest neighbor (the most similar data point). Therefore, neighboring data points in higher dimensions end up in the same neighborhood in lower dimensions, resulting in a compact and well-separated clusters where it is easier for the clustering algorithms to distinguish them.

We report the significance of these results in Table~\ref{tab:rq2_stat}, including the values of the Vargha and Delaney's $\hat{A}_{12}$ effect size and the $p$-values resulting from performing a Mann-Whitney U-test to compare the percentages of covered failure scenarios resulting from the pipelines using UMAP (Node $9$ in the decision tree in Figure~\ref{fig:rq2_average_tree}), and the other pipelines. 
Table~\ref{tab:rq2_stat} shows that the $p$-values, when comparing the pipelines using UMAP to the other pipelines, are always below $0.05$. This implies that in all the cases, differences are statistically significant with large effect sizes (above $0.77$).
\input{tables/VDA_results2}

\input{tables/RQ2_Coverages_top}

In Table~\ref{tab:rq2_top}, we report the pipelines that generated clusters covering at least $90\%$ of the failure scenarios across all case study subjects, along with the coverage obtained for each case study subject (complete results for all the pipelines are reported in Appendix~\ref{appendix:RQ2}, Table \ref{tab:rq2_appendix}). %
If the coverage is equal to $100\%$, all the failure scenarios are covered by the RCCs.
Unsurprisingly, the pipelines in Table~\ref{tab:rq2_top} belong to Node 7 in Figure \ref{fig:rq2_average_tree}: they rely on a non-fine-tuned transfer learning model for feature extraction, and UMAP for dimensionality reduction. Further, they all use DBSCAN for clustering. 
These pipelines consistently yielded the best results for all individual case studies (confirming the results obtained in RQ1).

Such findings are further supported by the results in Table~\ref{tab:rq1_appendix} and Table~\ref{tab:rq2_appendix}, where we notice that the combination of UMAP with DBSCAN always achieves higher purity and coverage (in bold) than its alternatives, regardless of the used feature extraction method.

 \subsection{RQ3: How is the quality of root cause clusters generated affected by infrequent failure scenarios?}
\label{sec:evaluation:RQ3}

\subsubsection{Design and measurements.} We study the effect of infrequent failure scenarios on the quality of the RCCs generated by the pipelines. \MAJOR{R1.15}{Indeed, infrequent scenarios may not be properly captured  by clustering algorithms. With K-means, the number of clusters depends on within-cluster SSD (see Section~\ref{sec:clust:kmeans}) but the exclusion of small clusters may lead to unnoticeable changes in the computed SSD. With DBSCAN, small clusters may be treated as noise. With HDBSCAN, small clusters, which have a limited persistence ($\epsilon$ cannot be higher than the number of datapoints, see Section~\ref{sec:clust:hdbscan}), may not be identified.}

We consider a failure scenario infrequent when it is observed in a low proportion of the images in the failure-inducing set.
To be practically useful, a good pipeline should be able to generate root-cause clusters even for infrequent failure scenarios; indeed, in safety-critical contexts, infrequent failure scenarios may lead to hazards and thus should be detected when testing the system.
For instance, if only five out of hundred failure-inducing images belong to a failure scenario and we have three failure scenarios in total, 
a \MINOR{R2.1}{reliable} pipeline should still generate an RCC containing only the images of the infrequent failure scenario. 
\subsubsection{\MAJOR{R1.18}{Methodology}}
We generate $10$ different failure-inducing sets for each case study subject (a total of $60$ failure-inducing sets). 
To construct a failure-inducing set, for each root cause that might affect the case study (see Table \ref{tab:dnns}, Page~\pageref{tab:dnns}), we generate a number $n$ of images affected by the injected root cause. We randomly select a number $n$ that is lower than the number of images selected for the same root cause in RQ1 (see Table \ref{tab:dnns}). Further, for classifier DNNs, we select a value higher than the number of classes of the corresponding case study (we enforce one root cause of failures for one image per class, at least); for  regression DNNs, we select a value above $2$. 
Since $n$ is randomly selected (uniform distribution), we obtained failure-inducing sets containing failure scenarios whose number vary.
\MAJOR{R1.14}{Table~\ref{tab:rq3_distribution}, Appendix~\ref{appendix:rq3}, provides the details for each case study;} \MAJOR{R1.19}{for instance, the number of images representing a failure scenario for each failure-inducing set of the HPD case study ($9$ classes) is randomly selected between $9$ and $90$.}

In addition, we also include a randomly selected number of images belonging to pre-existing failure scenarios, to mimic what happens in practice (see RQ1). The number of images belonging to pre-existing failure scenarios varies between two and the total number of injected failure scenario images.

Since we aim to study the effect of infrequent failure scenarios on the quality of the generated RCCs, we categorize our 290 failure scenarios 
into \emph{infrequent} and \emph{frequent}. Infrequent failure scenarios are the ones that include a proportion of injected images that is lower than the median proportion in all the generated failure-inducing sets (equals to $18\%$ in our study).
For example, noise is frequent in the dataset GD\_1 ($64>18$) but infrequent in the dataset OC\_2 ($4<18$). 

We consider only the best pipelines resulting from the experiments in RQ1 and RQ2 (i.e., having purity or coverage above 90\% as shown in Tables~\ref{tab:rq1_top} and~\ref{tab:rq2_top}); they are pipeline $26$ (\emph{VGG16/DB\-SCAN/UMAP/NoFT}), $44$ (\emph{ResNet50/DBSCAN/UMAP/NoFT}), $62$ (\emph{InceptionV3/DB\-SCAN/UMAP/NoFT}), $19$ (\emph{VGG16/K-means/None/NoFT}), $25$ (\emph{VGG16/K-means/UMAP/NoFT}), $39$ (\emph{ResNet50/HDBSCAN/None/NoFT}), 
$43$ (\emph{ResNet50/K-means/UMAP/NoFT}),
and $80$ (\emph{Xception/DBSCAN/UMAP/NoFT}). The first three pipelines (i.e., $26$, $44$, $62$) were the best for both RQ1 and RQ2, the next four (i.e., $19$, $25$, $39$, $43$) were selected based on RQ1 results while the latter (i.e., $80$) based those of RQ2.
We compute the purity and coverage of the RCCs generated by each of these pipelines, following the same procedures adopted for RQ1 and RQ2. We then compare the distribution of purity and coverage for infrequent and frequent failure scenarios. The most \MINOR{R2.1}{reliable} pipelines are the ones being affected the least, in terms of purity and coverage, by infrequent failure scenarios.

\subsubsection{Results.} 
\label{sec:rq3:results}
\begin{figure}[t]
 \centering
 \includegraphics[width=\textwidth]{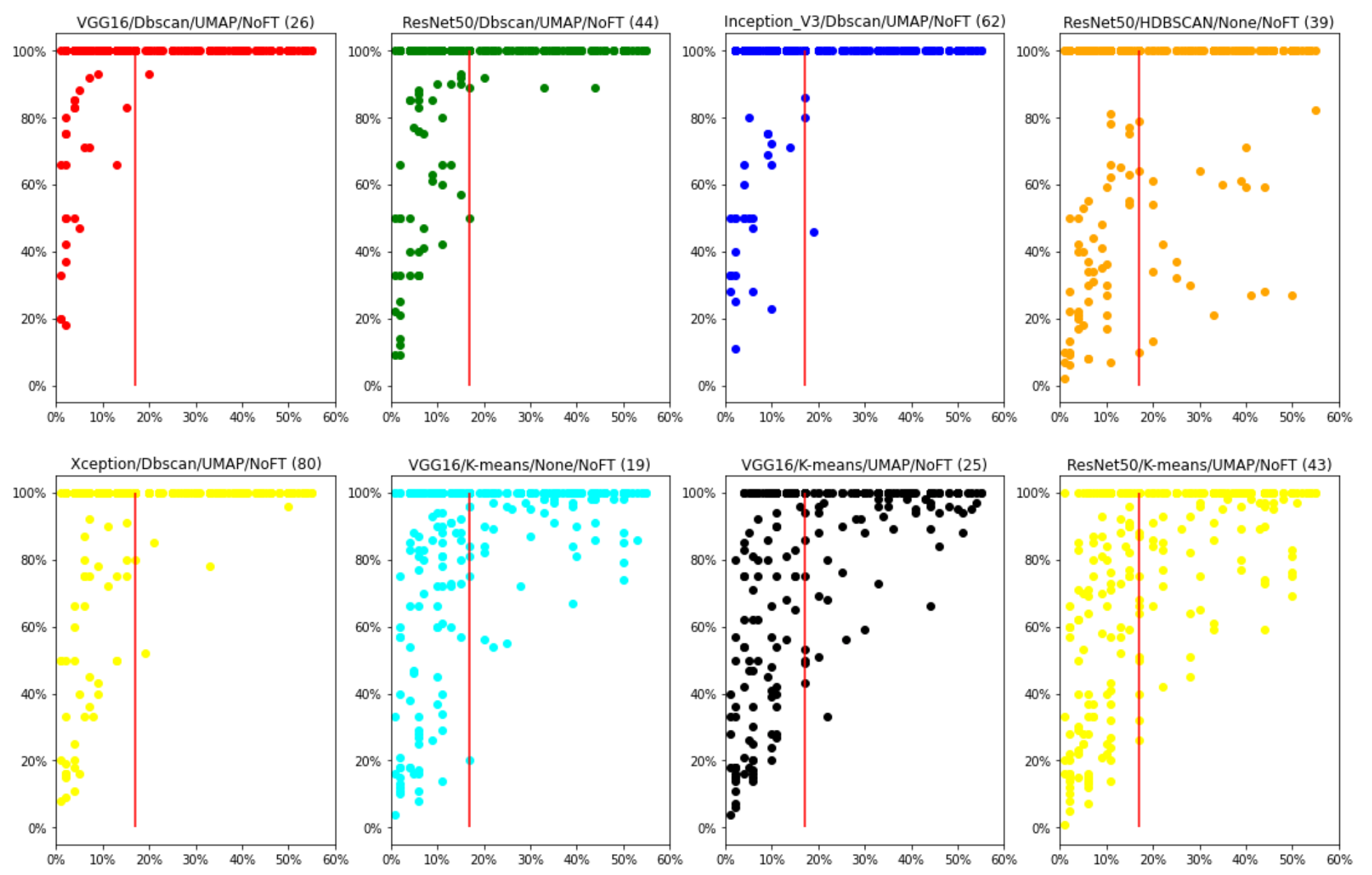}
 \caption{Purity of the clusters associated with frequent and infrequent failure scenarios. The x-axis captures the frequency of a failure scenario (i.e., proportion of failure-inducing images for a failure scenario).  Each data point is the average of all the RCCs associated to one distinct failure scenario. The red vertical line represents the median frequency of failure scenarios.}\label{fig:rq3_purity}
 \end{figure}

In Figure~\ref{fig:rq3_purity}, for each selected pipeline, we report the average purity across all the RCCs\footnote{As discussed in Section~\ref{sec:evaluation:RQ1:design}. The red vertical line represents the median frequency of failure scenarios. We say that an RCC is associated with (or captures) an injected failure scenario $f$ when the majority of the images in the cluster belong to scenario $f$.} with the injected failure scenarios having a certain frequency. The $x$-axis reports the proportion of images for failure scenarios whereas the $y$-axis reports the average purity of the RCCs associated to each failure scenario.

Figure~\ref{fig:rq3_purity} shows that when the frequency of the failure scenarios is below the median (infrequent scenario), the cluster purity obtained by pipelines tends to significantly lower and decrease rapidly as the frequency decreases.
This is expected because when a failure scenario is infrequent, the clustering algorithm tends to either cluster its images as noise or distribute them over the other clusters. 
For density-based clustering algorithms, images belonging to infrequent scenarios may not become core points when the identification of a core point requires more data points in their neighborhood. In such case, images belonging to infrequent scenarios will be either labeled as noise points or border points (belonging to other clusters). The same is true for K-means, where these points are usually spread across other clusters because they cannot form a cluster.

To strengthen our findings, in Table~\ref{tab:rq3_stat}, we report the results when comparing the purity of the selected pipelines for frequent and infrequent failure scenarios; further, we report the Vargha and Delaney's $\hat{A}_{12}$ effect size and the $p$-values resulting from performing a Mann-Whitney U-test. 
We notice that for all pipelines, the difference between frequent and infrequent scenarios are significant (p-value < 0.05). However, the effect sizes for Pipelines $26$, $62$, $45$, and $80$ are small, while they are
medium for Pipelines $19$ and $44$, which indicates that 
pipelines including DBSCAN (i.e., Pipelines $26$, $62$, $45$, and $80$) are much more \MINOR{R2.1}{reliable with} infrequent scenarios than others (i.e., the difference between frequent and infrequent scenarios is less pronounced).
Actually, the pipelines using DBSCAN fare better than the rest also in the general case. Indeed, almost all the injected failure scenarios with frequency above 18\% have 100\% purity (see Figure~\ref{fig:rq3_purity}); further for infrequent failure scenarios they include less data points below 100\% than the other pipelines. This is because DBSCAN tends to find clusters with different sizes if these clusters are dense enough; K-means, instead, derives clusters that are of similar size.

\input{tables/RQ3_stat}
\input{tables/RQ3_stat4} 
  
Further, we notice that the purity of the clusters generated by Pipeline $26$
(\emph{VGG16/Dbscan/UM\-AP/NoFT}), for infrequent failure scenarios, 
is higher (average is 94\%) than the purity of the clusters generated by the other pipelines; differences are significant (see Table~\ref{tab:rq3_stat4}), thus suggesting Pipeline $26$ might be the best choice.

\begin{figure}[t]
 \centering
 \includegraphics[width=\textwidth]{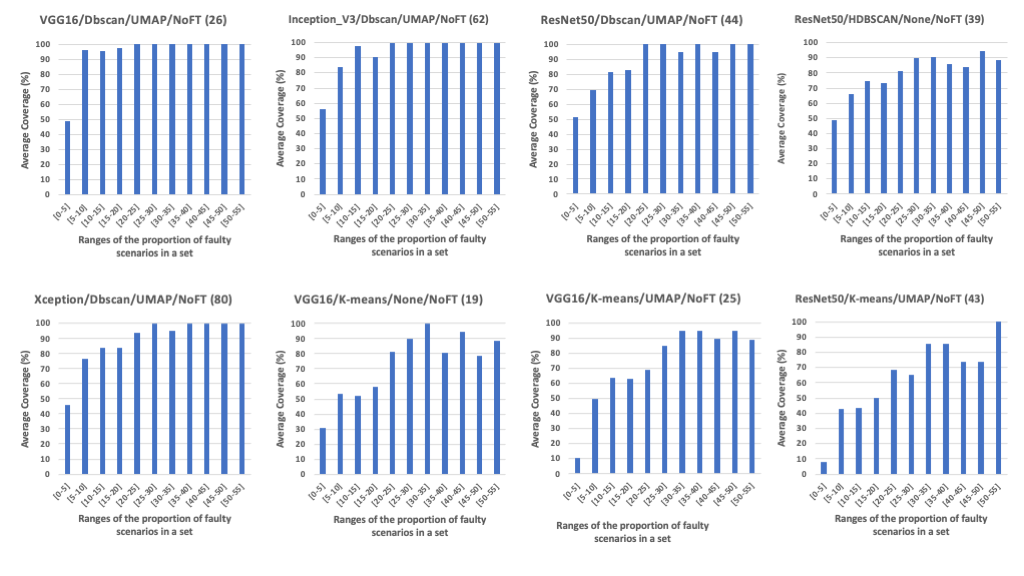}
 \caption{Comparing the percentage of coverage across different ranges of proportions of failure scenarios in each set.}
 \label{fig:rq3_coverage}
 \end{figure}

Concerning \emph{coverage}, Figure~\ref{fig:rq3_coverage} shows, for each pipeline, histograms with the average coverage obtained for failure scenarios having  proportions of failure inducing images within specific ranges. In general, we observe that coverage is higher for frequent scenarios. This is due to the correlation between pure clusters and coverage; the less pure the generated clusters, the fewer failure scenarios they cover. When the failure scenarios are infrequent, their images are distributed over the other clusters, reducing their purity and, thus, reducing the probability of these scenarios being covered. To demonstrate the significance of the difference between coverage results obtained with frequent and infrequent scenarios, we apply the Fisher's Exact test\footnote{The Fisher's Exact test~\cite{upton1992fisher} is a statistical test used to determine if there is a non-random association between two categorical variables~\cite{Fisher}.} to compare the coverage of frequent and infrequent scenarios for the clusters generated by the selected pipelines. We report the $p$-values resulting from the Fisher's Exact test in Table~\ref{tab:rq3_stat2} and observe that  
differences are statistically significant thus indicating that  pipelines perform better with frequent failure scenarios.
\input{tables/RQ3_stat2}

Further, Figure~\ref{fig:rq3_coverage} shows that pipeline $62$ (\emph{InceptionV3/DBSCAN/UMAP/NoFT}) is the one performing best with the least frequent scenarios (i.e., range 0-5\%) but no pipeline fares well in that range. Pipeline $26$ (\emph{VGG16/DBSCAN/UMAP/NoFT}) is the one performing best with infrequent scenarios in the range 5\% to 20\%; indeed, it is the only pipeline providing an average coverage above 90\% for that range.
To further demonstrate the significance of the difference in performance between Pipeline $26$ and the other pipelines, we apply Fisher's exact test to the coverage obtained for infrequent scenarios.
We report the $p$-values resulting from this test in Table~\ref{tab:rq3_stat3}. We notice that all the $p$-values are below $0.05$ except when Pipeline $26$ is compared to Pipeline $62$; 
indeed, the results of these two pipelines are similar as visible in Figure~\ref{fig:rq3_coverage}), even though Pipeline $26$
performs slightly better on average.

\input{tables/RQ3_stat3}

\MAJOR{R1.15}{In conclusion, infrequent failure scenarios affect both purity and coverage; pipelines tend to perform worse when the failure scenarios are infrequent (their frequency is below the median). However, some pipelines fare better than others. Our results suggest that the pipeline relying on a non-fine-tuned VGG16 model, with UMAP and DBSCAN (Pipeline $26$) is the best choice because it yields significantly higher purity and coverage than the other pipelines.}
Pipeline 26 is also less negatively affected by infrequent failure scenarios since its coverage is above 90\% when the frequency is above 5\%, which is not the case for all the other pipelines.

 \MAJOR{R3.7}{\subsection{RQ4: How do pipelines perform with failure scenarios that are not synthetically injected?}
\label{sec:pre-existing-results}}
\MAJORBEGIN{}
\subsubsection{Design and measurements.} 
Our objective is to determine if the best pipelines identified in RQ1, RQ2, and RQ3 perform best also with pre-existing failure scenarios.
As stated in Section~\ref{sec:pre-existing}, to address this research question we considered only the subject DNNs for which it is possible to determine the pre-existing failure scenarios each failure-inducing image may belong to; the selected DNNs are OC, GD, and HPD. The list of pre-existing failure scenarios is shown in Table~\ref{tab:pre-existing-distribution} (page~\pageref{tab:pre-existing-distribution}).

A pipeline should, ideally, identify all the pre-existing failure scenarios (i.e., generate at least one cluster for each pre-existing failure scenario thus maximizing coverage). Also, the generated clusters should be pure, that is, include only images belonging to a same pre-existing failure scenario.
Consequently, as for RQ1 to RQ3, we compare pipelines based on the purity and coverage of the generated clusters.

\subsubsection{Methodology}
For each subject DNN, we applied all our pipelines to the set of failure-inducing images in the original test set and belonging to a pre-existing failure scenario.

As per RQ1 to RQ3, we compute coverage and purity of each cluster as follows.
For each image, we know the pre-existing failure scenarios it belongs to. Therefore, for each generated cluster, we can determine the number of images belonging to each pre-existing failure scenario.
Each cluster is considered to cover the pre-existing failure scenario with the largest number of clustered images; indeed, being the most frequent, the commonalities in those images are likely to be noticed by the engineer inspecting the results. For each cluster, purity is computed as the
proportion of clustered images belonging to the scenario covered by the cluster.

We consider the pipelines leading to the best results for purity and coverage for RQ1, RQ2, and RQ3, and compare them with the pipelines leading to the best purity and coverage results when applied to the failure inducing images described above, across the three selected subject DNNs.

\subsubsection{Results} 

In Table~\ref{tab:pre-existing-results}, we report the pipelines leading to the best purity and coverage when applied to the datasets with injected (RQ1, RQ2, RQ3) and pre-existing failure scenarios (i.e., RQ4). The values in parentheses capture the ranking of a pipeline for each dataset. For both purity and coverage, for each RQ,  we rank our pipelines after sorting them in a decreasing order based on the average of the metric value computed for the OC, HPD, and GD DNNs; pipelines having the same average are assigned the same rank.

The results in Table~\ref{tab:pre-existing-results} show that the pipeline with the highest coverage for pre-existing failure scenarios is 
Pipeline 26 (see column \emph{Coverage-RQ4}), which confirms our findings for RQ3 (Section~\ref{sec:rq3:results}) where pipeline 26 leads to the highest coverage results when failure scenarios do not occur with the same frequency; the results observed for RQ4 can thus be explained by the fact that, in the original test set, failure scenarios do not have the same frequency. Further, Pipeline 26 achieves high purity with pre-existing failure scenarios; indeed it is ranked 4th in column \emph{Purity-RQ4}. Interestingly, a white-box pipeline (i.e., Pipeline 8, combining HUDD, DBSCAN and UMAP) leads to the highest purity for RQ4's dataset; however, it does not lead to the best coverage (only 91\%, ranked 7-th). Since in safety-critical systems one would prioritize the discovery of all  failure scenarios, Pipeline 26 should be a better option than Pipeline 8; indeed, Pipeline 26 achieves top coverage while having a very high purity (87\% VS 92\% of Pipeline 8). Further, for pre-existing failure scenarios, Pipeline 26 is the only pipeline with a purity rank up to 4 being among the best 10 pipelines for coverage. 

Pipeline 26 and Pipeline 80 are the only two pipelines being among the best ten for both purity and coverage, with pre-existing failure scenarios. Also, they are among the ten best pipelines for all the other datasets (i.e., RQ1, RQ2, and R3). 
More in general, the Pipelines 26, 44, 62, and 80, which are all the pipelines relying on 
 transfer learning, DBSCAN, and not using fine-tuning, lead to top ranked results.
However, only Pipeline 26 achieves the highest rank for more than one dataset, thus confirming it is a preferable choice as we suggested in Section~\ref{sec:rq3:results}.

Interestingly, four of the ten best-ranked pipelines for coverage with pre-existing failure scenarios include fine-tuning; however, they poorly perform in terms of purity. Based on our discussion in Section~\ref{sec:res:RQ1}, it is predictable that fine-tuning performs better with pre-existing failure scenarios; indeed, the failure-inducing images do not differ from the ones considered for fine-tuning (i.e., fine-tuning captures features that are present in the failure-inducing test set). However, the reason fine-tuning did not help achieve clusters with high purity is its reliance on a dataset with different scenarios occurring according to very different frequencies. Indeed, fine-tuning may overfit the features belonging to the most frequent scenarios, consequently the fine-tuned autoencoder may not extract relevant features for infrequent scenarios.
To conclude, fine-tuning seems not to be advisable because (1) failure scenarios, as shown in our experiment, are unlikely to include the same proportion of images, (2) it is not realistic to expect engineers to construct datasets with the same proportion of images for all failure scenarios, 
and (3) failure scenarios may largely differ from the images observed in the training set, which led to poor performance for fine-tuned pipelines in Section~\ref{sec:res:RQ1}.

\input{tables/pre-existing-results.tex}

\MAJOREND

\subsection{Discussion}

The results of RQ1 and RQ2 show that there is a family of pipelines leading to higher purity (i.e., they simplify the identification of root causes) and coverage (i.e., they enable the identification of all root causes). Such pipelines rely on transfer learning, UMAP for dimensionality reduction, DBSCAN for clustering, and are not using fine-tuning. Among such pipelines, considering that it is reasonable to expect unsafe scenarios to be infrequent, based on the results of RQ3, we suggest to use the pipeline relying on VGG16 (Pipeline 26) 
as transfer learning model. \MAJOR{3.2}{Pipeline 26 also leads to the best results when applied to pre-existing failure scenarios (RQ4), probably  due to infrequent pre-existing failure scenarios.}

\begin{figure}[H]
 \centering
 \includegraphics[width=0.8\textwidth]{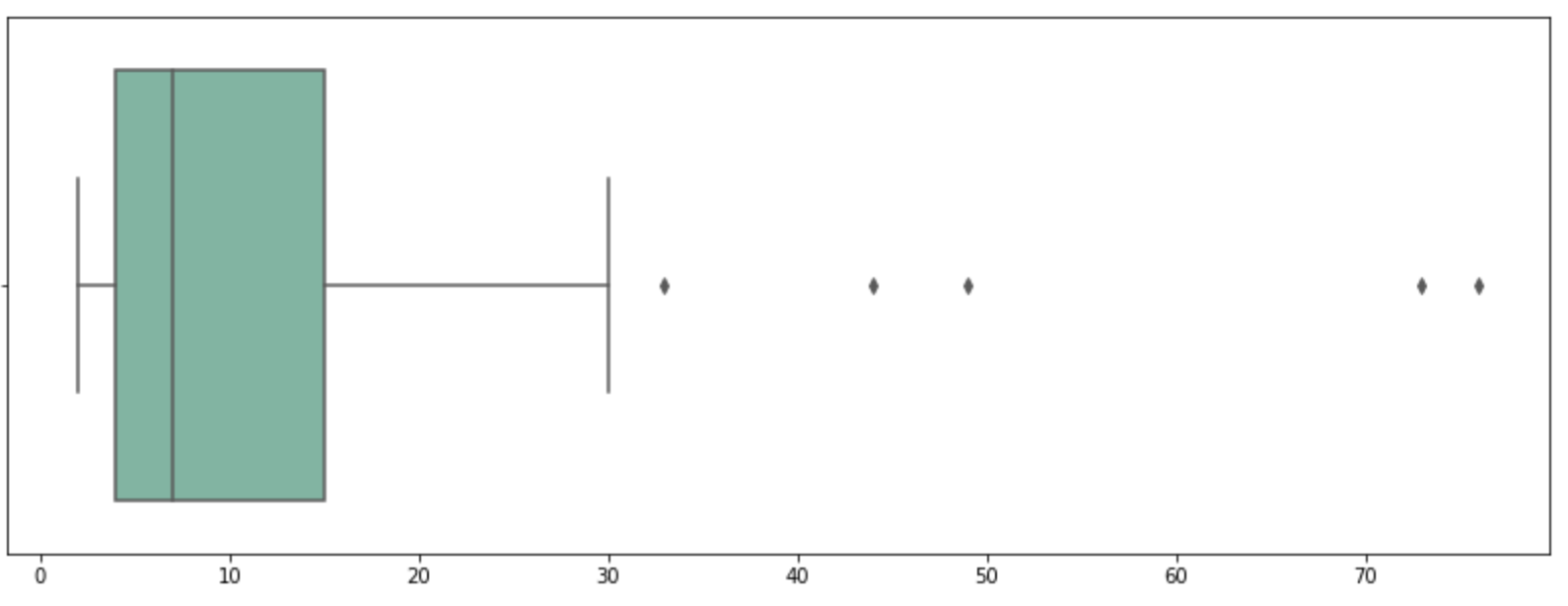}
 \caption{Box plot capturing the distribution of the size of the  clusters generated by the best pipeline (VGG16/Dbscan/UMAP/NoFT)}\label{fig:clustersize}
 \end{figure}

In our study, we focused on effectiveness, not cost; indeed, our main purpose is to identify the pipeline that generates clusters that do not confuse the end-user (i.e., they are pure) and is likely to help identify all the root causes of failures (i.e., they have high coverage). In contrast, cost is related to the number of clusters being inspected.
\MAJOR{R3.9}{To discuss such cost,
we report in Figure~\ref{fig:clustersize} a boxplot with the size of the clusters generated for RQ1, RQ2, RQ3, and RQ4 by Pipeline 26.
 As shown in Figure~\ref{fig:clustersize}, across all our experiments, the number of images per cluster ranges from 2 to 76, with 75\% of our clusters including at most 13 images (third quartile in Figure~\ref{fig:clustersize}). Based on such numbers, we can conclude that the effort required to inspect a cluster is limited (i.e., at most 13 images to be visualized for 75\% of our clusters); further, we have previously demonstrated through a user study that the inspection of five images per cluster is sufficient for a correct identification of the root cause of a DNN failure~\cite{attaoui2022black}.
Finally}, our root cause analysis toolset~\cite{HUDD:tool} includes the generation of animated gifs, one for each cluster, thus enabling the quick visualization of all the images in a cluster. 
\MAJOR{R3.5}{In conclusion, either with animated gifs, or when cluster images are inspected in sequence, we conjecture that the number of clusters' images does not strongly impact cost since clusters are typically small and small subsets of larger clusters are sufficient for a correct identification of failure root causes.} 

What is important, instead, is the purity of clusters as low purity makes it difficult for the end-user to determine commonalities among images.
Nevertheless, to further discuss cost, we measure the  number of clusters to be inspected for each pipeline considering the dataset used for RQ1 and RQ2. We count only clusters capturing the injected failure scenarios.
A lower number of clusters should indicate lower cost and, since a number of clusters higher than the number of failure scenarios to be discovered implies the presence of redundant clusters, we compute the degree of redundancy as: 

$$\mathit{redundancy\ ratio}=\frac{\mathit{number\ of\ clusters}}{\mathit{covered\ failure\ scenarios}}$$

  \begin{figure}[t]
 \centering
 \includegraphics[width=\textwidth]{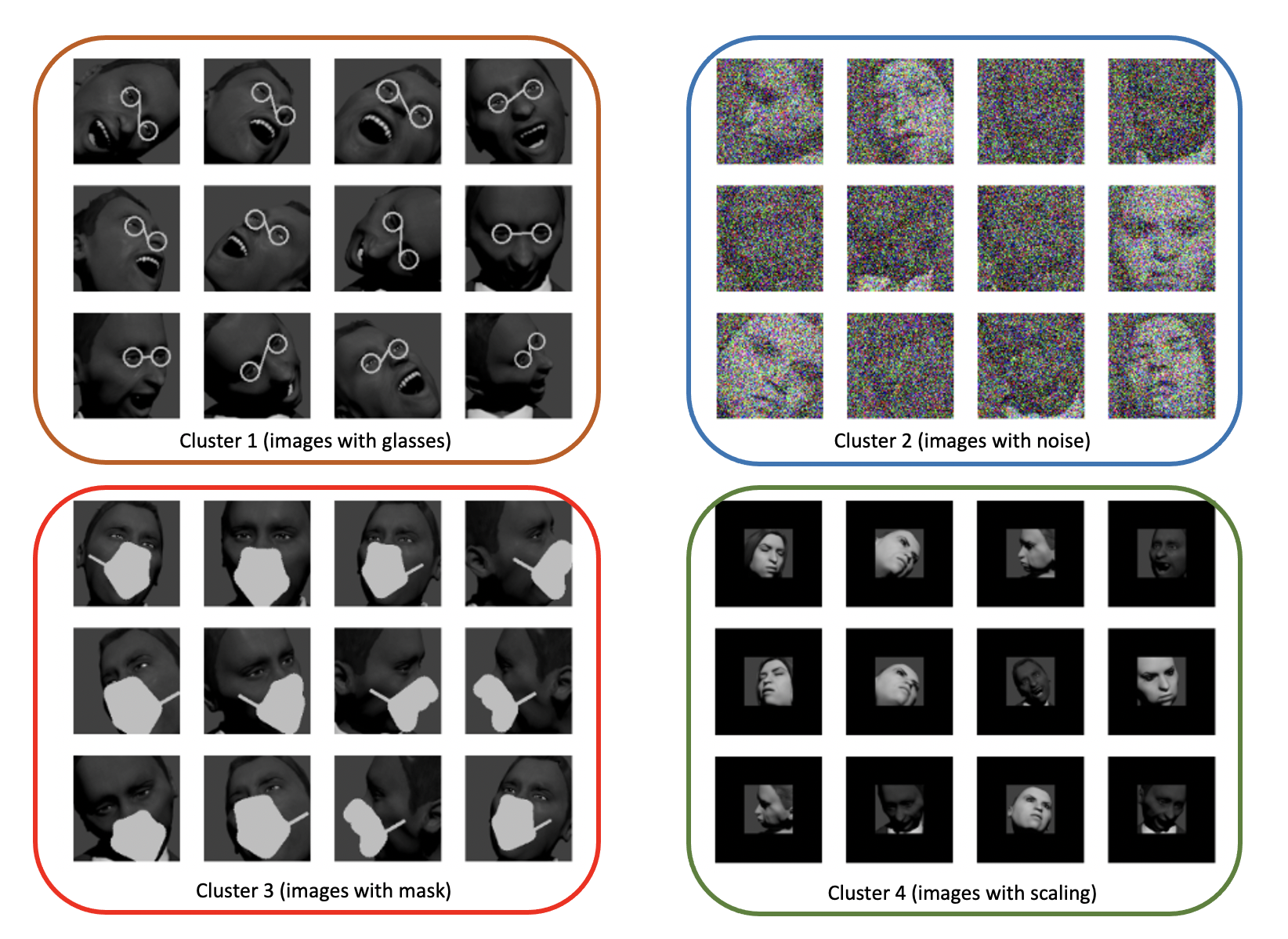}
 \caption{Examples of clusters generated by pipeline 26 for the HPD case study subject.}\label{fig:clusters_example}
 \end{figure}
Finally, to discuss how well each pipeline improves current practice in industry, we estimate the degree of savings with respect to the such practice, which entails the visual inspection of all images. To do so, we assume that inspecting a single cluster, \MAJOR{R3.9}{especially when using animated gifs,} is as inexpensive as visualizing one single image. Indeed, though clusters involve several images, through animation, they actually make it easier to quickly identify commonalities rather than guessing root causes from a single image.  Figure~\ref{fig:clusters_example} shows four example clusters where all the images present a commonality (i.e., the root cause of the DNN failure) that is easy to determine when visualizing all the images in a sequence.
Therefore, we estimate savings as:

$$\mathit{savings}=1-\frac{\mathit{number\ of\ clusters}}{\mathit{number\ of\ images}}$$
\input{tables/redundancy.tex}
Table~\ref{tab:redundancy} shows our results; it reports the number of RCCs generated for each case study DNN and across all of them. Further, it reports the \emph{percentage and number of failure scenarios} covered by each pipeline (used to compute redundancy and providing information about the effectiveness of a pipeline), along with \emph{redundancy ratio} and \emph{savings}. We report only the results for the best pipelines identified when addressing RQ1 to RQ2 because there is no reason to select pipelines that do not achieve high purity and coverage. 
 
The number of clusters generated by the selected pipelines ranges between 18 and 284. The pipelines leading to the lowest number of clusters are the ones including K-means:
\emph{ResNet50/K-means/UMAP/NoFT} (18),
\emph{VGG16/K-means/None/NoFT} (19), and \emph{VGG16/K-means/UMAP/NoFT} (24). Pipelines with DBSCAN and HDBSCAN lead to a much higher number of clusters. To discuss the practical impact of such a high number of clusters, we focus on the \emph{redundancy ratio}, which ranges between 1.12 and 11.8; the redundancy ratio indicates that the pipeline with the highest number of clusters (i.e., \emph{ResNet50/HDBSCAN/None/NoFT}), on average, presents 11 redundant clusters for each identified failure scenario. Given that, in the presence of pure clusters, understanding the scenario captured by one pipeline is quick with animated gifs, we consider that inspecting 11 redundant clusters per fault has a limited impact on cost.
Finally, if we focus on \emph{savings}, we can observe that respect to current practice, all the pipelines except (\emph{ResNet50/HDBSCAN/Only/NoFT}) lead to savings above 90\%, thus showing that their adoption is highly beneficial. 

Although the pipelines including K-means lead to the lowest cost, their coverage is particularly low for infrequent scenarios (see Table~\ref{tab:rq3_stat2}, with coverage below 35\% for the range [0-5], and below 60\% for the range [5-10]), which is bound to be a common situation in practice.
Since pipelines leading to a small number of clusters can be highly ineffective in realistic safety-critical contexts (i.e., when some failure scenarios are infrequent), assuming that redundant clusters are easy to manage, we conclude that the best choice are the pipelines that maximize purity and coverage, as discussed above (i.e., Pipeline $26$, \emph{VGG16/DBSCAN/UMAP/NoFT}). A possible tradeoff is Pipeline $80$ (\emph{Xception/DBSCAN/UMAP/NoFT}), which is among the best performing for RQ3 (e.g., coverage above 40\% for the range [0-5], and above 70\% for the range [5-10]) and leads to 3.6 redundant clusters only, on average.

%% file: tables/tableDNNs.tex
\begin{table}[tb]
\centering
\caption{Case Study Systems}
\footnotesize
\begin{tabular}{
|@{\hspace{1pt}}p{6mm}
|@{\hspace{1pt}}p{15mm}
|@{\hspace{1pt}}p{10mm}
|@{\hspace{1pt}}p{15mm}
|@{\hspace{1pt}}p{10mm}
|@{\hspace{1pt}}p{5mm}
|@{\hspace{1pt}}p{5mm}
|@{\hspace{1pt}}p{5mm}
|@{\hspace{1pt}}p{5mm}
|@{\hspace{1pt}}p{5mm}
|@{\hspace{1pt}}p{5mm}
|@{\hspace{1pt}}p{5mm}
|@{\hspace{1pt}}p{5mm}|
@{\hspace{1pt}}p{5mm}|
@{\hspace{1pt}}p{5mm}|
@{\hspace{1pt}}p{5mm}|
@{\hspace{1pt}}p{5mm}|
@{\hspace{1pt}}p{5mm}|
@{\hspace{1pt}}p{5mm}|
@{\hspace{1pt}}p{5mm}|
@{\hspace{1pt}}p{5mm}|
@{\hspace{1pt}}p{5mm}|}
\hline
\textbf{DNN}&		 \textbf{Data}	&\textbf{Training}& \textbf{Test}  & \textbf{Failure}  & \textbf{\#}& \textbf{\#} & \textbf{\# }& \textbf{\#}& \textbf{\# } & \textbf{\# } & \textbf{\# } & \textbf{\# }& \textbf{\# } & \textbf{\# }\\
&		 \textbf{Source}& \textbf{Set Size} & \textbf{Set Size (Accuracy)} & \textbf{inducing images} & \textbf{ M$^1$}& \textbf{ N$^2$} & \textbf{H$^3$}&  \textbf{B$^5$}& \textbf{SG$^6$}& \textbf{EG$^7$} & \textbf{EO$^8$}& \textbf{S$^9$}& \textbf{D$^{10}$}& \textbf{NF$^{11}$}	\\
\hline
\GD&	 UnityEyes & 61,063 &  132,630 (96\%) &5,371 &- &80 & -& 80  &- &- &- &80 & 80& 80\\
\CloseDNN&      UnityEyes &1,704	  &	4,232 (88\%) &506& -&20 &- &   20&- &- &- &  20& 20&20\\
\HPD&      Blender &16,013  &	2,825 (44\%) &1,580 & 90 &90 &90 & 90 &90 &90 & -& 90 & 90 & 90\\
SVIRO&Blender      &15,489  &	3,427 (74\%) &884 & -& 30& -& 30 & -& -& 30& 30& 30& 30\\
SAP&  Autopilot~\cite{sullychen} &33,808  &45,406 (84\%) &7,169& -& 90&- &  90 & -& -& -& 90& 90& 90\\
CPD&  Apollo~\cite{apollo}     &5,208  &	4,996 (91\%) &444\textbf{} & -& 90& -& 90 & -& -& -& 90& 90& 90\\
\hline
\end{tabular}

\label{tab:dnns}
\footnotesize{$^1$ Mask $^2$ Noise $^3$ Hand $^5$ Blurriness $^{6}$ SunGlasses $^7$ EyeGlasses $^8$ Everyday Object $^9$ Scaling $^{10}$ Darkness $^{11}$ No Injected Fault }\\

\end{table}%

%% file: tables/tableBoundaries.tex
\begin{table}[H]
\caption{Failure-inducing parameters considered to address RQ4}
\footnotesize
\begin{tabular}{
|@{\hspace{1pt}}p{7mm}
|@{\hspace{1pt}}p{3cm}
|@{\hspace{1pt}}p{8cm}|}
\hline
\textbf{DNN}&\textbf{Parameter}&\textbf{Unsafe values}\\
\hline
\multirow{7}{*}{\GD,\OC}&Gaze Angle&Values used to label the gaze angle in eight classes (i.e., 22.5$^{\circ}$, 67.5$^{\circ}$, 112.5$^{\circ}$, 157.5$^{\circ}$, 202.5$^{\circ}$, 247.5$^{\circ}$, 292.5$^{\circ}$, 337.5$^{\circ}$).\\
&Openness&Value used to label the gaze openness in two classes (i.e., 20 pixels) %
.
\\
&H\_Headpose&Values indicating a head turned completely left or right (i.e., 160$^{\circ}$, 220$^{\circ}$)\\
&V\_Headpose&Values indicating a head looking at the very top/bottom (i.e., 20$^{\circ}$, 340$^{\circ}$)\\
&DistToCenter&Value below which the eye is looking middle center (i.e., 11.5 pixels).\\
&PupilToBottom&Value below which the pupil is mostly under the eyelid (i.e., -16 pixels).\\
&TopToPupil&Value below which the pupil is mostly above the eyelid (i.e., -16 pixels).\\
\hline
\multirow{2}{*}{\HPD}&Headpose-X&Boundary cases (i.e.,-28.88$^{\circ}$,21.35$^{\circ}$), values used to label the headpose in nine classes (-10$^{\circ}$,10$^{\circ}$).\\
\cline{2-3}
&Headpose-Y&Boundary cases (i.e.,-88.10$^{\circ}$,74.17$^{\circ}$), values used to label the headpose in nine classes (-10$^{\circ}$,10$^{\circ}$).\\
\hline
\end{tabular}

\label{tab:boundary2}
\end{table}%

%% file: tables/pre-existing-distribution.tex
\begin{table}[tb]
\caption{Size of the failure-inducing set and the distribution of pre-existing failure scenario for each case study.}
\label{tab:pre-existing-distribution}
\footnotesize
\begin{tabular}{|c|c|c|c|}
\hline
                              & GD   & OC  & HPD \\ \hline

\# of failure inducing images & 4937 & 283 & 865 \\ \hline
Unrealistic               & /    & 279 & /   \\ \hline
Openness 25                   & 795  & 192 & /   \\ \hline
Border 337.5                  & 726  & /   & /   \\ \hline
Border 22.5                   & 619  & /   & /   \\ \hline
Border 67.5                   & 386  & /   & /   \\ \hline
Border 112.5                  & 564  & /   & /   \\ \hline
Border 157.5                  & 635  & /   & /   \\ \hline
Border 202.5                  & 685  & /   & /   \\ \hline
Border 247.5                  & 486  & /   & /   \\ \hline
Border 292.5                  & 665  & /   & /   \\ \hline
Headpose-X: -10               & /    & /   & 55  \\ \hline
Headpose-X: +10               & /    & /   & 319 \\ \hline
Headpose-X: -28               & /    & /   & 6   \\ \hline
Headpose-X: +21               & /    & /   & 16  \\ \hline
Headpose-Y: -10               & /    & /   & 431 \\ \hline
Headpose-Y: +10               & /    & /   & 163 \\ \hline
Headpose-Y: -88               & /    & /   & 0   \\ \hline
Headpose-Y: +74               & /    & /   & 14  \\ \hline
\end{tabular}
\end{table}

%% file: tables/VDA_results.tex
\begin{table}[H]
\centering
\caption{RQ1: p-values and and effect size values when comparing the results of the pipelines with the best purity of clusters (according to the decision tree) to the other pipelines.}
\label{tab:rq1_stat}
\begin{tabular}{|c|c|c|c|}
\hline
               & Node 3 & Node 4 & Node 6 \\ \hline
p-value        & 7e-11  & 2e-7   & 5e-6   \\ \hline
$\hat{A}_{12}$ & 1.00   & 1.00   & 0.80   \\ \hline
\end{tabular}
\end{table}

%% file: tables/RQ1_Purities_top.tex
\begin{table*}[ht]
\centering
\smaller
\footnotesize
\caption{RQ1: Pipelines with a purity greater than $90\%$. The last column represents the average of averages.}
\label{tab:rq1_top}
\makebox[0.9\textwidth]{\begin{tabular}{|c|c|c|c|c||c|c|c|c|c|c||c|}
\hline
\multicolumn{5}{|c||}{\textbf{Pipelines}} & \multicolumn{7}{c|}{\textbf{Avg. purity across RCCs}}
\\ \hline
\# & \textbf{FE} & \textbf{FT} & \textbf{DR} & \textbf{CA} & \textbf{GD} & \textbf{OC} & \textbf{HPD} & \textbf{SVIRO} & \textbf{SAP} & \textbf{CPD} & \textbf{Avg.} \\ \hline

19 & VGG-16 & NO & None & K-Means & 91.7\% & 92.1\% & 95.5\% & 82.5\% & 97.3\% & 99.7\% & 93.2\% \\ \hline

25 & VGG-16 & NO & UMAP & K-Means & 97.6\% & 84.4\% & 93.7\% & 82.4\% & 90.3\% & 97.8\% & 91.0\% \\ \hline

26 & VGG-16 & NO & UMAP & DBSCAN & 99.0\% & 93.0\% & 99.6\% & 79.7\% & 98.1\% & 96.6\% & 94.3\% \\ \hline

39 & ResNet-50 & NO & None & HDBSCAN & 96.4\% & 100.0\% & 100.0\% & 78.8\% & 87.5\% & 100.0\% & 93.8\% \\ \hline

43 & ResNet-50 & NO & UMAP & K-Means & 99.4\% & 93.0\% & 82.3\% & 79.6\% & 99.6\% & 97.7\% & 91.9\% \\ \hline

44 & ResNet-50 & NO & UMAP & DBSCAN & 100.0\% & 95.8\% & 95.8\% & 79.0\% & 99.7\% & 99.3\% & 94.9\% \\ \hline

62 & Inception-V3 & NO & UMAP & DBSCAN & 93.4\% & 95.2\% & 98.1\% & 76.6\% & 97.4\% & 83.1\% & 90.7\% \\ \hline

\end{tabular}}

\footnotesize{$^{FE}$ Feature Extraction $^{FT}$ Fine-tuning $^{DR}$ Dimensionality Reduction $^{CA}$ Clustering Algorithm}\\
\end{table*}

%% file: tables/VDA_results2.tex
\begin{longtable}[c]{|c|c|c|c|c|}
\caption{RQ2: p-values and and effect size values when comparing the results of the pipelines with the best coverage of the faulty scenarios (according to the decision tree) to the other pipelines.}
\label{tab:rq2_stat}\\
\hline
               & Node 3                      & Node 4                      & Node 6                      & Node 8                      \\ \hline
\endhead
p-value        & 1e-5 & 1e-5 & 4e-5 & 8e-3 \\ \hline
$\hat{A}_{12}$ & 0.95                        & 1.00                        & 0.91                        & 0.77                        \\ \hline
\end{longtable}

%% file: tables/RQ2_Coverages_top.tex
\begin{table*}[ht]
\centering
\smaller
\footnotesize
\caption{RQ2: Pipelines with a coverage greater than $90\%$. The last column represents the average of averages.}
\label{tab:rq2_top}
\makebox[0.9\textwidth]{\begin{tabular}{|c|c|c|c|c||c|c|c|c|c|c||c|}
\hline
\multicolumn{5}{|c||}{\textbf{Pipelines}} & \multicolumn{7}{c|}{\textbf{Percentage of covered faulty scenarios}}
\\ \hline
\# & \textbf{FE} & \textbf{FT} & \textbf{DR} & \textbf{CA} & \textbf{GD} & \textbf{OC} & \textbf{HPD} & \textbf{SVIRO} & \textbf{SAP} & \textbf{CPD} & \textbf{Avg.} \\ \hline

26 & VGG-16 & None & UMAP & DBSCAN & 100.0\% & 100.0\% & 100.0\% & 80.0\% & 100.0\% & 100.0\% & 96.7\% \\ \hline

44 & ResNet-50 & None & UMAP & DBSCAN & 100.0\% & 100.0\% & 100.0\% & 60.0\% & 100.0\% & 100.0\% & 93.3\% \\ \hline

62 & Inception-V3 & None & UMAP & DBSCAN & 100.0\% & 100.0\% & 100.0\% & 100.0\% & 100.0\% & 100.0\% & 100.0\% \\ \hline

80 & Xception & None & UMAP & DBSCAN & 100.0\% & 100.0\% & 100.0\% & 40.0\% & 100.0\% & 100.0\% & 90.0\% \\ \hline

\end{tabular}}

\footnotesize{$^{FE}$ Feature Extraction $^{FT}$ Fine-tuning $^{DR}$ Dimensionality Reduction $^{CA}$ Clustering Algorithm}\\

\end{table*}

%% file: tables/RQ3_stat.tex
\begin{table}[t]
\caption{RQ3: p-values and effect size values when comparing the purity of the best pipelines with the frequent and infrequent failure scenarios.}
\label{tab:rq3_stat}
\begin{tabular}[c]{|p{5cm}|c|c|c|c|c|c|c|c|}
\hline
\textbf{Pipelines}      & 26      & 44    & 62 & 39 & 80  & 19    & 25 & 43     \\ \hline
\textbf{Average Purity for infrequent failure scenarios} & 94\% & 87\% & 91\% & 79\% & 87\% & 76\% & 70\% & 65\%\\ \hline
\textbf{Average Purity for frequent failure scenarios} & 100\% & 100\% & 100\% & 92\% & 99\% & 96\% & 96\% & 93\%\\ \hline
\textbf{p-value}        & 4e-6 & 2e-10 & 1e-6    &  2e-9  & 8e-9 & 8e-5 & 2e-10 & 3e-14 \\ \hline
\textbf{$\hat{A}_{12}$} & 0.58                        & 0.64                         & 0.59                           & 0.60                        & 0.63                        & 0.68 & 0.70 & 0.75                        \\ \hline
\end{tabular}
\end{table}

%% file: tables/RQ3_stat4.tex
\begin{table}[t]
\caption{RQ3: p-values and effect size values when comparing the best pipeline in Table~\ref{tab:rq3_stat} (i.e., Pipeline $26$, \emph{VGG16/Dbscan/UMAP/NoFT}) to the other pipelines based on the average purity of the clusters associated to infrequent failure scenarios.}
\label{tab:rq3_stat4}
\begin{tabular}[t]{|p{5cm}|c|c|c|c|c|c|c|}
\hline
\textbf{Pipelines }        & 44    & 62    & 39 & 80    & 19     & 25 & 43   \\ \hline
\textbf{p-value}        & 0.002 & 0.51 & 4e-5   &  0.006  & 3e-12 & 2e-14 & 4e-21  \\ \hline
\textbf{$\hat{A}_{12}$} & 0.57                        & 0.51                         & 0.60                          & 0.56                        & 0.69                       & 0.71 & 0.77                        \\ \hline
\end{tabular}
\end{table}

%% file: tables/RQ3_stat2.tex
\begin{table}[t]
\caption{RQ3: Fisher exact test values when comparing the coverage of the lowly represented and highly represented faulty scenarios by the clusters generated by the best pipelines.}
\label{tab:rq3_stat2}
\begin{tabular}[t]{|p{5cm}|c|c|c|c|c|c|c|c|}
\hline
Pipelines         & 26    & 44    & 62    & 39 &  80    & 19  & 25 & 43    \\ \hline

\textbf{Average Coverage for infrequent failure scenarios} & 85\% & 71\% & 82\% & 66\% & 73\% & 51\% & 46\% & 34\% \\ \hline
\textbf{Average Coverage for frequent failure scenarios} & 100\% & 98\% & 99\% & 86\% & 98\% & 86\% & 87\% & 77\%\\ \hline
\textbf{Fisher's Exact tes}t & 1e-5 & 1e-5 & 1e-5 &  1e-5 & 1e-5 & 2e-4 & 1e-5 & 1e-5 \\ \hline
\end{tabular}
\end{table}

%% file: tables/RQ3_stat3.tex
\begin{table}[t]
\caption{RQ3: Fisher exact test values when comparing the best pipeline "VGG16/Dbscan/UMAP/NoFT" to the other pipelines based on the coverage of the infrequent failure scenarios.}
\label{tab:rq3_stat3}
\begin{tabular}[t]{|p{5cm}|c|c|c|c|c|c|c|}
\hline
\textbf{Pipelines }        & 44    & 62    & 39 & 80    & 19     & 25 & 43   \\ \hline
\textbf{Fisher's Exact test} & 4e-2 & 0.55 & 1e-5 & 2e-3 & 0.018 &  1e-5 & 1e-5\\ \hline
\end{tabular}
\end{table}

%% file: tables/pre-existing-results.tex
\begin{table}[t]
\caption{Purity and Coverage of the best ten pipelines on datasets with injected and pre-existing failure scenarios; the values between parentheses indicate the rank of a pipeline for a RQ's dataset. 
We selected the top-ten ranked pipelines based on each of the datasets considered for RQ1, RQ2, RQ3, and RQ4.}
\label{tab:pre-existing-results}
\footnotesize
\begin{tabular}{|c|cccc|ccc|ccc|}
\hline
\textbf{}   & \multicolumn{4}{c|}{\textbf{Best Pipelines}}                                                                         & \multicolumn{3}{c|}{\textbf{Purity}}                                                                   & \multicolumn{3}{c|}{\textbf{Coverage}}                                                                 \\ \hline
\textbf{\#} & \multicolumn{1}{c|}{\textbf{FE}} & \multicolumn{1}{c|}{\textbf{FT}} & \multicolumn{1}{c|}{\textbf{CA}} & \textbf{DR} & \multicolumn{1}{c|}{\textbf{RQ1}} & \multicolumn{1}{c|}{\textbf{RQ3}} & \textbf{RQ4} & \multicolumn{1}{c|}{\textbf{RQ2}} & \multicolumn{1}{c|}{\textbf{RQ3}} & \textbf{RQ4} \\ \hline
5           & \multicolumn{1}{c|}{HUDD}        & \multicolumn{1}{c|}{NoFT}        & \multicolumn{1}{c|}{Dbscan}      & PCA         & \multicolumn{1}{c|}{55 (60)}      & \multicolumn{1}{c|}{57 (65)}      & 85 (10)                        & \multicolumn{1}{c|}{25 (54)}      & \multicolumn{1}{c|}{32 (60)}      & 86 (9)                         \\ \hline
8           & \multicolumn{1}{c|}{HUDD}        & \multicolumn{1}{c|}{NoFT}        & \multicolumn{1}{c|}{Dbscan}      & UMAP        & \multicolumn{1}{c|}{99 (5)}       & \multicolumn{1}{c|}{85 (13)}      & \textbf{92 (1)}                          & \multicolumn{1}{c|}{96 (9)}       & \multicolumn{1}{c|}{68 (14)}      & 91 (7)                         \\ \hline
17          & \multicolumn{1}{c|}{LRP}         & \multicolumn{1}{c|}{NoFT}        & \multicolumn{1}{c|}{Dbscan}      & UMAP        & \multicolumn{1}{c|}{99 (5)}       & \multicolumn{1}{c|}{84 (15)}      & 86 (8)                          & \multicolumn{1}{c|}{92 (10)}      & \multicolumn{1}{c|}{64 (21)}      & 70 (15)                        \\ \hline
19          & \multicolumn{1}{c|}{VGG16}       & \multicolumn{1}{c|}{NoFT}        & \multicolumn{1}{c|}{K-means}     & None        & \multicolumn{1}{c|}{94 (14)}      & \multicolumn{1}{c|}{91 (6)}       & 85 (10)                        & \multicolumn{1}{c|}{79 (15)}      & \multicolumn{1}{c|}{79 (8)}       & 51 (52)                        \\ \hline
20          & \multicolumn{1}{c|}{VGG16}       & \multicolumn{1}{c|}{NoFT}        & \multicolumn{1}{c|}{Dbscan}      & None        & \multicolumn{1}{c|}{78 (24)}      & \multicolumn{1}{c|}{76 (26)}      & 87 (3)                         & \multicolumn{1}{c|}{58 (22)}      & \multicolumn{1}{c|}{56 (25)}      & 62 (25)                        \\ \hline
22          & \multicolumn{1}{c|}{VGG16}       & \multicolumn{1}{c|}{NoFT}        & \multicolumn{1}{c|}{K-means}     & PCA         & \multicolumn{1}{c|}{84 (19)}      & \multicolumn{1}{c|}{81 (19)}      & 87 (4)                         & \multicolumn{1}{c|}{54 (34)}      & \multicolumn{1}{c|}{57 (23)}      & 47 (61)                        \\ \hline
25          & \multicolumn{1}{c|}{VGG16}       & \multicolumn{1}{c|}{NoFT}        & \multicolumn{1}{c|}{K-means}     & UMAP        & \multicolumn{1}{c|}{99 (5)}       & \multicolumn{1}{c|}{90 (8)}       & 87 (4)                         & \multicolumn{1}{c|}{\textbf{100 (1)}}      & \multicolumn{1}{c|}{76 (11)}      & 51 (52)                        \\ \hline
26          & \multicolumn{1}{c|}{VGG16}       & \multicolumn{1}{c|}{NoFT}        & \multicolumn{1}{c|}{Dbscan}      & UMAP        & \multicolumn{1}{c|}{\textbf{100 (1)}}      & \multicolumn{1}{c|}{97 (3)}       & 87 (4)                         & \multicolumn{1}{c|}{\textbf{100 (1)}}      & \multicolumn{1}{c|}{93 (4)}       & \textbf{100 (1)}                        \\ \hline
27          & \multicolumn{1}{c|}{VGG16}       & \multicolumn{1}{c|}{NoFT}        & \multicolumn{1}{c|}{HDBSCAN}     & PCA         & \multicolumn{1}{c|}{61 (50)}      & \multicolumn{1}{c|}{81 (19)}      & 90 (2)                         & \multicolumn{1}{c|}{37 (47)}      & \multicolumn{1}{c|}{70 (13)}      & 34 (94)                        \\ \hline
35          & \multicolumn{1}{c|}{VGG16}       & \multicolumn{1}{c|}{FT}          & \multicolumn{1}{c|}{Dbscan}      & UMAP        & \multicolumn{1}{c|}{74 (30)}      & \multicolumn{1}{c|}{72 (29)}      & 53 (67)                        & \multicolumn{1}{c|}{42 (43)}      & \multicolumn{1}{c|}{10 (85)}      & 77 (10)                        \\ \hline
43          & \multicolumn{1}{c|}{ResNet50}    & \multicolumn{1}{c|}{NoFT}        & \multicolumn{1}{c|}{K-means}     & UMAP        & \multicolumn{1}{c|}{97 (10)}      & \multicolumn{1}{c|}{86 (12)}      & 62 (49)                        & \multicolumn{1}{c|}{92 (10)}      & \multicolumn{1}{c|}{71 (12)}      & 47 (61)                        \\ \hline
44          & \multicolumn{1}{c|}{ResNet50}    & \multicolumn{1}{c|}{NoFT}        & \multicolumn{1}{c|}{Dbscan}      & UMAP        & \multicolumn{1}{c|}{\textbf{100 (1)}}      & \multicolumn{1}{c|}{97 (3)}       & 87 (4)                         & \multicolumn{1}{c|}{\textbf{100 (1)}}      & \multicolumn{1}{c|}{94 (3)}       & 95 (3)                         \\ \hline
53          & \multicolumn{1}{c|}{ResNet50}    & \multicolumn{1}{c|}{FT}          & \multicolumn{1}{c|}{Dbscan}      & UMAP        & \multicolumn{1}{c|}{76 (26)}      & \multicolumn{1}{c|}{66 (48)}      & 52 (68)                        & \multicolumn{1}{c|}{42 (43)}      & \multicolumn{1}{c|}{32 (60)}      & 86 (9)                         \\ \hline
55          & \multicolumn{1}{c|}{InceptionV3} & \multicolumn{1}{c|}{NoFT}        & \multicolumn{1}{c|}{K-means}     & None        & \multicolumn{1}{c|}{99 (5)}       & \multicolumn{1}{c|}{91 (6)}       & 65 (39)                        & \multicolumn{1}{c|}{\textbf{100 (1)}}      & \multicolumn{1}{c|}{81 (6)}       & 47 (61)                        \\ \hline
61          & \multicolumn{1}{c|}{InceptionV3} & \multicolumn{1}{c|}{NoFT}        & \multicolumn{1}{c|}{K-means}     & UMAP        & \multicolumn{1}{c|}{\textbf{100 (1)}}      & \multicolumn{1}{c|}{93 (5)}       & 65 (39)                        & \multicolumn{1}{c|}{\textbf{100 (1)}}      & \multicolumn{1}{c|}{85 (5)}       & 47 (61)                        \\ \hline
62          & \multicolumn{1}{c|}{InceptionV3} & \multicolumn{1}{c|}{NoFT}        & \multicolumn{1}{c|}{Dbscan}      & UMAP        & \multicolumn{1}{c|}{\textbf{100 (1)}}      & \multicolumn{1}{c|}{\textbf{98 (1)}}       & 84 (12)                        & \multicolumn{1}{c|}{\textbf{100 (1)}}      & \multicolumn{1}{c|}{\textbf{97 (1)}}       & 94 (6)                         \\ \hline
71          & \multicolumn{1}{c|}{InceptionV3} & \multicolumn{1}{c|}{FT}          & \multicolumn{1}{c|}{Dbscan}      & UMAP        & \multicolumn{1}{c|}{70 (37)}      & \multicolumn{1}{c|}{70 (37)}      & 55 (63)                        & \multicolumn{1}{c|}{42 (43)}      & \multicolumn{1}{c|}{38 (57)}      & 95 (3)                         \\ \hline
73          & \multicolumn{1}{c|}{Xception}    & \multicolumn{1}{c|}{NoFT}        & \multicolumn{1}{c|}{K-means}     & None        & \multicolumn{1}{c|}{97 (10)}      & \multicolumn{1}{c|}{89 (9)}       & 66 (35)                        & \multicolumn{1}{c|}{\textbf{100 (1)}}      & \multicolumn{1}{c|}{77 (10)}      & 51 (52)                        \\ \hline
75          & \multicolumn{1}{c|}{Xception}    & \multicolumn{1}{c|}{NoFT}        & \multicolumn{1}{c|}{HDBSCAN}     & None        & \multicolumn{1}{c|}{96 (12)}      & \multicolumn{1}{c|}{88 (11)}      & 63 (46)                        & \multicolumn{1}{c|}{92 (10)}      & \multicolumn{1}{c|}{81 (6)}       & 55 (48)                        \\ \hline
79          & \multicolumn{1}{c|}{Xception}    & \multicolumn{1}{c|}{NoFT}        & \multicolumn{1}{c|}{K-means}     & UMAP        & \multicolumn{1}{c|}{89 (17)}      & \multicolumn{1}{c|}{89 (9)}       & 64 (42)                        & \multicolumn{1}{c|}{92 (10)}      & \multicolumn{1}{c|}{79 (8)}       & 43 (72)                        \\ \hline
80          & \multicolumn{1}{c|}{Xception}    & \multicolumn{1}{c|}{NoFT}        & \multicolumn{1}{c|}{Dbscan}      & UMAP        & \multicolumn{1}{c|}{99 (5)}       & \multicolumn{1}{c|}{\textbf{98 (1)}}       & 86 (8)                         & \multicolumn{1}{c|}{\textbf{100 (1)}}      & \multicolumn{1}{c|}{95 (2)}       & 97 (2)                         \\ \hline
89          & \multicolumn{1}{c|}{Xception}    & \multicolumn{1}{c|}{FT}          & \multicolumn{1}{c|}{Dbscan}      & UMAP        & \multicolumn{1}{c|}{55 (60)}      & \multicolumn{1}{c|}{65 (50)}      & 55 (63)                        & \multicolumn{1}{c|}{8 (76)}       & \multicolumn{1}{c|}{46 (43)}      & 95 (3)                         \\ \hline
98          & \multicolumn{1}{c|}{AE}          & \multicolumn{1}{c|}{NoFT}        & \multicolumn{1}{c|}{Dbscan}      & UMAP        & \multicolumn{1}{c|}{91 (16)}      & \multicolumn{1}{c|}{77 (24)}      & 83 (13)                        & \multicolumn{1}{c|}{67 (18)}      & \multicolumn{1}{c|}{54 (27)}      & 76 (12)                        \\ \hline
\end{tabular}
\end{table}

%% file: tables/redundancy.tex
\begin{table}[]
\centering
\caption{The number of redundant clusters generated by the best pipelines for each case study subject and across all of them. The last columns represent the number and the percentage of failure scenarios covered by the pipelines, the redundancy ratio, and the savings.}
\label{tab:redundancy}
\footnotesize
\begin{tabular}{|@{\hspace{0.5mm}}l|ccccccc|c|c|c@{\hspace{0.5mm}}|}
\hline
\multirow{2}{*}{\textbf{Pipelines}} & \multicolumn{7}{c|}{\textbf{Number of generated clusters}}                                                                                                              & \multirow{2}{*}{\textbf{CFS}} & \multirow{2}{*}{\textbf{RR}} & \multirow{2}{*}{\textbf{S}} \\ \cline{2-8}
                                    & \multicolumn{1}{c|}{GD} & \multicolumn{1}{c|}{HPD} & \multicolumn{1}{c|}{OC} & \multicolumn{1}{c|}{SVIRO} & \multicolumn{1}{c|}{CPD} & \multicolumn{1}{c|}{SAP} & TOTAL &                                                          &                                            &                                  \\ \hline
VGG16/K-means/None/NoFT             & \multicolumn{1}{c|}{3}  & \multicolumn{1}{c|}{5}   & \multicolumn{1}{c|}{3}  & \multicolumn{1}{c|}{3}     & \multicolumn{1}{c|}{3}   & \multicolumn{1}{c|}{2}   & 19    & 17 (59\%)                                                & 1,12                                       & 0,99                             \\ \hline
VGG16/K-means/UMAP/NoFT             & \multicolumn{1}{c|}{4}  & \multicolumn{1}{c|}{8}   & \multicolumn{1}{c|}{2}  & \multicolumn{1}{c|}{4}     & \multicolumn{1}{c|}{3}   & \multicolumn{1}{c|}{3}   & 24    & 20 (69\%)                                                & 1,20                                       & 0,99                             \\ \hline
ResNet50/K-means/UMAP/NoFT          & \multicolumn{1}{c|}{3}  & \multicolumn{1}{c|}{4}   & \multicolumn{1}{c|}{3}  & \multicolumn{1}{c|}{2}     & \multicolumn{1}{c|}{2}   & \multicolumn{1}{c|}{4}   & 18    & 15 (52\%)                                                & 1,20                                       & 0,99                             \\ \hline
VGG16/Dbscan/UMAP/NoFT              & \multicolumn{1}{c|}{26} & \multicolumn{1}{c|}{77}  & \multicolumn{1}{c|}{13} & \multicolumn{1}{c|}{8}     & \multicolumn{1}{c|}{13}  & \multicolumn{1}{c|}{37}  & 174   & 28 (97\%)                                                & 6,21                                       & 0,91                             \\ \hline
ResNet50/Dbscan/UMAP/NoFT           & \multicolumn{1}{c|}{42} & \multicolumn{1}{c|}{51}  & \multicolumn{1}{c|}{5}  & \multicolumn{1}{c|}{10}    & \multicolumn{1}{c|}{27}  & \multicolumn{1}{c|}{44}  & 179   & 27 (93\%)                                                & 6,63                                       & 0,91                             \\ \hline
Inception\_V3/Dbscan/UMAP/NoFT      & \multicolumn{1}{c|}{28} & \multicolumn{1}{c|}{60}  & \multicolumn{1}{c|}{7}  & \multicolumn{1}{c|}{9}     & \multicolumn{1}{c|}{15}  & \multicolumn{1}{c|}{42}  & 161   & 29 (100\%)                                               & 5,55                                       & 0,92                             \\ \hline
Xception/Dbscan/UMAP/NoFT           & \multicolumn{1}{c|}{33} & \multicolumn{1}{c|}{30}  & \multicolumn{1}{c|}{9}  & \multicolumn{1}{c|}{2}     & \multicolumn{1}{c|}{9}   & \multicolumn{1}{c|}{14}  & 97    & 25 (86\%)                                                & 3,88                                       & 0,95                             \\ \hline
ResNet50/HDBSCAN/None/NoFT          & \multicolumn{1}{c|}{14} & \multicolumn{1}{c|}{171} & \multicolumn{1}{c|}{15} & \multicolumn{1}{c|}{7}     & \multicolumn{1}{c|}{74}  & \multicolumn{1}{c|}{3}   & 284   & 24 (83\%)                                                & 11,83                                      & 0,86                             \\ \hline
\end{tabular}\\
 \textbf{Legend:}  CFS: Covered failure scenarios (percentage \%); RR: Redundancy Ratio; S: Savings. 
\end{table}

%% file: threats.tex
\subsection{Threats to validity} 
We discuss internal, conclusion, construct, and external validity according to conventional practices~\cite{Wohlin2012}.

\subsubsection{Internal validity}

Since 72 of our 99 pipelines use a Transfer Learning pre-trained model to extract features, a possible internal threat is that this model can negatively affect our results if inadequate. Indeed, clustering relies on the similarity computed on the extracted features. \MAJOR{R1.16}{However, since every pre-trained model is integrated into at least one of the best pipelines identified in our experiments (see Table~\ref{tab:pre-existing-results}), we conclude that they are suitable. Also, to mitigate the risk that our purity metric might not reflect what is perceived by the end-user as a pure cluster, we relied on the same purity metric adopted in our previous work~\cite{attaoui2022black} to conduct an empirical study with human subjects, which demonstrated that end-users can understand the root causes of failures by looking at a small random subset of images in each cluster. Further, we visually inspected a random subset of our clusters to check their consistency.} Such consistency suggests that the features extracted by the models contain enough information to cluster the images based on their similarity. 

Another potential threat might be that the dataset (with the injected faults) was created with the proposed approach in mind. Therefore, there might be a risk of bias. To mitigate this risk, all the methods used in our pipelines (feature extraction methods, clustering algorithms, dimensionality reduction techniques) are independent of the data. These methods do not require any a priori knowledge on the data. We also publish our data to further mitigate this risk. All the experiments can be reproduced with any injected faulty scenario.

\subsubsection{External validity}
To alleviate the threats related to the choice of the case study DNNs, we use six well-studied datasets with diverse complexity. Four out of six subject DNNs implement tasks motivated by IEE business needs. These DNNs address problems that are quite common in the automotive industry. The other two DNNs are also related to the automotive industry and were used in many Kaggle challenges~\cite{pku, car-object-detection}. 

Although our pipelines were only tested on case study DNNs related to the automotive industry, we believe they will perform well with other datasets. This is because the models used for the feature extraction were pre-trained on ImageNet, which means that the model can capture features related to $1,000$ classes, including humans, animals, and objects. As for AE, it can learn the aspects of any dataset during training and provide high-quality clusters. 
Finally, for HUDD and LRP, the extraction of heatmap-based features is performed on well-known layer types that are part of any DNN model, regardless of the task at hand (i.e., they can be extended to DNNs that were not studied in this work).

\subsubsection{Construct validity}

The construct considered in our work is effectiveness. We measure the effectiveness through complementary indicators as follows: 

For RQ1, we evaluate the effectiveness of our pipelines by computing the purity of the generated clusters. The purity of a cluster is measured as the maximum proportion of images representing one faulty scenario in this cluster.

For RQ2, we evaluate the effectiveness of our pipelines based on the coverage of the injected faulty scenarios by the root cause clusters. A faulty scenario is covered by a cluster if at least $90\%$ of the images in this cluster represent such faulty scenario.

Finally, for RQ3, we consider both the purity and the coverage to measure the robustness of the top-performing pipelines to rare faulty scenarios.

\subsubsection{Conclusion validity}
Conclusion validity addresses threats that impact the ability to conclude appropriately. To mitigate such threats and to avoid violating parametric assumptions in our statistical analysis, we rely on a non-parametric test and effect size measure (i.e., Mann Whitney U-test and the Vargha and Delaney’s $\hat{A}_{12}$ statistics, respectively) to assess the statistical significance of differences in our results. 
Additionally, we applied the Fisher's exact test when comparing coverage results related to different distributions of faulty scenarios (i.e., RQ3), which is commonly used in similar contexts.
All results were reported based on both purity and coverage parameters, and six datasets were analyzed during our experiments.

\subsection{Data Availability}

All our implementations, the failure-inducing sets, the generated root-cause clusters and the data generated to address our research questions are available online~\cite{replicability}.

%% file: related.tex
\section{Related Work}
\label{sec:related}

Our paper is related to the literature on  DNN debugging and applications of transfer learning to perform root cause analysis~\cite{pan2021unsupervised,ter2022comparing}.

Heatmap-based approaches~\cite{Petsiuk2018rise,Dabkowski17,Montavon2019,Selvaraju17,Zeiler14,DB15a,Zhou16} explain the DNN's prediction of an image by highlighting which region of that image influenced the most the DNN output. For example, Grad-CAM generates a heatmap from the gradient flowing into the last layer. The heatmap is then superposed on the original image to highlight the regions of the image that activated the DNN and influenced the decision~\cite{Selvaraju17}. The main limitation of these approaches is that they require the inspection of all the heatmaps generated for the images processed by the DNN (e.g., error-inducing images) and, different from our pipelines, do not provide engineers with guidance for their inspection (i.e., one cluster for each failure root cause).
SHAP (SHapley Additive exPlanation)~\cite{lundberg2017unified} generates explanations by calculating the contribution of each feature to predictions, thus explaining what features are the most important for each prediction. In the case of an image CNN, SHAP considers a group of pixels as a feature and calculates their contribution to the decision made by the DNN.
Like heatmap-based approaches, SHAP does not provide guidance for the investigation of multiple failure-inducing images.

DeepJanus~\cite{riccio2020model} helps identify misbehaviors in a Deep Learning system by finding a set of pairs of inputs that are similar to each other and that trigger different behaviors of the Deep Learning system. This set of pairs represents the border between the input regions where the DNN behaves as expected or fails. 
Different from our work,  DeepJanus characterizes the behaviour of a DNN that can be tested with a simulator but cannot provide explanations for failures observed with real-world images.

Some DNN testing approaches explain the input regions where DNN errors are observed by relying on decision trees constructed using the simulator parameters used to generate test input images~\cite{Haq:2021,abdessalem2018testing}. Although decision trees are an effective mean to provide explanations for failures detected during simulator-based testing, they cannot be applied to provide explanations for failures observed with real-world images. To overcome such a limitation, we have recently developed SEDE~\cite{SEDE:Hazem}, a technique that applies HUDD to failure-inducing real-world images to generate root cause clusters and then relies on evolutionary algorithms to drive the generation, for each RCC, of similar images using simulators. The simulator parameter values used to generate such images are then fed into PART~\cite{PART}, a tree-based rule learning algorithm to characterize each RCCs in terms of simulator parameters (i.e., it generates expressions that constrain simulator parameters). The work in this paper is complementary to SEDE since the latter can be applied to clusters generated with the best pipeline (i.e., Pipeline $26$).

Pan et al.~\cite{pan2021unsupervised} combine Transfer Learning with clustering to find root causes of hardware failures. The proposed method uses different clustering algorithms (K-means~\cite{mcqueen1967smc}, decision tree clustering~\cite{liu2000clustering}, hierarchical clustering~\cite{kohn2014hierarchical}) on hardware test data to cluster failures likely due to the same causes. 
Different from their work, we aim to explain failures in DNNs that process images (i.e., our feature space is much larger).
Ter Burg et al.~\cite{ter2022comparing} explain DNNs based on a transfer learning model that has been fine-tuned to detect geometric shapes connecting face landmarks.
Such shapes are treated as features and the contribution of each feature is computed by relying on SHAP. The output should help end-users determine what influenced the DNN output. Unfortunately, similar to heatmap-based approaches, this approach does not support the explanation of multiple failures but require engineers to process them one by one.

To conclude, our previous works (i.e., HUDD~\cite{fahmysupporting} and SAFE~\cite{attaoui2022black}) have been the first to apply clustering algorithms to white-box and black-box feature extraction approaches to explain failure causes in DNN-based systems. This study is the first to systematically assess and compare the performance of alternative white-box and black-box feature extraction approaches, dimensionality reduction techniques, and clustering algorithms using a wide variety of practical, realistic failure scenarios.

%% file: conclusion.tex
\section{Conclusion}
\label{sec:conclusion}

In this paper, we presented a large-scale empirical evaluation of $99$ different pipelines for root cause analysis of DNN failures. 
Our pipelines receive as input a set of images leading to DNN failures and generate as output cluster of images sharing similar characteristics. As demonstrated by our previous work, by visualizing the images in each cluster, an engineer can notice commonalities across the images in each cluster; such commonalities represent the root causes of failures, help characterize failure scenarios and, thus, support engineers in improving the system (e.g., by selecting additional similar images to retrain the DNN or by introducing countermeasures in the system).

We considered $99$ pipelines resulting from the combination of  five methods for feature extraction, two techniques for dimensionality reduction and three clustering algorithms. Our methods for feature extraction include  white-box (i.e., heatmap generation techniques) and black-box approaches (i.e., fine-tuned and non-finetuned transfer learning models). Additionally, we rely on PCA and UMAP for dimensionality reduction and K-means, DBSCAN, and HDBSCAN for clustering.

We evaluated our pipelines in terms of clusters' purity and coverage of failures based on a controlled set of failure scenarios artificially injected into our datasets and widely varying in terms of frequency, thus analyzing the impact of rare scenarios on our best pipelines. Further, we assess the performance of our clustering pipelines in identifying failure scenarios not artificially injected but naturally present in our datasets.
Based on six case study subjects in the automotive domain,  
our results show that the best results are obtained with a pipeline relying on VGG-16 as transfer learning model, not using fine tuning, leveraging UMAP as a
dimensionality reduction technique, and using  DBSCAN as clustering algorithm. When the failure scenarios are equally distributed, the best pipeline achieved a purity of 94.3\% (i.e., almost all the images in RCCs present the same failure scenario) and a coverage of 96.7\%. The same pipeline also performs well with rare failure scenarios; indeed, when images belonging to failure scenarios represent between 5 and 10\% of the total number of images, it still can cover 90\% of the failure scenarios with a cluster purity above 70\%. Finally, we observed that the pipeline performing the best with injected failure scenarios also leads to the best results with 
failure scenarios already present in the datasets; specifically, it achieves 100\% coverage and an average purity of 87\% across clusters.

%% file: appendix.tex
\clearpage
\appendix

\section{Additional material for RQ1}
\label{appendix:RQ1}

\input{tables/RQ1_Purities}

\clearpage
\section{Additional material for RQ2}
\label{appendix:RQ2}
\input{tables/RQ2_Coverages}

\clearpage
\section{Additional material for RQ3}
\label{appendix:rq3}
\input{tables/RQ3_distribution}

%% file: tables/RQ1_Purities.tex
\setlength{\doublerulesep}{0pt}
\footnotesize
\begin{longtable}[c]{|c|c|c|c|c||c|c|c|c|c|c||c|}

\caption{Comparing the clusters generated by the different pipelines based on the average of the purity across root cause clusters. The last column represents the average of averages.}
\label{tab:rq1_appendix}\\ \hline
\multicolumn{5}{|c||}{\textbf{Pipelines}} & \multicolumn{7}{c|}{\textbf{Case Study Subjects}}
\\ \hline
\# & \textbf{FE} & \textbf{FT} & \textbf{DR} & \textbf{CA} & \textbf{GD} & \textbf{OC} & \textbf{HPD} & \textbf{SVIRO} & \textbf{SAP} & \textbf{CPD} & \textbf{Avg.} \\ \hline\hline\hline
\endhead
1 & HUDD & None & None & K-Means & 51.3\% & 36.6\% & 40.7\% & 62.4\% & 78.9\% & 39.9\% & 51.6\% \\ \hline

2 & HUDD & None & None & DBSCAN & 56.2\% & 53.4\% & 43.1\% & 53.0\% & 80.6\% & 63.7\% & 58.3\% \\ \hline

3 & HUDD & None & None & HDBScan & 68.5\% & 61.7\% & 43.7\% & 45.4\% & 51.2\% & 27.4\% & 49.6\% \\ \hline

4 & HUDD & None & PCA & K-Means & 49.1\% & 56.4\% & 40.8\% & 74.8\% & 79.7\% & 27.4\% & 54.7\% \\ \hline

5 & HUDD & None & PCA & DBSCAN & 43.4\% & 54.6\% & 48.7\% & 48.3\% & 80.6\% & 27.4\% & 50.5\% \\ \hline

6 & HUDD & None & PCA & HDBScan & 68.5\% & 61.7\% & 35.5\% & 45.4\% & 74.1\% & 26.9\% & 52.0\% \\ \hline

7 & HUDD & None & UMAP & K-Means & 56.7\% & 47.6\% & 42.3\% & 54.3\% & 61.1\% & 32.2\% & 49.0\% \\ \hline

8 & HUDD & None & UMAP & DBSCAN & 69.6\% & 58.7\% & 68.3\% & 59.3\% & 68.7\% & 53.9\% & \textbf{63.1\%} \\ \hline

9 & HUDD & None & UMAP & HDBScan & 68.5\% & 61.7\% & 33.3\% & 45.4\% & 74.1\% & 27.4\% & 51.7\% \\ \hline\hline\hline

10 & LRP & None & None & K-Means & 42.9\% & 36.6\% & 56.5\% & 33.8\% & 82.8\% & 73.6\% & 54.4\% \\ \hline

11 & LRP & None & None & DBSCAN & 39.5\% & 28.7\% & 69.8\% & 50.7\% & 96.5\% & 35.4\% & 53.4\% \\ \hline

12 & LRP & None & None & HDBScan & 69.0\% & 58.3\% & 56.2\% & 47.8\% & 42.4\% & 27.3\% & 50.2\% \\ \hline

13 & LRP & None & PCA & K-Means & 54.2\% & 56.4\% & 54.9\% & 35.7\% & 82.9\% & 73.6\% & 59.6\% \\ \hline

14 & LRP & None & PCA & DBSCAN & 47.2\% & 20.6\% & 71.8\% & 48.9\% & 79.7\% & 35.4\% & 50.6\% \\ \hline

15 & LRP & None & PCA & HDBScan & 52.5\% & 58.3\% & 25.5\% & 47.8\% & 46.8\% & 26.2\% & 42.8\% \\ \hline

16 & LRP & None & UMAP & K-Means & 55.9\% & 47.6\% & 54.7\% & 31.8\% & 67.0\% & 32.2\% & 48.2\% \\ \hline

17 & LRP & None & UMAP & DBSCAN & 67.0\% & 62.8\% & 68.9\% & 49.7\% & 80.3\% & 33.5\% & \textbf{60.4\%} \\ \hline

18 & LRP & None & UMAP & HDBScan & 69.0\% & 58.3\% & 56.2\% & 47.8\% & 51.6\% & 26.4\% & 51.5\% \\ \hline\hline\hline

19 & VGG-16 & None & None & K-Means & 91.7\% & 92.1\% & 95.5\% & 82.5\% & 97.3\% & 99.7\% & 93.2\% \\ \hline

20 & VGG-16 & None & None & DBSCAN & 87.0\% & 85.9\% & 96.7\% & 57.3\% & 98.0\% & 100.0\% & 87.5\% \\ \hline

21 & VGG-16 & None & None & HDBSCAN & 52.6\% & 99.0\% & 30.7\% & 73.4\% & 77.8\% & 54.5\% & 64.7\% \\ \hline

22 & VGG-16 & None & PCA & K-Means & 90.5\% & 87.6\% & 58.3\% & 87.7\% & 94.2\% & 92.2\% & 85.1\% \\ \hline

23 & VGG-16 & None & PCA & DBSCAN & 90.7\% & 94.9\% & 81.0\% & 71.6\% & 96.0\% & 91.8\% & 87.7\% \\ \hline

24 & VGG-16 & None & PCA & HDBSCAN & 45.6\% & 95.1\% & 56.2\% & 93.5\% & 100.0\% & 76.0\% & 77.7\% \\ \hline

25 & VGG-16 & None & UMAP & K-Means & 97.6\% & 84.4\% & 93.7\% & 82.4\% & 90.3\% & 97.8\% & 91.0\% \\ \hline

26 & VGG-16 & None & UMAP & DBSCAN & 99.0\% & 93.0\% & 99.6\% & 79.7\% & 98.1\% & 96.6\% & \textbf{94.3\%} \\ \hline

27 & VGG-16 & None & UMAP & HDBSCAN & 78.0\% & 96.7\% & 56.2\% & 88.9\% & 44.1\% & 79.9\% & 74.0\% \\ \hline

28 & VGG-16 & FT & None & K-Means & 26.2\% & 33.9\% & 15.8\% & 24.3\% & 27.9\% & 25.5\% & 25.6\% \\ \hline

29 & VGG-16 & FT & None & DBSCAN & 26.4\% & 38.5\% & 18.2\% & 32.8\% & 29.6\% & 25.2\% & 28.4\% \\ \hline

30 & VGG-16 & FT & None & HDBSCAN & 23.4\% & 51.1\% & 14.3\% & 48.1\% & 25.7\% & 53.2\% & 36.0\% \\ \hline

31 & VGG-16 & FT & PCA & K-Means & 25.4\% & 37.7\% & 16.7\% & 24.5\% & 26.8\% & 25.8\% & 26.1\% \\ \hline

32 & VGG-16 & FT & PCA & DBSCAN & 29.2\% & 51.4\% & 23.4\% & 46.6\% & 32.3\% & 29.0\% & 35.3\% \\ \hline

33 & VGG-16 & FT & PCA & HDBSCAN & 22.7\% & 43.7\% & 13.8\% & 41.5\% & 25.3\% & 23.3\% & 28.4\% \\ \hline

34 & VGG-16 & FT & UMAP & K-Means & 26.3\% & 33.6\% & 18.0\% & 25.3\% & 26.2\% & 26.2\% & 25.9\% \\ \hline

35 & VGG-16 & FT & UMAP & DBSCAN & 45.4\% & 42.8\% & 27.0\% & 39.1\% & 43.8\% & 44.0\% & 40.4\% \\ \hline

36 & VGG-16 & FT & UMAP & HDBSCAN & 23.5\% & 40.4\% & 14.1\% & 36.6\% & 22.0\% & 24.2\% & 26.8\% \\ \hline\hline\hline

37 & ResNet-50 & None & None & K-Means & 84.2\% & 84.6\% & 74.0\% & 61.2\% & 86.0\% & 83.7\% & 78.9\% \\ \hline

38 & ResNet-50 & None & None & DBSCAN & 63.5\% & 84.6\% & 87.5\% & 72.6\% & 75.5\% & 72.0\% & 76.0\% \\ \hline

39 & ResNet-50 & None & None & HDBSCAN & 96.4\% & 100.0\% & 100.0\% & 78.8\% & 87.5\% & 100.0\% & 93.8\% \\ \hline

40 & ResNet-50 & None & PCA & K-Means & 67.4\% & 79.6\% & 61.3\% & 53.4\% & 85.8\% & 75.6\% & 70.5\% \\ \hline

41 & ResNet-50 & None & PCA & DBSCAN & 79.7\% & 72.9\% & 51.1\% & 45.0\% & 89.8\% & 80.3\% & 69.8\% \\ \hline

42 & ResNet-50 & None & PCA & HDBSCAN & 40.8\% & 79.6\% & 56.2\% & 32.0\% & 42.5\% & 49.7\% & 50.2\% \\ \hline

43 & ResNet-50 & None & UMAP & K-Means & 99.4\% & 93.0\% & 82.3\% & 79.6\% & 99.6\% & 97.7\% & 91.9\% \\ \hline

44 & ResNet-50 & None & UMAP & DBSCAN & 100.0\% & 95.8\% & 95.8\% & 79.0\% & 99.7\% & 99.3\% & \textbf{94.9\%} \\ \hline

45 & ResNet-50 & None & UMAP & HDBSCAN & 82.6\% & 87.3\% & 38.8\% & 60.0\% & 30.5\% & 69.4\% & 61.4\% \\ \hline

46 & ResNet-50 & FT & None & K-Means & 26.7\% & 37.4\% & 19.3\% & 30.0\% & 26.2\% & 25.6\% & 27.5\% \\ \hline

47 & ResNet-50 & FT & None & DBSCAN & 47.2\% & 40.9\% & 32.1\% & 33.7\% & 35.2\% & 39.4\% & 38.1\% \\ \hline

48 & ResNet-50 & FT & None & HDBSCAN & 55.0\% & 46.7\% & 15.4\% & 45.2\% & 26.4\% & 24.9\% & 35.6\% \\ \hline

49 & ResNet-50 & FT & PCA & K-Means & 29.5\% & 37.1\% & 17.8\% & 39.5\% & 26.6\% & 26.2\% & 29.5\% \\ \hline

50 & ResNet-50 & FT & PCA & DBSCAN & 40.1\% & 45.6\% & 23.8\% & 41.5\% & 39.4\% & 39.4\% & 38.3\% \\ \hline

51 & ResNet-50 & FT & PCA & HDBSCAN & 23.7\% & 50.7\% & 15.7\% & 48.2\% & 24.6\% & 23.3\% & 31.1\% \\ \hline

52 & ResNet-50 & FT & UMAP & K-Means & 25.5\% & 34.6\% & 17.3\% & 25.7\% & 27.5\% & 25.6\% & 26.0\% \\ \hline

53 & ResNet-50 & FT & UMAP & DBSCAN & 37.8\% & 54.8\% & 35.9\% & 48.4\% & 41.7\% & 50.6\% & 44.9\% \\ \hline

54 & ResNet-50 & FT & UMAP & HDBSCAN & 23.9\% & 44.5\% & 15.1\% & 23.3\% & 23.7\% & 24.2\% & 25.8\% \\ \hline\hline\hline

55 & Inception-V3 & None & None & K-Means & 84.6\% & 86.0\% & 95.1\% & 69.2\% & 91.2\% & 87.8\% & 85.7\% \\ \hline

56 & Inception-V3 & None & None & DBSCAN & 100.0\% & 63.2\% & 80.9\% & 17.9\% & 98.4\% & 60.4\% & 70.1\% \\ \hline

57 & Inception-V3 & None & None & HDBSCAN & 62.6\% & 96.0\% & 99.8\% & 77.9\% & 62.6\% & 95.6\% & 82.4\% \\ \hline

58 & Inception-V3 & None & PCA & K-Means & 66.5\% & 74.5\% & 80.9\% & 68.6\% & 88.9\% & 75.7\% & 75.8\% \\ \hline

59 & Inception-V3 & None & PCA & DBSCAN & 97.9\% & 83.8\% & 86.0\% & 96.5\% & 92.0\% & 82.2\% & 89.7\% \\ \hline

60 & Inception-V3 & None & PCA & HDBSCAN & 92.5\% & 86.1\% & 53.4\% & 70.8\% & 69.4\% & 34.3\% & 67.8\% \\ \hline

61 & Inception-V3 & None & UMAP & K-Means & 94.1\% & 87.4\% & 95.0\% & 67.0\% & 90.1\% & 71.3\% & 84.2\% \\ \hline

62 & Inception-V3 & None & UMAP & DBSCAN & 93.4\% & 95.2\% & 98.1\% & 76.6\% & 97.4\% & 83.1\% & \textbf{90.7\%} \\ \hline

63 & Inception-V3 & None & UMAP & HDBSCAN & 74.0\% & 83.3\% & 55.9\% & 80.1\% & 65.8\% & 70.0\% & 71.5\% \\ \hline

64 & Inception-V3 & FT & None & K-Means & 26.6\% & 34.4\% & 15.6\% & 23.6\% & 26.7\% & 25.2\% & 25.4\% \\ \hline

65 & Inception-V3 & FT & None & DBSCAN & 50.0\% & 38.3\% & 24.7\% & 17.0\% & 51.0\% & 44.1\% & 37.5\% \\ \hline

66 & Inception-V3 & FT & None & HDBSCAN & 49.9\% & 48.2\% & 15.4\% & 44.4\% & 53.0\% & 53.8\% & 44.1\% \\ \hline

67 & Inception-V3 & FT & PCA & K-Means & 24.4\% & 31.5\% & 16.7\% & 25.6\% & 27.2\% & 25.2\% & 25.1\% \\ \hline

68 & Inception-V3 & FT & PCA & DBSCAN & 32.1\% & 50.6\% & 29.1\% & 38.5\% & 35.1\% & 38.6\% & 37.3\% \\ \hline

69 & Inception-V3 & FT & PCA & HDBSCAN & 46.7\% & 44.3\% & 15.1\% & 45.2\% & 25.0\% & 22.9\% & 33.2\% \\ \hline

70 & Inception-V3 & FT & UMAP & K-Means & 26.1\% & 31.0\% & 17.1\% & 25.8\% & 25.2\% & 27.2\% & 25.4\% \\ \hline

71 & Inception-V3 & FT & UMAP & DBSCAN & 45.7\% & 50.4\% & 23.4\% & 45.1\% & 44.4\% & 52.5\% & 43.6\% \\ \hline

72 & Inception-V3 & FT & UMAP & HDBSCAN & 47.2\% & 26.2\% & 15.3\% & 37.8\% & 24.7\% & 25.0\% & 29.4\% \\ \hline\hline\hline

73 & Xception & None & None & K-Means & 85.2\% & 90.4\% & 88.8\% & 68.4\% & 95.2\% & 72.6\% & 83.4\% \\ \hline

74 & Xception & None & None & DBSCAN & 60.3\% & 35.2\% & 82.6\% & 34.7\% & 73.6\% & 57.0\% & 57.2\% \\ \hline

75 & Xception & None & None & HDBSCAN & 88.9\% & 91.7\% & 85.6\% & 71.5\% & 100.0\% & 92.2\% & 88.3\% \\ \hline

76 & Xception & None & PCA & K-Means & 75.5\% & 73.3\% & 67.2\% & 57.6\% & 83.6\% & 44.4\% & 66.9\% \\ \hline

77 & Xception & None & PCA & DBSCAN & 78.8\% & 87.7\% & 83.6\% & 71.2\% & 93.6\% & 35.2\% & 75.0\% \\ \hline

78 & Xception & None & PCA & HDBSCAN & 75.8\% & 84.5\% & 47.5\% & 75.5\% & 62.0\% & 32.7\% & 63.0\% \\ \hline

79 & Xception & None & UMAP & K-Means & 93.1\% & 90.2\% & 89.8\% & 66.3\% & 92.6\% & 66.7\% & 83.1\% \\ \hline

80 & Xception & None & UMAP & DBSCAN & 94.7\% & 92.7\% & 98.7\% & 62.4\% & 99.0\% & 85.4\% & \textbf{88.8\%} \\ \hline

81 & Xception & None & UMAP & HDBSCAN & 77.9\% & 86.3\% & 36.7\% & 78.1\% & 99.4\% & 62.4\% & 73.5\% \\ \hline

82 & Xception & FT & None & K-Means & 25.0\% & 35.6\% & 17.1\% & 27.3\% & 25.4\% & 26.2\% & 26.1\% \\ \hline

83 & Xception & FT & None & DBSCAN & 32.3\% & 34.9\% & 22.7\% & 17.8\% & 20.2\% & 55.2\% & 30.5\% \\ \hline

84 & Xception & FT & None & HDBSCAN & 34.7\% & 46.2\% & 15.9\% & 33.4\% & 25.2\% & 54.0\% & 34.9\% \\ \hline

85 & Xception & FT & PCA & K-Means & 27.0\% & 35.6\% & 17.0\% & 26.2\% & 25.4\% & 36.8\% & 28.0\% \\ \hline

86 & Xception & FT & PCA & DBSCAN & 44.3\% & 31.1\% & 21.4\% & 28.2\% & 35.4\% & 35.2\% & 32.6\% \\ \hline

87 & Xception & FT & PCA & HDBSCAN & 46.8\% & 56.3\% & 14.8\% & 37.6\% & 24.9\% & 26.6\% & 34.5\% \\ \hline

88 & Xception & FT & UMAP & K-Means & 26.4\% & 32.8\% & 17.1\% & 28.7\% & 27.0\% & 24.6\% & 26.1\% \\ \hline

89 & Xception & FT & UMAP & DBSCAN & 36.9\% & 42.8\% & 26.7\% & 50.4\% & 46.2\% & 45.5\% & 41.4\% \\ \hline

90 & Xception & FT & UMAP & HDBSCAN & 23.1\% & 26.2\% & 14.9\% & 40.9\% & 24.7\% & 26.4\% & 26.1\% \\ \hline\hline\hline

91 & AE & None & None & K-Means & 40.9\% & 56.1\% & 47.0\% & 41.6\% & 72.5\% & 73.0\% & 55.2\% \\ \hline

92 & AE & None & None & DBSCAN & 35.2\% & 60.8\% & 11.5\% & 16.9\% & 20.3\% & 21.1\% & 27.6\% \\ \hline

93 & AE & None & None & HDBSCAN & 36.9\% & 67.8\% & 0.0\% & 67.5\% & 0.0\% & 63.0\% & 39.2\% \\ \hline

94 & AE & None & PCA & K-Means & 62.5\% & 55.1\% & 51.7\% & 52.9\% & 60.9\% & 77.2\% & 60.0\% \\ \hline

95 & AE & None & PCA & DBSCAN & 20.4\% & 73.3\% & 68.2\% & 63.6\% & 89.1\% & 76.7\% & 65.2\% \\ \hline

96 & AE & None & PCA & HDBSCAN & 73.5\% & 60.8\% & 25.7\% & 68.1\% & 76.1\% & 27.8\% & 55.3\% \\ \hline

97 & AE & None & UMAP & K-Means & 43.7\% & 46.2\% & 36.4\% & 37.6\% & 69.2\% & 66.0\% & 49.9\% \\ \hline

98 & AE & None & UMAP & DBSCAN & 65.3\% & 61.8\% & 59.0\% & 51.9\% & 69.1\% & 70.6\% & \textbf{62.9\%} \\ \hline

99 & AE & None & UMAP & HDBSCAN & 28.4\% & 60.5\% & 45.4\% & 47.2\% & 41.6\% & 41.7\% & 44.1\% \\\hline

\end{longtable}

%% file: tables/RQ2_Coverages.tex
\setlength{\doublerulesep}{0pt}
\footnotesize
\begin{longtable}[c]{|c|c|c|c|c||c|c|c|c|c|c||c|}

\caption{Percentage of faulty scenarios covered by the root cause clusters generated for each pipeline. The last column represents the average of averages.}
\label{tab:rq2_appendix}\\ \hline
\multicolumn{5}{|c||}{\textbf{Pipelines}} & \multicolumn{7}{c|}{\textbf{Case Study Subjects}}
\\ \hline
\# & \textbf{FE} & \textbf{FT} & \textbf{DR} & \textbf{CA} & \textbf{GD} & \textbf{OC} & \textbf{HPD} & \textbf{SVIRO} & \textbf{SAP} & \textbf{CPD} & \textbf{Avg.} \\ \hline\hline\hline
\endhead
1 & HUDD & None & None & K-Means & 0.0\% & 0.0\% & 0.0\% & 20.0\% & 25.0\% & 0.0\% & 7.5\% \\ \hline

2 & HUDD & None & None & DBSCAN & 25.0\% & 25.0\% & 0.0\% & 80.0\% & 25.0\% & 25.0\% & 30.0\% \\ \hline

3 & HUDD & None & None & HDBScan & 100.0\% & 25.0\% & 0.0\% & 0.0\% & 0.0\% & 0.0\% & 20.8\% \\ \hline

4 & HUDD & None & PCA & K-Means & 0.0\% & 25.0\% & 0.0\% & 40.0\% & 50.0\% & 0.0\% & 19.2\% \\ \hline

5 & HUDD & None & PCA & DBSCAN & 0.0\% & 25.0\% & 0.0\% & 20.0\% & 50.0\% & 0.0\% & 15.8\% \\ \hline

6 & HUDD & None & PCA & HDBScan & 100.0\% & 25.0\% & 0.0\% & 0.0\% & 100.0\% & 0.0\% & 37.5\% \\ \hline

7 & HUDD & None & UMAP & K-Means & 25.0\% & 0.0\% & 0.0\% & 0.0\% & 0.0\% & 0.0\% & 4.2\% \\ \hline

8 & HUDD & None & UMAP & DBSCAN & 100.0\% & 50.0\% & 75.0\% & 60.0\% & 75.0\% & 100.0\% & \textbf{76.7\%} \\ \hline

9 & HUDD & None & UMAP & HDBScan & 100.0\% & 25.0\% & 0.0\% & 0.0\% & 100.0\% & 0.0\% & 37.5\% \\ \hline\hline\hline

10 & LRP & None & None & K-Means & 0.0\% & 0.0\% & 12.5\% & 0.0\% & 50.0\% & 25.0\% & 14.6\% \\ \hline

11 & LRP & None & None & DBSCAN & 0.0\% & 0.0\% & 37.5\% & 20.0\% & 50.0\% & 0.0\% & 17.9\% \\ \hline

12 & LRP & None & None & HDBScan & 100.0\% & 25.0\% & 12.5\% & 20.0\% & 0.0\% & 0.0\% & 26.2\% \\ \hline

13 & LRP & None & PCA & K-Means & 25.0\% & 25.0\% & 12.5\% & 0.0\% & 50.0\% & 25.0\% & 22.9\% \\ \hline

14 & LRP & None & PCA & DBSCAN & 0.0\% & 0.0\% & 12.5\% & 20.0\% & 50.0\% & 0.0\% & 13.8\% \\ \hline

15 & LRP & None & PCA & HDBScan & 0.0\% & 25.0\% & 0.0\% & 20.0\% & 0.0\% & 0.0\% & 7.5\% \\ \hline

16 & LRP & None & UMAP & K-Means & 25.0\% & 0.0\% & 25.0\% & 0.0\% & 50.0\% & 0.0\% & 16.7\% \\ \hline

17 & LRP & None & UMAP & DBSCAN & 75.0\% & 100.0\% & 75.0\% & 40.0\% & 100.0\% & 0.0\% & \textbf{65.0\%} \\ \hline

18 & LRP & None & UMAP & HDBScan & 100.0\% & 25.0\% & 12.5\% & 20.0\% & 0.0\% & 0.0\% & 26.2\% \\ \hline\hline\hline

19 & VGG-16 & None & None & K-Means & 50.0\% & 50.0\% & 87.5\% & 80.0\% & 100.0\% & 100.0\% & 77.9\% \\ \hline

20 & VGG-16 & None & None & DBSCAN & 50.0\% & 50.0\% & 75.0\% & 20.0\% & 100.0\% & 100.0\% & 65.8\% \\ \hline

21 & VGG-16 & None & None & HDBSCAN & 0.0\% & 75.0\% & 0.0\% & 60.0\% & 50.0\% & 0.0\% & 30.8\% \\ \hline

22 & VGG-16 & None & PCA & K-Means & 50.0\% & 50.0\% & 12.5\% & 60.0\% & 100.0\% & 75.0\% & 57.9\% \\ \hline

23 & VGG-16 & None & PCA & DBSCAN & 50.0\% & 50.0\% & 37.5\% & 40.0\% & 75.0\% & 75.0\% & 54.6\% \\ \hline

24 & VGG-16 & None & PCA & HDBSCAN & 0.0\% & 75.0\% & 12.5\% & 100.0\% & 75.0\% & 50.0\% & 52.1\% \\ \hline

25 & VGG-16 & None & UMAP & K-Means & 75.0\% & 75.0\% & 87.5\% & 80.0\% & 75.0\% & 100.0\% & 82.1\% \\ \hline

26 & VGG-16 & None & UMAP & DBSCAN & 100.0\% & 100.0\% & 100.0\% & 80.0\% & 100.0\% & 100.0\% & \textbf{96.7\%} \\ \hline

27 & VGG-16 & None & UMAP & HDBSCAN & 50.0\% & 75.0\% & 12.5\% & 60.0\% & 0.0\% & 50.0\% & 41.2\% \\ \hline

28 & VGG-16 & FT & None & K-Means & 0.0\% & 0.0\% & 0.0\% & 0.0\% & 0.0\% & 0.0\% & 0.0\% \\ \hline

29 & VGG-16 & FT & None & DBSCAN & 0.0\% & 0.0\% & 0.0\% & 0.0\% & 0.0\% & 0.0\% & 0.0\% \\ \hline

30 & VGG-16 & FT & None & HDBSCAN & 0.0\% & 25.0\% & 0.0\% & 20.0\% & 0.0\% & 100.0\% & 24.2\% \\ \hline

31 & VGG-16 & FT & PCA & K-Means & 0.0\% & 0.0\% & 0.0\% & 0.0\% & 0.0\% & 0.0\% & 0.0\% \\ \hline

32 & VGG-16 & FT & PCA & DBSCAN & 0.0\% & 25.0\% & 0.0\% & 20.0\% & 0.0\% & 0.0\% & 7.5\% \\ \hline

33 & VGG-16 & FT & PCA & HDBSCAN & 0.0\% & 0.0\% & 0.0\% & 0.0\% & 0.0\% & 0.0\% & 0.0\% \\ \hline

34 & VGG-16 & FT & UMAP & K-Means & 0.0\% & 0.0\% & 0.0\% & 0.0\% & 0.0\% & 0.0\% & 0.0\% \\ \hline

35 & VGG-16 & FT & UMAP & DBSCAN & 25.0\% & 0.0\% & 0.0\% & 0.0\% & 50.0\% & 25.0\% & 16.7\% \\ \hline

36 & VGG-16 & FT & UMAP & HDBSCAN & 0.0\% & 0.0\% & 0.0\% & 0.0\% & 0.0\% & 0.0\% & 0.0\% \\ \hline\hline\hline

37 & ResNet-50 & None & None & K-Means & 50.0\% & 50.0\% & 25.0\% & 40.0\% & 50.0\% & 50.0\% & 44.2\% \\ \hline

38 & ResNet-50 & None & None & DBSCAN & 25.0\% & 50.0\% & 37.5\% & 40.0\% & 50.0\% & 25.0\% & 37.9\% \\ \hline

39 & ResNet-50 & None & None & HDBSCAN & 50.0\% & 100.0\% & 100.0\% & 60.0\% & 75.0\% & 100.0\% & 80.8\% \\ \hline

40 & ResNet-50 & None & PCA & K-Means & 25.0\% & 50.0\% & 12.5\% & 20.0\% & 50.0\% & 25.0\% & 30.4\% \\ \hline

41 & ResNet-50 & None & PCA & DBSCAN & 50.0\% & 50.0\% & 12.5\% & 0.0\% & 50.0\% & 25.0\% & 31.2\% \\ \hline

42 & ResNet-50 & None & PCA & HDBSCAN & 0.0\% & 100.0\% & 12.5\% & 0.0\% & 0.0\% & 0.0\% & 18.8\% \\ \hline

43 & ResNet-50 & None & UMAP & K-Means & 100.0\% & 75.0\% & 50.0\% & 40.0\% & 100.0\% & 100.0\% & 77.5\% \\ \hline

44 & ResNet-50 & None & UMAP & DBSCAN & 100.0\% & 100.0\% & 100.0\% & 60.0\% & 100.0\% & 100.0\% & \textbf{93.3\%} \\ \hline

45 & ResNet-50 & None & UMAP & HDBSCAN & 50.0\% & 75.0\% & 0.0\% & 20.0\% & 0.0\% & 25.0\% & 28.3\% \\ \hline

46 & ResNet-50 & FT & None & K-Means & 0.0\% & 0.0\% & 0.0\% & 0.0\% & 0.0\% & 0.0\% & 0.0\% \\ \hline

47 & ResNet-50 & FT & None & DBSCAN & 0.0\% & 0.0\% & 0.0\% & 0.0\% & 0.0\% & 0.0\% & 0.0\% \\ \hline

48 & ResNet-50 & FT & None & HDBSCAN & 50.0\% & 0.0\% & 0.0\% & 0.0\% & 0.0\% & 0.0\% & 8.3\% \\ \hline

49 & ResNet-50 & FT & PCA & K-Means & 0.0\% & 0.0\% & 0.0\% & 20.0\% & 0.0\% & 0.0\% & 3.3\% \\ \hline

50 & ResNet-50 & FT & PCA & DBSCAN & 0.0\% & 0.0\% & 0.0\% & 20.0\% & 0.0\% & 0.0\% & 3.3\% \\ \hline

51 & ResNet-50 & FT & PCA & HDBSCAN & 0.0\% & 0.0\% & 0.0\% & 40.0\% & 0.0\% & 0.0\% & 6.7\% \\ \hline

52 & ResNet-50 & FT & UMAP & K-Means & 0.0\% & 0.0\% & 0.0\% & 0.0\% & 0.0\% & 0.0\% & 0.0\% \\ \hline

53 & ResNet-50 & FT & UMAP & DBSCAN & 0.0\% & 50.0\% & 0.0\% & 40.0\% & 25.0\% & 100.0\% & 35.8\% \\ \hline

54 & ResNet-50 & FT & UMAP & HDBSCAN & 0.0\% & 0.0\% & 0.0\% & 0.0\% & 0.0\% & 0.0\% & 0.0\% \\ \hline\hline\hline

55 & Inception-V3 & None & None & K-Means & 75.0\% & 75.0\% & 87.5\% & 40.0\% & 100.0\% & 50.0\% & 71.2\% \\ \hline

56 & Inception-V3 & None & None & DBSCAN & 75.0\% & 25.0\% & 50.0\% & 0.0\% & 100.0\% & 25.0\% & 45.8\% \\ \hline

57 & Inception-V3 & None & None & HDBSCAN & 25.0\% & 100.0\% & 87.5\% & 100.0\% & 25.0\% & 100.0\% & 72.9\% \\ \hline

58 & Inception-V3 & None & PCA & K-Means & 50.0\% & 50.0\% & 37.5\% & 40.0\% & 75.0\% & 25.0\% & 46.2\% \\ \hline

59 & Inception-V3 & None & PCA & DBSCAN & 50.0\% & 75.0\% & 62.5\% & 40.0\% & 75.0\% & 25.0\% & 54.6\% \\ \hline

60 & Inception-V3 & None & PCA & HDBSCAN & 25.0\% & 75.0\% & 12.5\% & 60.0\% & 25.0\% & 0.0\% & 32.9\% \\ \hline

61 & Inception-V3 & None & UMAP & K-Means & 75.0\% & 75.0\% & 87.5\% & 40.0\% & 75.0\% & 50.0\% & 67.1\% \\ \hline

62 & Inception-V3 & None & UMAP & DBSCAN & 100.0\% & 100.0\% & 100.0\% & 100.0\% & 100.0\% & 100.0\% & \textbf{100.0\%} \\ \hline

63 & Inception-V3 & None & UMAP & HDBSCAN & 25.0\% & 100.0\% & 12.5\% & 100.0\% & 25.0\% & 25.0\% & 47.9\% \\ \hline

64 & Inception-V3 & FT & None & K-Means & 0.0\% & 0.0\% & 0.0\% & 0.0\% & 0.0\% & 0.0\% & 0.0\% \\ \hline

65 & Inception-V3 & FT & None & DBSCAN & 0.0\% & 0.0\% & 0.0\% & 0.0\% & 25.0\% & 0.0\% & 4.2\% \\ \hline

66 & Inception-V3 & FT & None & HDBSCAN & 75.0\% & 25.0\% & 0.0\% & 40.0\% & 75.0\% & 100.0\% & 52.5\% \\ \hline

67 & Inception-V3 & FT & PCA & K-Means & 0.0\% & 0.0\% & 0.0\% & 0.0\% & 0.0\% & 0.0\% & 0.0\% \\ \hline

68 & Inception-V3 & FT & PCA & DBSCAN & 0.0\% & 25.0\% & 0.0\% & 0.0\% & 0.0\% & 0.0\% & 4.2\% \\ \hline

69 & Inception-V3 & FT & PCA & HDBSCAN & 25.0\% & 0.0\% & 0.0\% & 20.0\% & 0.0\% & 0.0\% & 7.5\% \\ \hline

70 & Inception-V3 & FT & UMAP & K-Means & 0.0\% & 0.0\% & 0.0\% & 0.0\% & 0.0\% & 0.0\% & 0.0\% \\ \hline

71 & Inception-V3 & FT & UMAP & DBSCAN & 25.0\% & 50.0\% & 0.0\% & 20.0\% & 25.0\% & 75.0\% & 32.5\% \\ \hline

72 & Inception-V3 & FT & UMAP & HDBSCAN & 50.0\% & 0.0\% & 0.0\% & 0.0\% & 0.0\% & 0.0\% & 8.3\% \\ \hline\hline\hline

73 & Xception & None & None & K-Means & 50.0\% & 75.0\% & 87.5\% & 40.0\% & 100.0\% & 75.0\% & 71.2\% \\ \hline

74 & Xception & None & None & DBSCAN & 25.0\% & 0.0\% & 37.5\% & 0.0\% & 25.0\% & 25.0\% & 18.8\% \\ \hline

75 & Xception & None & None & HDBSCAN & 100.0\% & 100.0\% & 62.5\% & 60.0\% & 50.0\% & 100.0\% & 78.8\% \\ \hline

76 & Xception & None & PCA & K-Means & 50.0\% & 50.0\% & 37.5\% & 0.0\% & 75.0\% & 0.0\% & 35.4\% \\ \hline

77 & Xception & None & PCA & DBSCAN & 50.0\% & 75.0\% & 50.0\% & 0.0\% & 75.0\% & 0.0\% & 41.7\% \\ \hline

78 & Xception & None & PCA & HDBSCAN & 100.0\% & 50.0\% & 0.0\% & 80.0\% & 25.0\% & 0.0\% & 42.5\% \\ \hline

79 & Xception & None & UMAP & K-Means & 75.0\% & 75.0\% & 62.5\% & 40.0\% & 100.0\% & 50.0\% & 67.1\% \\ \hline

80 & Xception & None & UMAP & DBSCAN & 100.0\% & 100.0\% & 100.0\% & 40.0\% & 100.0\% & 100.0\% & \textbf{90.0\%} \\ \hline

81 & Xception & None & UMAP & HDBSCAN & 50.0\% & 75.0\% & 0.0\% & 60.0\% & 100.0\% & 25.0\% & 51.7\% \\ \hline

82 & Xception & FT & None & K-Means & 0.0\% & 0.0\% & 0.0\% & 0.0\% & 0.0\% & 0.0\% & 0.0\% \\ \hline

83 & Xception & FT & None & DBSCAN & 0.0\% & 0.0\% & 0.0\% & 0.0\% & 0.0\% & 25.0\% & 4.2\% \\ \hline

84 & Xception & FT & None & HDBSCAN & 0.0\% & 0.0\% & 0.0\% & 0.0\% & 0.0\% & 100.0\% & 16.7\% \\ \hline

85 & Xception & FT & PCA & K-Means & 0.0\% & 0.0\% & 0.0\% & 0.0\% & 0.0\% & 0.0\% & 0.0\% \\ \hline

86 & Xception & FT & PCA & DBSCAN & 0.0\% & 0.0\% & 0.0\% & 0.0\% & 0.0\% & 0.0\% & 0.0\% \\ \hline

87 & Xception & FT & PCA & HDBSCAN & 75.0\% & 50.0\% & 0.0\% & 0.0\% & 0.0\% & 0.0\% & 20.8\% \\ \hline

88 & Xception & FT & UMAP & K-Means & 0.0\% & 0.0\% & 0.0\% & 0.0\% & 0.0\% & 0.0\% & 0.0\% \\ \hline

89 & Xception & FT & UMAP & DBSCAN & 0.0\% & 0.0\% & 0.0\% & 60.0\% & 50.0\% & 0.0\% & 18.3\% \\ \hline

90 & Xception & FT & UMAP & HDBSCAN & 0.0\% & 0.0\% & 0.0\% & 0.0\% & 0.0\% & 0.0\% & 0.0\% \\ \hline\hline\hline

91 & AE & None & None & K-Means & 0.0\% & 25.0\% & 12.5\% & 20.0\% & 50.0\% & 50.0\% & 26.2\% \\ \hline

92 & AE & None & None & DBSCAN & 0.0\% & 25.0\% & 0.0\% & 0.0\% & 0.0\% & 0.0\% & 4.2\% \\ \hline

93 & AE & None & None & HDBSCAN & 0.0\% & 25.0\% & 0.0\% & 40.0\% & 0.0\% & 25.0\% & 15.0\% \\ \hline

94 & AE & None & PCA & K-Means & 0.0\% & 25.0\% & 12.5\% & 0.0\% & 50.0\% & 50.0\% & 22.9\% \\ \hline

95 & AE & None & PCA & DBSCAN & 0.0\% & 25.0\% & 0.0\% & 20.0\% & 50.0\% & 50.0\% & 24.2\% \\ \hline

96 & AE & None & PCA & HDBSCAN & 100.0\% & 50.0\% & 0.0\% & 80.0\% & 50.0\% & 0.0\% & 46.7\% \\ \hline

97 & AE & None & UMAP & K-Means & 0.0\% & 0.0\% & 0.0\% & 0.0\% & 25.0\% & 50.0\% & 12.5\% \\ \hline

98 & AE & None & UMAP & DBSCAN & 50.0\% & 50.0\% & 37.5\% & 40.0\% & 50.0\% & 100.0\% & \textbf{54.6\%} \\ \hline

99 & AE & None & UMAP & HDBSCAN & 0.0\% & 25.0\% & 0.0\% & 20.0\% & 0.0\% & 0.0\% & 7.5\% \\ \hline
\end{longtable}

%% file: tables/RQ3_distribution.tex
\begin{longtable}[c]{|c|c||cccccccccc|}
\caption{Distribution of faults for the different failure inducing sets for each case study subject.}
\label{tab:rq3_distribution}\\
\hline
\multirow{2}{*}{\textbf{Case Study}} & \multirow{2}{*}{\textbf{Dataset}}  & \multicolumn{10}{c|}{\textbf{Injected Failure Scenarios}}                                                                                                                                                                                                       \\ \cline{3-12}

     &     & \multicolumn{1}{c|}{\textbf{N}}  & \multicolumn{1}{c|}{\textbf{B}}  & \multicolumn{1}{c|}{\textbf{S}}  & \multicolumn{1}{c|}{\textbf{D}}  & \multicolumn{1}{c|}{\textbf{M}}  & \multicolumn{1}{c|}{\textbf{H}}  & \multicolumn{1}{c|}{\textbf{SG}} & \multicolumn{1}{c|}{\textbf{EG}} & \multicolumn{1}{c|}{\textbf{EO}} & \textbf{NF} \\ \hline\hline\hline
\endhead
\multirow{10}{*}{GD} & GD\_1     & \multicolumn{1}{c|}{64} & \multicolumn{1}{c|}{40} & \multicolumn{1}{c|}{48} & \multicolumn{1}{c|}{72} & \multicolumn{1}{c|}{-}  & \multicolumn{1}{c|}{-}  & \multicolumn{1}{c|}{-}  & \multicolumn{1}{c|}{-}  & \multicolumn{1}{c|}{-}  & 56 \\  \cline{2-12}
& GD\_2     & \multicolumn{1}{c|}{48} & \multicolumn{1}{c|}{32} & \multicolumn{1}{c|}{24} & \multicolumn{1}{c|}{16} & \multicolumn{1}{c|}{-}  & \multicolumn{1}{c|}{-}  & \multicolumn{1}{c|}{-}  & \multicolumn{1}{c|}{-}  & \multicolumn{1}{c|}{-}  & 64 \\ \cline{2-12}
& GD\_3     & \multicolumn{1}{c|}{24} & \multicolumn{1}{c|}{64} & \multicolumn{1}{c|}{16} & \multicolumn{1}{c|}{40} & \multicolumn{1}{c|}{-}  & \multicolumn{1}{c|}{-}  & \multicolumn{1}{c|}{-}  & \multicolumn{1}{c|}{-}  & \multicolumn{1}{c|}{-}  & 8  \\ \cline{2-12}
& GD\_4     & \multicolumn{1}{c|}{72} & \multicolumn{1}{c|}{16} & \multicolumn{1}{c|}{24} & \multicolumn{1}{c|}{48} & \multicolumn{1}{c|}{-}  & \multicolumn{1}{c|}{-}  & \multicolumn{1}{c|}{-}  & \multicolumn{1}{c|}{-}  & \multicolumn{1}{c|}{-}  & 40 \\ \cline{2-12}
& GD\_5     & \multicolumn{1}{c|}{40} & \multicolumn{1}{c|}{24} & \multicolumn{1}{c|}{72} & \multicolumn{1}{c|}{16} & \multicolumn{1}{c|}{-}  & \multicolumn{1}{c|}{-}  & \multicolumn{1}{c|}{-}  & \multicolumn{1}{c|}{-}  & \multicolumn{1}{c|}{-}  & 8  \\ \cline{2-12}
& GD\_6     & \multicolumn{1}{c|}{56} & \multicolumn{1}{c|}{32} & \multicolumn{1}{c|}{48} & \multicolumn{1}{c|}{16} & \multicolumn{1}{c|}{-}  & \multicolumn{1}{c|}{-}  & \multicolumn{1}{c|}{-}  & \multicolumn{1}{c|}{-}  & \multicolumn{1}{c|}{-}  & 24 \\ \cline{2-12}
& GD\_7     & \multicolumn{1}{c|}{64} & \multicolumn{1}{c|}{32} & \multicolumn{1}{c|}{8}  & \multicolumn{1}{c|}{56} & \multicolumn{1}{c|}{-}  & \multicolumn{1}{c|}{-}  & \multicolumn{1}{c|}{-}  & \multicolumn{1}{c|}{-}  & \multicolumn{1}{c|}{-}  & 24 \\ \cline{2-12}
& GD\_8     & \multicolumn{1}{c|}{72} & \multicolumn{1}{c|}{24} & \multicolumn{1}{c|}{16} & \multicolumn{1}{c|}{8}  & \multicolumn{1}{c|}{-}  & \multicolumn{1}{c|}{-}  & \multicolumn{1}{c|}{-}  & \multicolumn{1}{c|}{-}  & \multicolumn{1}{c|}{-}  & 64 \\ \cline{2-12}
& GD\_9     & \multicolumn{1}{c|}{40} & \multicolumn{1}{c|}{64} & \multicolumn{1}{c|}{48} & \multicolumn{1}{c|}{16} & \multicolumn{1}{c|}{-}  & \multicolumn{1}{c|}{-}  & \multicolumn{1}{c|}{-}  & \multicolumn{1}{c|}{-}  & \multicolumn{1}{c|}{-}  & 56 \\ \cline{2-12}
& GD\_10    & \multicolumn{1}{c|}{56} & \multicolumn{1}{c|}{8}  & \multicolumn{1}{c|}{64} & \multicolumn{1}{c|}{24} & \multicolumn{1}{c|}{-}  & \multicolumn{1}{c|}{-}  & \multicolumn{1}{c|}{-}  & \multicolumn{1}{c|}{-}  & \multicolumn{1}{c|}{-}  & 48 \\ \hline
\multirow{10}{*}{OC} & OC\_1     & \multicolumn{1}{c|}{18} & \multicolumn{1}{c|}{6}  & \multicolumn{1}{c|}{10} & \multicolumn{1}{c|}{4}  & \multicolumn{1}{c|}{-}  & \multicolumn{1}{c|}{-}  & \multicolumn{1}{c|}{-}  & \multicolumn{1}{c|}{-}  & \multicolumn{1}{c|}{-}  & 8  \\ \cline{2-12}
& OC\_2     & \multicolumn{1}{c|}{4}  & \multicolumn{1}{c|}{18} & \multicolumn{1}{c|}{12} & \multicolumn{1}{c|}{2}  & \multicolumn{1}{c|}{-}  & \multicolumn{1}{c|}{-}  & \multicolumn{1}{c|}{-}  & \multicolumn{1}{c|}{-}  & \multicolumn{1}{c|}{-}  & 14 \\ \cline{2-12}
& OC\_3     & \multicolumn{1}{c|}{2}  & \multicolumn{1}{c|}{14} & \multicolumn{1}{c|}{10} & \multicolumn{1}{c|}{6}  & \multicolumn{1}{c|}{-}  & \multicolumn{1}{c|}{-}  & \multicolumn{1}{c|}{-}  & \multicolumn{1}{c|}{-}  & \multicolumn{1}{c|}{-}  & 16 \\ \cline{2-12}
& OC\_4     & \multicolumn{1}{c|}{2}  & \multicolumn{1}{c|}{4}  & \multicolumn{1}{c|}{6}  & \multicolumn{1}{c|}{10} & \multicolumn{1}{c|}{-}  & \multicolumn{1}{c|}{-}  & \multicolumn{1}{c|}{-}  & \multicolumn{1}{c|}{-}  & \multicolumn{1}{c|}{-}  & 8  \\ \cline{2-12}
& OC\_5     & \multicolumn{1}{c|}{6}  & \multicolumn{1}{c|}{4}  & \multicolumn{1}{c|}{16} & \multicolumn{1}{c|}{18} & \multicolumn{1}{c|}{-}  & \multicolumn{1}{c|}{-}  & \multicolumn{1}{c|}{-}  & \multicolumn{1}{c|}{-}  & \multicolumn{1}{c|}{-}  & 10 \\ \cline{2-12}
& OC\_6     & \multicolumn{1}{c|}{8}  & \multicolumn{1}{c|}{6}  & \multicolumn{1}{c|}{10} & \multicolumn{1}{c|}{12} & \multicolumn{1}{c|}{-}  & \multicolumn{1}{c|}{-}  & \multicolumn{1}{c|}{-}  & \multicolumn{1}{c|}{-}  & \multicolumn{1}{c|}{-}  & 16 \\ \cline{2-12}
& OC\_7     & \multicolumn{1}{c|}{18} & \multicolumn{1}{c|}{8}  & \multicolumn{1}{c|}{16} & \multicolumn{1}{c|}{6}  & \multicolumn{1}{c|}{-}  & \multicolumn{1}{c|}{-}  & \multicolumn{1}{c|}{-}  & \multicolumn{1}{c|}{-}  & \multicolumn{1}{c|}{-}  & 2  \\ \cline{2-12}
& OC\_8     & \multicolumn{1}{c|}{16} & \multicolumn{1}{c|}{18} & \multicolumn{1}{c|}{14} & \multicolumn{1}{c|}{10} & \multicolumn{1}{c|}{-}  & \multicolumn{1}{c|}{-}  & \multicolumn{1}{c|}{-}  & \multicolumn{1}{c|}{-}  & \multicolumn{1}{c|}{-}  & 4  \\ \cline{2-12}
& OC\_9     & \multicolumn{1}{c|}{14} & \multicolumn{1}{c|}{16} & \multicolumn{1}{c|}{4}  & \multicolumn{1}{c|}{10} & \multicolumn{1}{c|}{-}  & \multicolumn{1}{c|}{-}  & \multicolumn{1}{c|}{-}  & \multicolumn{1}{c|}{-}  & \multicolumn{1}{c|}{-}  & 2  \\ \cline{2-12}
& OC\_10    & \multicolumn{1}{c|}{10} & \multicolumn{1}{c|}{2}  & \multicolumn{1}{c|}{14} & \multicolumn{1}{c|}{8}  & \multicolumn{1}{c|}{-}  & \multicolumn{1}{c|}{-}  & \multicolumn{1}{c|}{-}  & \multicolumn{1}{c|}{-}  & \multicolumn{1}{c|}{-}  & 18 \\ \hline
\multirow{10}{*}{HPD} & HPD\_1    & \multicolumn{1}{c|}{45} & \multicolumn{1}{c|}{72} & \multicolumn{1}{c|}{54} & \multicolumn{1}{c|}{9}  & \multicolumn{1}{c|}{36} & \multicolumn{1}{c|}{63} & \multicolumn{1}{c|}{81} & \multicolumn{1}{c|}{18} & \multicolumn{1}{c|}{-}  & 27 \\ \cline{2-12}
& HPD\_2    & \multicolumn{1}{c|}{27} & \multicolumn{1}{c|}{81} & \multicolumn{1}{c|}{45} & \multicolumn{1}{c|}{18} & \multicolumn{1}{c|}{54} & \multicolumn{1}{c|}{63} & \multicolumn{1}{c|}{72} & \multicolumn{1}{c|}{36} & \multicolumn{1}{c|}{-}  & 9  \\ \cline{2-12}
& HPD\_3    & \multicolumn{1}{c|}{54} & \multicolumn{1}{c|}{81} & \multicolumn{1}{c|}{27} & \multicolumn{1}{c|}{63} & \multicolumn{1}{c|}{18} & \multicolumn{1}{c|}{45} & \multicolumn{1}{c|}{9}  & \multicolumn{1}{c|}{36} & \multicolumn{1}{c|}{-}  & 72 \\ \cline{2-12}
& HPD\_4    & \multicolumn{1}{c|}{36} & \multicolumn{1}{c|}{18} & \multicolumn{1}{c|}{63} & \multicolumn{1}{c|}{72} & \multicolumn{1}{c|}{9}  & \multicolumn{1}{c|}{81} & \multicolumn{1}{c|}{54} & \multicolumn{1}{c|}{27} & \multicolumn{1}{c|}{-}  & 45 \\ \cline{2-12}
& HPD\_5    & \multicolumn{1}{c|}{27} & \multicolumn{1}{c|}{63} & \multicolumn{1}{c|}{18} & \multicolumn{1}{c|}{72} & \multicolumn{1}{c|}{36} & \multicolumn{1}{c|}{9}  & \multicolumn{1}{c|}{45} & \multicolumn{1}{c|}{81} & \multicolumn{1}{c|}{-}  & 54 \\ \cline{2-12}
& HPD\_6    & \multicolumn{1}{c|}{45} & \multicolumn{1}{c|}{36} & \multicolumn{1}{c|}{54} & \multicolumn{1}{c|}{63} & \multicolumn{1}{c|}{81} & \multicolumn{1}{c|}{9}  & \multicolumn{1}{c|}{72} & \multicolumn{1}{c|}{27} & \multicolumn{1}{c|}{-}  & 18 \\ \cline{2-12}
& HPD\_7    & \multicolumn{1}{c|}{63} & \multicolumn{1}{c|}{45} & \multicolumn{1}{c|}{81} & \multicolumn{1}{c|}{36} & \multicolumn{1}{c|}{27} & \multicolumn{1}{c|}{72} & \multicolumn{1}{c|}{18} & \multicolumn{1}{c|}{54} & \multicolumn{1}{c|}{-}  & 9  \\ \cline{2-12}
& HPD\_8    & \multicolumn{1}{c|}{72} & \multicolumn{1}{c|}{9}  & \multicolumn{1}{c|}{63} & \multicolumn{1}{c|}{27} & \multicolumn{1}{c|}{36} & \multicolumn{1}{c|}{18} & \multicolumn{1}{c|}{81} & \multicolumn{1}{c|}{54} & \multicolumn{1}{c|}{-}  & 45 \\ \cline{2-12}
& HPD\_9    & \multicolumn{1}{c|}{72} & \multicolumn{1}{c|}{63} & \multicolumn{1}{c|}{18} & \multicolumn{1}{c|}{27} & \multicolumn{1}{c|}{45} & \multicolumn{1}{c|}{9}  & \multicolumn{1}{c|}{81} & \multicolumn{1}{c|}{54} & \multicolumn{1}{c|}{-}  & 36 \\ \cline{2-12}
& HPD\_10   & \multicolumn{1}{c|}{54} & \multicolumn{1}{c|}{81} & \multicolumn{1}{c|}{63} & \multicolumn{1}{c|}{27} & \multicolumn{1}{c|}{45} & \multicolumn{1}{c|}{72} & \multicolumn{1}{c|}{18} & \multicolumn{1}{c|}{9}  & \multicolumn{1}{c|}{-}  & 36 \\ \hline
\multirow{10}{*}{SVIRO} & SVIRO\_1  & \multicolumn{1}{c|}{6}  & \multicolumn{1}{c|}{12} & \multicolumn{1}{c|}{18} & \multicolumn{1}{c|}{21} & \multicolumn{1}{c|}{-}  & \multicolumn{1}{c|}{-}  & \multicolumn{1}{c|}{-}  & \multicolumn{1}{c|}{-}  & \multicolumn{1}{c|}{24} & 15 \\ \cline{2-12}
& SVIRO\_2  & \multicolumn{1}{c|}{9}  & \multicolumn{1}{c|}{24} & \multicolumn{1}{c|}{6}  & \multicolumn{1}{c|}{12} & \multicolumn{1}{c|}{-}  & \multicolumn{1}{c|}{-}  & \multicolumn{1}{c|}{-}  & \multicolumn{1}{c|}{-}  & \multicolumn{1}{c|}{15} & 3  \\ \cline{2-12}
& SVIRO\_3  & \multicolumn{1}{c|}{15} & \multicolumn{1}{c|}{18} & \multicolumn{1}{c|}{21} & \multicolumn{1}{c|}{3}  & \multicolumn{1}{c|}{-}  & \multicolumn{1}{c|}{-}  & \multicolumn{1}{c|}{-}  & \multicolumn{1}{c|}{-}  & \multicolumn{1}{c|}{27} & 9  \\ \cline{2-12}
& SVIRO\_4  & \multicolumn{1}{c|}{21} & \multicolumn{1}{c|}{9}  & \multicolumn{1}{c|}{12} & \multicolumn{1}{c|}{24} & \multicolumn{1}{c|}{-}  & \multicolumn{1}{c|}{-}  & \multicolumn{1}{c|}{-}  & \multicolumn{1}{c|}{-}  & \multicolumn{1}{c|}{6}  & 27 \\ \cline{2-12}
& SVIRO\_5  & \multicolumn{1}{c|}{27} & \multicolumn{1}{c|}{21} & \multicolumn{1}{c|}{3}  & \multicolumn{1}{c|}{9}  & \multicolumn{1}{c|}{-}  & \multicolumn{1}{c|}{-}  & \multicolumn{1}{c|}{-}  & \multicolumn{1}{c|}{-}  & \multicolumn{1}{c|}{18} & 24 \\ \cline{2-12}
& SVIRO\_6  & \multicolumn{1}{c|}{3}  & \multicolumn{1}{c|}{27} & \multicolumn{1}{c|}{24} & \multicolumn{1}{c|}{6}  & \multicolumn{1}{c|}{-}  & \multicolumn{1}{c|}{-}  & \multicolumn{1}{c|}{-}  & \multicolumn{1}{c|}{-}  & \multicolumn{1}{c|}{21} & 12 \\ \cline{2-12}
& SVIRO\_7  & \multicolumn{1}{c|}{24} & \multicolumn{1}{c|}{6}  & \multicolumn{1}{c|}{15} & \multicolumn{1}{c|}{18} & \multicolumn{1}{c|}{-}  & \multicolumn{1}{c|}{-}  & \multicolumn{1}{c|}{-}  & \multicolumn{1}{c|}{-}  & \multicolumn{1}{c|}{3}  & 21 \\ \cline{2-12}
& SVIRO\_8  & \multicolumn{1}{c|}{15} & \multicolumn{1}{c|}{3}  & \multicolumn{1}{c|}{27} & \multicolumn{1}{c|}{24} & \multicolumn{1}{c|}{-}  & \multicolumn{1}{c|}{-}  & \multicolumn{1}{c|}{-}  & \multicolumn{1}{c|}{-}  & \multicolumn{1}{c|}{12} & 6  \\ \cline{2-12}
& SVIRO\_9  & \multicolumn{1}{c|}{12} & \multicolumn{1}{c|}{15} & \multicolumn{1}{c|}{3}  & \multicolumn{1}{c|}{27} & \multicolumn{1}{c|}{-}  & \multicolumn{1}{c|}{-}  & \multicolumn{1}{c|}{-}  & \multicolumn{1}{c|}{-}  & \multicolumn{1}{c|}{9}  & 18 \\ \cline{2-12}
& SVIRO\_10 & \multicolumn{1}{c|}{18} & \multicolumn{1}{c|}{21} & \multicolumn{1}{c|}{9}  & \multicolumn{1}{c|}{15} & \multicolumn{1}{c|}{-}  & \multicolumn{1}{c|}{-}  & \multicolumn{1}{c|}{-}  & \multicolumn{1}{c|}{-}  & \multicolumn{1}{c|}{21} & 18 \\ \hline
\multirow{10}{*}{CPD} & CPD\_1    & \multicolumn{1}{c|}{87} & \multicolumn{1}{c|}{74} & \multicolumn{1}{c|}{55} & \multicolumn{1}{c|}{28} & \multicolumn{1}{c|}{-}  & \multicolumn{1}{c|}{-}  & \multicolumn{1}{c|}{-}  & \multicolumn{1}{c|}{-}  & \multicolumn{1}{c|}{-}  & 44 \\ \cline{2-12}
& CPD\_2    & \multicolumn{1}{c|}{43} & \multicolumn{1}{c|}{56} & \multicolumn{1}{c|}{27} & \multicolumn{1}{c|}{22} & \multicolumn{1}{c|}{-}  & \multicolumn{1}{c|}{-}  & \multicolumn{1}{c|}{-}  & \multicolumn{1}{c|}{-}  & \multicolumn{1}{c|}{-}  & 74 \\ \cline{2-12}
& CPD\_3    & \multicolumn{1}{c|}{6}  & \multicolumn{1}{c|}{32} & \multicolumn{1}{c|}{4}  & \multicolumn{1}{c|}{62} & \multicolumn{1}{c|}{-}  & \multicolumn{1}{c|}{-}  & \multicolumn{1}{c|}{-}  & \multicolumn{1}{c|}{-}  & \multicolumn{1}{c|}{-}  & 35 \\ \cline{2-12}
& CPD\_4    & \multicolumn{1}{c|}{49} & \multicolumn{1}{c|}{22} & \multicolumn{1}{c|}{88} & \multicolumn{1}{c|}{34} & \multicolumn{1}{c|}{-}  & \multicolumn{1}{c|}{-}  & \multicolumn{1}{c|}{-}  & \multicolumn{1}{c|}{-}  & \multicolumn{1}{c|}{-}  & 5  \\ \cline{2-12}
& CPD\_5    & \multicolumn{1}{c|}{24} & \multicolumn{1}{c|}{69} & \multicolumn{1}{c|}{37} & \multicolumn{1}{c|}{57} & \multicolumn{1}{c|}{-}  & \multicolumn{1}{c|}{-}  & \multicolumn{1}{c|}{-}  & \multicolumn{1}{c|}{-}  & \multicolumn{1}{c|}{-}  & 86 \\ \cline{2-12}
& CPD\_6    & \multicolumn{1}{c|}{13} & \multicolumn{1}{c|}{69} & \multicolumn{1}{c|}{58} & \multicolumn{1}{c|}{54} & \multicolumn{1}{c|}{-}  & \multicolumn{1}{c|}{-}  & \multicolumn{1}{c|}{-}  & \multicolumn{1}{c|}{-}  & \multicolumn{1}{c|}{-}  & 25 \\ \cline{2-12}
& CPD\_7    & \multicolumn{1}{c|}{3}  & \multicolumn{1}{c|}{32} & \multicolumn{1}{c|}{51} & \multicolumn{1}{c|}{9}  & \multicolumn{1}{c|}{-}  & \multicolumn{1}{c|}{-}  & \multicolumn{1}{c|}{-}  & \multicolumn{1}{c|}{-}  & \multicolumn{1}{c|}{-}  & 59 \\ \cline{2-12}
& CPD\_8    & \multicolumn{1}{c|}{77} & \multicolumn{1}{c|}{62} & \multicolumn{1}{c|}{12} & \multicolumn{1}{c|}{53} & \multicolumn{1}{c|}{-}  & \multicolumn{1}{c|}{-}  & \multicolumn{1}{c|}{-}  & \multicolumn{1}{c|}{-}  & \multicolumn{1}{c|}{-}  & 4  \\ \cline{2-12}
& CPD\_9    & \multicolumn{1}{c|}{85} & \multicolumn{1}{c|}{27} & \multicolumn{1}{c|}{78} & \multicolumn{1}{c|}{30} & \multicolumn{1}{c|}{-}  & \multicolumn{1}{c|}{-}  & \multicolumn{1}{c|}{-}  & \multicolumn{1}{c|}{-}  & \multicolumn{1}{c|}{-}  & 62 \\ \cline{2-12}
& CPD\_10   & \multicolumn{1}{c|}{65} & \multicolumn{1}{c|}{46} & \multicolumn{1}{c|}{66} & \multicolumn{1}{c|}{89} & \multicolumn{1}{c|}{-}  & \multicolumn{1}{c|}{-}  & \multicolumn{1}{c|}{-}  & \multicolumn{1}{c|}{-}  & \multicolumn{1}{c|}{-}  & 40 \\ \hline

\multirow{10}{*}{SAP} & SAP\_1    & \multicolumn{1}{c|}{22} & \multicolumn{1}{c|}{33} & \multicolumn{1}{c|}{54} & \multicolumn{1}{c|}{48} & \multicolumn{1}{c|}{-}  & \multicolumn{1}{c|}{-}  & \multicolumn{1}{c|}{-}  & \multicolumn{1}{c|}{-}  & \multicolumn{1}{c|}{-}  & 72 \\ \cline{2-12}
& SAP\_2    & \multicolumn{1}{c|}{75} & \multicolumn{1}{c|}{22} & \multicolumn{1}{c|}{48} & \multicolumn{1}{c|}{17} & \multicolumn{1}{c|}{-}  & \multicolumn{1}{c|}{-}  & \multicolumn{1}{c|}{-}  & \multicolumn{1}{c|}{-}  & \multicolumn{1}{c|}{-}  & 57 \\ \cline{2-12}
& SAP\_3    & \multicolumn{1}{c|}{22} & \multicolumn{1}{c|}{4}  & \multicolumn{1}{c|}{57} & \multicolumn{1}{c|}{42} & \multicolumn{1}{c|}{-}  & \multicolumn{1}{c|}{-}  & \multicolumn{1}{c|}{-}  & \multicolumn{1}{c|}{-}  & \multicolumn{1}{c|}{-}  & 81 \\ \cline{2-12}

& SAP\_4    & \multicolumn{1}{c|}{74} & \multicolumn{1}{c|}{21} & \multicolumn{1}{c|}{40} & \multicolumn{1}{c|}{36} & \multicolumn{1}{c|}{-}  & \multicolumn{1}{c|}{-}  & \multicolumn{1}{c|}{-}  & \multicolumn{1}{c|}{-}  & \multicolumn{1}{c|}{-}  & 42 \\ \cline{2-12}
& SAP\_5    & \multicolumn{1}{c|}{15} & \multicolumn{1}{c|}{14} & \multicolumn{1}{c|}{86} & \multicolumn{1}{c|}{74} & \multicolumn{1}{c|}{-}  & \multicolumn{1}{c|}{-}  & \multicolumn{1}{c|}{-}  & \multicolumn{1}{c|}{-}  & \multicolumn{1}{c|}{-}  & 51 \\ \cline{2-12}
& SAP\_6    & \multicolumn{1}{c|}{73} & \multicolumn{1}{c|}{60} & \multicolumn{1}{c|}{2}  & \multicolumn{1}{c|}{83} & \multicolumn{1}{c|}{-}  & \multicolumn{1}{c|}{-}  & \multicolumn{1}{c|}{-}  & \multicolumn{1}{c|}{-}  & \multicolumn{1}{c|}{-}  & 72 \\ \cline{2-12}
& SAP\_7    & \multicolumn{1}{c|}{58} & \multicolumn{1}{c|}{57} & \multicolumn{1}{c|}{47} & \multicolumn{1}{c|}{83} & \multicolumn{1}{c|}{-}  & \multicolumn{1}{c|}{-}  & \multicolumn{1}{c|}{-}  & \multicolumn{1}{c|}{-}  & \multicolumn{1}{c|}{-}  & 43 \\ \cline{2-12}
& SAP\_8    & \multicolumn{1}{c|}{6}  & \multicolumn{1}{c|}{75} & \multicolumn{1}{c|}{26} & \multicolumn{1}{c|}{16} & \multicolumn{1}{c|}{-}  & \multicolumn{1}{c|}{-}  & \multicolumn{1}{c|}{-}  & \multicolumn{1}{c|}{-}  & \multicolumn{1}{c|}{-}  & 70 \\ \cline{2-12}
& SAP\_9    & \multicolumn{1}{c|}{89} & \multicolumn{1}{c|}{86} & \multicolumn{1}{c|}{66} & \multicolumn{1}{c|}{32} & \multicolumn{1}{c|}{-}  & \multicolumn{1}{c|}{-}  & \multicolumn{1}{c|}{-}  & \multicolumn{1}{c|}{-}  & \multicolumn{1}{c|}{-}  & 68 \\ \cline{2-12}
& SAP\_10   & \multicolumn{1}{c|}{67} & \multicolumn{1}{c|}{77} & \multicolumn{1}{c|}{14} & \multicolumn{1}{c|}{4}  & \multicolumn{1}{c|}{-}  & \multicolumn{1}{c|}{-}  & \multicolumn{1}{c|}{-}  & \multicolumn{1}{c|}{-}  & \multicolumn{1}{c|}{-}  & 55 \\ \hline
\end{longtable}